\documentclass[a4paper,11pt]{article}

\pdfoutput=1

\usepackage{jheppub} 
\usepackage{graphicx}    
\usepackage{dcolumn}     
\usepackage{bm}          
\usepackage{amssymb}     
\usepackage{subfigure}
\usepackage{verbatim}
\usepackage{environ}
\usepackage{pythonhighlight}
\usepackage{fancyvrb}
\usepackage[english]{babel}
\usepackage{xcolor}
\usepackage{graphicx}
\usepackage{amsmath}
\usepackage{dsfont}
\usepackage[makeroom]{cancel}
\usepackage{url}
\usepackage{hyperref}
\usepackage{slashed}
\usepackage{mathabx}

\newcommand{\as}{\alpha_s}

\newcommand{\eps}{\epsilon}

\usepackage{xspace}

\newcommand{\eq}[1]{eq.~(\ref{#1})}
\newcommand{\mad}{\textsc{MadNkLO}\xspace}
\newcommand{\aMC}{\textsc{MadGraph5\_aMC@NLO}\xspace}

\def\({\left(} 
\def\){\right)}

\newcommand{\beq}{\begin{eqnarray}}
\newcommand{\eeq}{\end{eqnarray}}
\newcommand{\nnb}{\nonumber}

\newcommand{\Li}{\mbox{Li}}

\newcommand{\bS}[1]{{\bf S}_{#1}}
\newcommand{\bC}[1]{{\bf C}_{#1}}
\newcommand{\bHC}[1]{{\bf HC}_{#1}}
\newcommand{\Bn}{B}

\newcommand{\Rl}{R}
\newcommand{\mc}{\mathcal}
\newcommand{\Norm}{\mc{N}_1}
\newcommand{\hc}{{\rm \, hc}}

\newcommand{\bbS}[1]{\overline{\bf S}_{#1}}
\newcommand{\bbC}[1]{\overline{\bf C}_{#1}}

\newcommand{\bbHC}[1]{\overline{\bf HC}_{#1}}

\newcommand{\sig}{\sigma}

\newcommand{\gam}{\gamma}

\newcommand{\npo}{{n+1}}

\newcommand{\varsi}{\varsigma}

\newcommand{\appn}[1]{Appendix~\ref{#1}}

\newcommand{\one}{\, (\mathbf{1})}
\newcommand{\two}{\, (\mathbf{2})}
\newcommand{\otwo}{\, (\mathbf{12})}

\newcommand{\de}{\partial}

\def\eq#1{Eq.~(\ref{#1})}

\newcommand{\Z}[1]{\mc{Z}_{#1}}
\newcommand{\bZ}[1]{\bar{\mc{Z}}_{#1}}

\newcommand{\kk}[2]{\bar k_{#1}^{(#2)}}
\newcommand{\sk}[2]{\bar s_{#1}^{(#2)}}

\newcommand{\RR}{RR}

\newcommand{\bB}{\bar{\Bn}}
\newcommand{\bR}{\bar{\Rl}}

\newcommand{\zg}{({\rm 0g})}
\newcommand{\og}{({\rm 1g})}
\newcommand{\tg}{({\rm 2g})}
\newcommand{\qg}{({\rm 1g},{\rm qg})}
\newcommand{\gq}{({\rm 1g},{\rm gq})}
\newcommand{\Fm}{{\rm M}}
\newcommand{\F}{{\rm F}}
\newcommand{\I}{{\rm I}}
\newcommand{\inF}[1]{\theta_{{#1}\in\F}}
\newcommand{\inI}[1]{\theta_{{#1}\in\I}}
\newcommand{\inFm}[1]{\theta_{{#1}\in\Fm}}
\newcommand{\inFI}[1]{\theta_{{#1}\in\F\!,\I}}
\newcommand{\inFM}[1]{\theta_{{#1}\in\F\!,\Fm}}
\newcommand{\sFm}{\scriptscriptstyle{\Fm}}
\newcommand{\sF}{\scriptscriptstyle{\F}}
\newcommand{\sI}{\scriptscriptstyle{\I}}
\newcommand{\sX}{\scriptscriptstyle{\rm X}}

\newcommand{\bs}{\bar{s}}

\title{Advances in Local Analytic Sector Subtraction: massive NLO and elements of NNLO automation\sloppy}

\author[a]{Gloria Bertolotti,}
\author[b,c,d]{Giovanni Limatola,}
\author[d]{Paolo Torrielli,}
\author[d]{and Sandro Uccirati}

\affiliation[a]{Department of Physics and Astronomy, University of Sussex, Brighton BN1 9QH, UK}
\affiliation[b]{II. Institute for Theoretical Physics, Hamburg University, Luruper Chaussee 149,\\ D-22761 Hamburg, Germany}
\affiliation[c]{Deutsches Elektronen-Synchrotron DESY, Notkestr. 85, 22607 Hamburg, Germany}
\affiliation[d]{Dipartimento di Fisica, Universit\`a di Torino, and INFN, Sezione di Torino, Via P.~Giuria 1,\\I-10125 Torino, Italy}

\emailAdd{g.bertolotti@sussex.ac.uk}
\emailAdd{giovanni.limatola@desy.de}
\emailAdd{paolo.torrielli@unito.it}
\emailAdd{sandro.uccirati@unito.it}

\abstract{
In this article we present a number of developments within the scheme of Local Analytic Sector Subtraction for infrared divergences in QCD.
First, we extend the scheme to deal with next-to-leading-order (NLO) singularities related to massive QCD particles in the final state.
Then, we document a new implementation of the NLO subtraction scheme in the \mad automated numerical framework, which is constructed to host higher-order subtraction algorithms. In particular, we discuss improvements in its performances with respect to the original implementation.
Finally we describe the \mad implementation of the first elements relevant to Local Analytic Sector Subtraction at next-to-next-to-leading order (NNLO), in the case of massless QCD partons in the final state.
}

\preprint{
\begin{flushright}
DESY-25-042
\end{flushright}
}

\allowdisplaybreaks

\begin{document} 


\maketitle
\flushbottom


\section{Introduction}
\label{sec:intro}

The quest for a deeper understanding of fundamental particle interactions and the identification of potential new phenomena at colliders crucially rely on the availability of high-accuracy theoretical predictions for complex scattering processes.

As the accuracy required for the modern precision-physics collider programme well exceeds the leading-order approximation in the Standard Model perturbation theory, computations inevitably have to deal with the presence of infrared (IR) divergences. These arise in separate contributions to scattering cross sections, and are guaranteed to cancel \cite{Kinoshita:1962ur,Lee:1964is} when such contributions are properly combined for the prediction of physical observables.

The problem of IR divergences at next-to-leading order (NLO) in QCD perturbation theory was solved in full generality already in the late `90s \cite{Frixione:1995ms,Frixione:1997np,Catani:1996jh,Catani:1996vz,Nagy:2003qn} by means of IR subtraction.
Within this method, the universal long-distance behaviour of scattering amplitudes is leveraged to design functions, the so-called IR counterterms, which approximate real-radiation squared matrix elements in all of their IR-singular limits. The difference between matrix elements and counterterms is then by construction integrable point by point in the phase space, and lends itself to a stable numerical evaluation. On the other hand, the counterterms, subtracted from real-radiation matrix elements, are analytically integrated in the radiation phase space, and then added back to virtual loop contributions, ensuring in turn their IR finiteness.
The availability of numerical frameworks automating NLO IR subtraction schemes \cite{Campbell:1999ah,Gleisberg:2007md,Frederix:2008hu,Czakon:2009ss,Hasegawa:2009tx,Frederix:2009yq,Alioli:2010xd} crucially spurred a theoretical accuracy revolution in the early 2000s, which has been instrumental for the success of the precision-physics programme of the first runs of the Large Hadron Collider (LHC).

At next-to-NLO (NNLO) in QCD, IR subtraction is a very active field of research. On one side, NNLO accuracy is required nowadays for a wider and wider class of processes, to cope with the LHC precision needs. On the other hand, the complexity of the IR problem increases significantly with respect to the NLO case, owing to the much more involved structure of matrix elements and of their IR divergences. 
Several viable NNLO subtraction schemes have been formulated \cite{Frixione:2004is,GehrmannDeRidder:2005cm,Currie:2013vh,Somogyi:2005xz,Somogyi:2006da,Somogyi:2006db,Czakon:2010td,Czakon:2011ve,Anastasiou:2003gr,Caola:2017dug,Catani:2007vq,Grazzini:2017mhc,Boughezal:2011jf,Boughezal:2015dva,Boughezal:2015aha,Gaunt:2015pea,Cacciari:2015jma,Sborlini:2016hat,Herzog:2018ily,Magnea:2018hab,
Magnea:2018ebr,Capatti:2019ypt,TorresBobadilla:2020ekr} in the last two decades. This has progressively extended the scope of NNLO QCD results, nowadays reaching processes with five external light particles at Born level (see for instance \cite{Kallweit:2020gcp,Chawdhry:2021hkp,Czakon:2021mjy,Badger:2023mgf}), and even allowing the first differential predictions at next-to-NNLO (N$^3$LO) in QCD for simple processes \cite{Dreyer:2016oyx,Dreyer:2018qbw,Currie:2018fgr,Chen:2021isd,Chen:2021vtu,Billis:2021ecs,Camarda:2021ict,Chen:2022cgv,Chen:2022lwc}. Nevertheless, a fully general solution of the NNLO subtraction problem has not been reached yet, with plenty of room for improvement in the universality and versatility of the subtraction algorithms.

The method of Local Analytic Sector Subtraction \cite{Magnea:2018hab,Magnea:2020trj,
Bertolotti:2022ohq,Bertolotti:2022aih}
was designed to minimise the complexity of IR counterterms, and especially of their analytic integration over the radiative phase space. In \cite{Magnea:2018hab} the features of the method were introduced, while the main elements for analytic integration at NNLO were laid down in \cite{Magnea:2020trj}, for the case of massless QCD in the final state. In \cite{Bertolotti:2022aih}, such features enabled the first analytic proof of the cancellation of NNLO IR singularities for generic processes with massless QCD particles in the final state\footnote{The cancellation of NNLO singularities for the production of a generic number of gluons in quark-antiquark annihilation was proven with the nested soft-collinear subtraction scheme in \cite{Devoto:2023rpv}.}. In \cite{Bertolotti:2022ohq} the case of QCD in the initial state was considered at NLO, and the first implementation of the subtraction scheme was presented in the \mad automated framework \cite{Lionetti:2018gko,Hirschi:2019fkz,Becchetti:2020wof}.

The present article builds on the ingredients of \cite{Bertolotti:2022ohq,Bertolotti:2022aih}, presenting developments in different directions within the environment of Local Analytic Sector Subtraction. First, in Section \ref{sec:massive NLO}, we consider the subtraction of singularities due to QCD radiation from massive final-state particles at NLO. After defining and analytically integrating the relevant counterterms, we show how the resulting poles organise to match the known virtual one-loop singularities in presence of massive particles.
Appendices \ref{app: mappings}--\ref{app: IntMassiveApp} collect several technical details related to the analytic integration of massive counterterms. In particular, different parametrisations of the massive phase space are studied, with the aim of providing ingredients that will allow the integration of massive counterterms at NNLO, which is left for a future development. In Section \ref{sec:implementation} we document a new implementation of the Local Analytic Sector Subtraction scheme in the automated \mad framework. New features have been introduced to overcome the technical limitations of the original implementation of \cite{Bertolotti:2022ohq}, especially concerning numerical phase-space integration and execution time. We show a validation of this new implementation at NLO in Section \ref{sec:numNLO}, both for processes with massless and with massive QCD particles in the final state. \mad is constructed to streamline the generation of all the universal ingredients entering the computation of scattering cross sections at higher perturbative orders. In Section \ref{sec:massless NNLO} we report the \mad implementation of the first elements for Local Analytic Sector Subtraction at NNLO, in the case of massless final-state QCD particles. In particular, we focus on finiteness tests relevant to a production channel in the $e^+e^-\to jj$ process at NNLO.
Some details on NNLO subtraction are collected in Appendix \ref{app:NNLOnum}.
Finally, Section \ref{sec:end} contains conclusions and future perspectives.

We note that the Local Analytic Sector Subtraction scheme for massless final-state radiation at NNLO is also being implemented independently of \mad, in the numerical framework \cite{Chargeishvili:2024ggm,Chargeishvili:2024xuc,Kardos:2024guf}, where all NNLO final-state analytic integrals of \cite{Magnea:2020trj} were also independently verified.


\section{NLO subtraction with massive particles}
\label{sec:massive NLO}

In \cite{Bertolotti:2022ohq} the NLO Local Analytic Sector Subtraction for 
massless QCD partons in the final as well as in the initial state was described in detail.
In this section we extend the NLO subtraction procedure to the case of massive 
QCD particles in the final state. 

In general, an NLO QCD real squared matrix element $R$ is singular when the extra radiated parton with respect to Born level, labelled with $i$, is a soft gluon, namely it has vanishing energy, $k_i^0\to0$. We denote with $\bS{i}$ the operation of singling out such a soft configuration.
Moreover, if $R$ features at least two massless QCD partons $i$ and $j$, it develops a collinear singularity whenever their relative angle $\theta_{ij}$ approaches 0. The corresponding operator extracting this limit is denoted with $\bC{ij}$.

We will discuss in detail these two types of configurations in the next subsections, after introducing phase-space partitions to disentangle different singular regions.


\subsection{Phase-space partitions and counterterm definition}

We handle the presence of multiple singular regions in the real matrix elements by introducing partition (or sector) functions $\mc Z_{ij}$ \cite{Frixione:1995ms,Bertolotti:2022aih} for each pair $ij$ of massless 
QCD partons. Since sector functions are determined uniquely by massless partons, the presence 
of massive particles in the final state does not prevent us from using the same definitions as in  \cite{Bertolotti:2022aih} for partitions. 
Setting $q^\mu$ as the total momentum of the initial state, and 
$s_{qa} = 2 \, q\cdot k_a$, with $k_a$ the momentum of particle $a$, we thus define sectors as
\beq
\label{eq:Zofsigma}
\Z{ij}
\, \equiv \,
\frac{\sig_{ij}+\sig_{ji}}{\sum\limits_{k}\sum\limits_{l<k}(\sig_{kl}+\sig_{lk})}
\, ,
\qquad\quad
\sig_{ab}
\, \equiv \,
\inF{a} \, \inFI{b\,}\,
\frac{s_{qb}}{s_{ab}}
\, ,
\qquad\quad
\sum\limits_{i}\sum\limits_{j<i}
\Z{ij}
\, = \,
1
\, ,
\eeq
where $s_{ab}=2 \, k_a \cdot k_b$, and the sums run over all massless QCD partons present in the considered scattering process at real-emission level. We stress that $\Z{ij}$ is symmetric under label exchange.
The symbol $\theta_{\cal C}$ is $1$ ($0$) if condition $\cal C$ is (is not) fulfilled, so that 
\beq
&&
\inFm{a}  \, \equiv \, \mbox{$a$ is a massive particle in the final state,}
\nnb\\
&&
\inF{a} \,\, (\inI{a}) \, \equiv \, \mbox{$a$ is a massless particle in the final (initial) state,}
\nnb
\eeq
and $\theta_{a \, \in {\rm X,Y}} \equiv \theta_{a \, \in {\rm X}} + \theta_{a \, \in {\rm Y}}$, for $\rm X, Y = F, M, I$.

By construction, when acting with singular soft or collinear limits on a given sector function $\Z{ij}$, only $\bS{i}$, $\bS{j}$, and $\bC{ij}$ return a non-vanishing result. 
In other words, a partition $\Z{ij}$ is built to dampen all singularities of 
the real matrix element except the mentioned ones, together with their composite soft-collinear configurations. The action of singular limits on $\Z{ij}$  yields
\beq
\label{eq:sec fun lim}
&&
\bS{i} \,\Z{ij}
\, = \, 
\frac{\sig_{ij}}{\sum\limits_{k\ne i} \sig_{ik}}
\, ,
\qquad
\sum_{j\ne i} \, \bS{i} \, \mc Z_{ij}
=
1
\, ,
\qquad
\bC{ij} \, \Z{ij}
\, = \, 
1
\, ,
\qquad
\bS{i} \, \bC{ij} \, \Z{ij}
\, = \, 
1
\, .
\qquad
\eeq
In each phase-space sector $\Z{ij}$ the structure of real-radiation singularities is trivial, as it stems uniquely from at most one collinear and two soft singularities. This makes it straightforward to define an infrared counterterm relevant for partition $\Z{ij}$ as
\beq
\label{eq:cnt ij def}
K_{ij}
& \equiv &
\Big[\,
\bbS{i} + \bbS{j} + \bbC{ij}\big(1-\bbS{i}-\bbS{j}\big)
\Big] \, R \, \Z{ij}
\nnb\\
& \equiv &
\Big[\, \bbS{i}+\bbS{j}+\bbHC{ij} \Big] \, R \, \Z{ij}
\, .
\eeq
We highlight that the notation ($\bbS{i}$, $\bbS{j}$, and $\bbC{ij}$) used in \eq{eq:cnt ij def} for soft and collinear counterterms differs by means of a bar from the one ($\bS{i}$, $\bS{j}$, and $\bC{ij}$) used for the corresponding strict limit operations. As will be detailed in the next subsections, barred limits acting on $R$ differ from unbarred ones only by subleading-power terms. The latter guarantee counterterm kinematics to be on shell and Born momentum conserving for all radiative configurations. For this reason we also refer to the barred limits defining the counterterm as \emph{improved} limits. At variance with the case of matrix elements, the action of improved limits on NLO sector functions is defined to be identical to that of unimproved limits, i.e.~$\overline{\bf L} \, \Z{ij} = {\bf L} \, \Z{ij}$, for all soft, collinear, or soft-collinear limits $\bf L$.
Moreover, in \eq{eq:cnt ij def}, for any improved limit $\overline{\bf L}$, it is understood that $\overline{\bf L} \, R \, \Z{ij} = (\overline{\bf L} \, R) \, (\overline{\bf L} \, \Z{ij})  = (\overline{\bf L} \, R) \, ({\bf L} \, \Z{ij})$. 
Finally, the notation $\bbHC{}$ indicates a singular collinear configuration 
with all of its soft singularities subtracted off, namely a hard-collinear 
improved limit.

The subtracted real radiation is then given by 
\beq
R^{\, \rm sub}_{ij}
\, \equiv \,
R \, \mc Z_{ij}
-
K_{ij}
\, ,
\qquad\quad
R^{\, \rm sub}
\, \equiv \,
\sum_{i}\sum_{j<i} \, R^{\, \rm sub}_{ij}
\, = \,
R - K
\, ,
\eeq
where $R^{\, \rm sub}_{ij}$ and, in turn, $R^{\, \rm sub}$ are free of 
non-integrable singularities over the whole radiative phase space. 
The counterterm $K$ is defined as 
\beq
\label{eq:K_def}
K
\, \equiv \,
\sum_{i}\sum_{j<i} \, K_{ij}
\, = \,
\sum_{i} \, \inF{i}\,\bbS{i}\,R
+
\sum_{i}\sum_{j<i} \, \inF{i}\,\inFI{j}\,\bbHC{ij} \, R
\, ,
\eeq
where in the latter step we have used the sum rules  in \eq{eq:sec fun lim} 
to get rid of partition functions. 
The counterterm is a collection of soft and hard-collinear improved limits 
acting on the real matrix element $R$, to be integrated in the radiative phase space to match the explicit poles of the virtual 
matrix element $V$.

We note that the counterterm outlined in \eq{eq:K_def} is structurally identical to the one relevant to the fully massless case (see for instance Eq.~(2.31) of~\cite{Bertolotti:2022aih}). Physics-wise, its differences are only due to the fact that a soft gluon can now correlate massive colour sources, while a collinear splitting may involve the momentum of a massive particle in its parametrisation.
In the following sections we define such massive soft and hard-collinear counterterms, and perform their analytic integration over the radiative phase space. 


\subsection{Soft counterterm with massive particles}
\label{sec:soft massive}

The soft limit $\bS{i}$ of the real matrix element $R$ is
\beq
\label{eq:SiR}
\bS{i} \, R
\, = \,
- \, \Norm
\sum_{k \neq i}
\sum_{l \neq i} \,
\mc I_{kl}^{(i)} \,
\Bn_{kl} ( \{ k \}_{\slashed i} ) 
\, ,
\qquad\qquad
\mc I_{kl}^{(i)}
\, = \, 
\inF{i} \, f_i^g \, \frac{s_{kl}}{s_{ik} \, s_{il}}
\, ,
\eeq
where in the eikonal kernel $\mc I_{kl}^{(i)}$ the symbols $\inF{i} \, f_i^g$ 
specify that the contribution is null in QCD unless the soft parton is 
a final-state gluon. In particular, $f_a^x$ is meant as a Kronecker symbol prescribing the flavour of $a$ to be equal to $x$, with $x=q, \bar q, g$.

Indices $k$ and $l$ in \eq{eq:SiR} run over all initial-state and final-state, massless as well as massive QCD particles. The soft limit of QCD amplitudes is independent of 
the mass of the colour source, however mass effects appear at the 
squared-amplitude level, where the diagonal terms $\mc I^{(i)}_{aa}$, proportional 
to $s_{aa} = 2 \, m_a^2$, are non vanishing only for massive radiators. 
The soft kinematics $\{k\}_\slashed{i}$ represents the set of real-radiation 
momenta after removal of soft momentum $k_i^\mu$. 
The colour-correlated Born matrix element is schematically defined as
\beq
\Bn_{kl} 
\, = \, 
\langle {\mc A}_n^{(0)} | {\bf T}_k \cdot {\bf T}_l | {\mc A}_n^{(0)} \rangle 
\, ,
\eeq
where $|{\mc A}_n^{(0)}\rangle$ is the Born amplitude written as a ket in colour space  \cite{Catani:1996vz}, undergoing non-trivial  transformations under the action of the SU($N_c$) generators ${\bf T}_a$. Lastly, the normalisation coefficient $\Norm$ reads
\beq
\label{norm1}
\Norm 
\, = \, 
8 \pi \as \left( \frac{\mu^2 e^{\gam_E}}{4 \pi} \right)^{\eps} 
\, ,
\eeq
with $\as$ the strong coupling, $\mu$ the renormalisation scale, $\eps=2-d/2$ the dimensional regulator, and $\gam_E\sim0.57721$ the Euler-Mascheroni constant.

Exploiting colour algebra, together with the symmetry of the eikonal and of the colour-connected Born under exchange of indices, the soft limit in \eq{eq:SiR} can be equivalently rewritten as
\beq
\label{eq:SiR2}
\bS{i} \, R
\, = \,
- \, 
2 \, 
\Norm
\sum_{k \neq i}
\sum_{\substack{l \neq i \\ l < k }}
\mc E_{kl}^{(i)} \,
\Bn_{kl} ( \{ k \}_{\slashed i}) 
\, ,
\qquad\qquad
\mc E_{kl}^{(i)}
\, \equiv \, 
\mc I_{kl}^{(i)}
-
\frac12
\mc I_{kk}^{(i)}
-
\frac12
\mc I_{ll}^{(i)}
\, .
\eeq

In order to construct the corresponding counterterm $\bbS{i} \, R$, we 
introduce massive dipole mappings \cite{Catani:2002hc} of real-radiation 
momenta $\{k\}$ onto Born-level ones $\{\bar k\}$. 
Given three momenta $k_a$, $k_b$, $k_c$, with $k_a^2=0$, the mappings are 
defined as
\beq
\label{eq:mapping}
\{k\}
\, \to \, 
\{\bar k\}^{(abc)}
\, \equiv \,
\Big\{ 
\{k\}_{\slashed{a}\slashed{b}\slashed{c}}, 
\kk{b}{abc}, \kk{c}{abc} 
\Big\}
\, ,
\eeq
so that the Born momenta $\bar{k}_b^{(abc)} \equiv \bar k_b$ and $\bar{k}_c^{(abc)} \equiv \bar k_c$ 
satisfy mass-shell conditions $\bar{k}_b^2 = m_b^2$, $\bar{k}_c^2 = m_c^2$, as well as Born momentum conservation at each point of the real-emission phase space $\Phi_{\npo}$\footnote{Here and in the following, by $\Phi_j$ we indicate the phase space for $j$ QCD particles 
plus an arbitrary number of colourless particles.}.
The soft counterterm is then given by 
\beq
\bbS{i} \, R 
& \equiv &
\bbS{i}^{(0)} R 
+
\bbS{i}^{(\Fm)} R 
\, ,
\eeq
where $\bbS{i}^{(0)}R$ is the massless soft counterterm as in~\cite{Bertolotti:2022ohq} (see Eq.~(2.44)), while 
$\bbS{i}^{({\rm M})}R$ is the massive soft counterterm, defined as 
\beq
\label{eq:SiRdef}
\bbS{i}^{(\Fm)} R 
& = &
- \, 
2 \, \Norm
\sum_{k \neq i}
\sum_{\substack{l \neq i \\ l < k }} \,
\mc E_{kl}^{(i)} \,
\Big[\;
\Big(
\inFM{k} \, \inFm{l}
+
\inFm{k} \, \inFI{l}
\Big) \,
\bar{B}^{(ikl)}_{kl} 
\nnb\\[-4mm]
&&
\hspace{31mm}
+ \,
\inI{k} \, \inFm{l} \,
\bar{B}^{(ilk)}_{kl} 
\;\Big] 
\, ,
\eeq
where we have set $\bar{B}^{(iab)}_{kl} \equiv B_{kl}(\{\bar k\}^{(iab)})$.
The kinematics $\{\bar k\}^{(ikl)}$ and $\{\bar k\}^{(ilk)}$ entering the Born matrix elements, together with the properties of the mappings and the corresponding radiative phase spaces are discussed in full detail in~\appn{app: mappings}.


\subsection{Hard-collinear counterterm with a massive recoiler}
\label{sec:hard-collinear massive}

In the collinear $\bC{ij}$ limit, the real squared matrix element 
factorises as the product of a spin-correlated Born matrix element 
$B_{\mu\nu}$, times a universal Altarelli-Parisi  kernel 
$P_{ab}^{\mu\nu}$, which solely depends on the flavour of collinear 
partons $i$ and $j$: 
\beq
\label{eq:CijR1} 
\bC{ij} \, R
& = & 
\frac{\Norm}{s_{ij}} \, 
\bigg\{
\inF{i} \, \inF{j} \,
P_{ij, \sF}^{\mu\nu}(z_i) \,
B_{\mu\nu}
\big( \{k\}_{\slashed i \slashed j}, k_{[ij]} \big)
\\
&&
\hspace{10mm}
+ \,
\Big[
\inF{i} \, \inI{j} \,
\frac{P_{[ij]i, \sI}^{\mu\nu}(x_{[ij]})}{x_{[ij]}}\,
B_{\mu\nu}
\big( \{k\}_{\slashed i \slashed j}, x_{[ij]} \, k_{j} \big)
+
(i\leftrightarrow j)
\Big]
\bigg\}
\, ,
\nnb
\eeq
with ($\rm X = F, I$)
\beq
\label{eq:APkernels_munu}
P_{ij,{\sX}}^{\mu\nu}(z)
& = &
-
P_{ij,{\sX}}(z) \, g^{\mu\nu}
\, + \, 
Q_{ij,{\sX}}(z) \, T^{\mu\nu}_{\sX}
\, ,
\qquad
T^{\mu\nu}_{\sX}
\, \equiv \,
- g^{\mu\nu} 
+ 
(d-2) 
\frac{\widetilde{k}_{\sX}^\mu\,\widetilde{k}_{\sX}^\nu}{\widetilde{k}_{\sX}^2}
\, . 
\eeq
The Altarelli-Parisi kernels relevant for final-state and initial-state splittings are collected in \appn{app: coll ker}.
The notation used in \eq{eq:CijR1} for the Born kinematics means that $i$ and $j$ have 
been replaced by $[ij]$, 
while the spin indices $\mu$, $\nu$ correspond to the polarisations of $[ij]$. 
Momentum $\widetilde{k}^\mu_{\sX}$ is the 
transverse momentum of $k_i$ with respect to the collinear direction, and its expression for final-state radiation ($\rm X=F$) and initial-state radiation ($\rm X=I$) is again reported in \appn{app: coll ker}.
The quantities $z_i$ and $z_j=1-z_i$ quantify the longitudinal momentum fraction of the sibling partons $i$ and $j$ relative to the parent $[ij]$ in a final-state splitting. 
Similarly $x_i$, and $x_{[ij]}=1-x_i$ are the momentum fractions of siblings $i$, and $[ij]$ relative to the initial-state parent $j$.
In a Lorentz-invariant formulation, one can define 
\beq
z_i = \frac{s_{ir}}{s_{ir}\!+\!s_{jr}}
\, ,
\qquad\qquad
x_{i} = \frac{s_{ir}}{s_{jr}} 
\, ,
\eeq
whence the collinear limit acquires a dependence on the momentum $k^\mu_r$ of a recoiler particle, $r\neq i,j\equiv r_{ij}$, serving as a reference direction to define longitudinal fractions. The case of a massless final-state recoiler has been treated in \cite{Magnea:2018hab,Bertolotti:2022aih} within Local Analytic Sector Subtraction, and we refrain from discussing it further. In the following we shall then focus solely on massive recoilers.

In the soft-collinear $\bS{i} \, \bC{ij}$ limit, the Altarelli-Parisi kernel reduces to a ratio of longitudinal momentum fractions, and it becomes insensitive to spin correlations:
\beq
\label{eq:SiCijR1} 
\bS{i} \, \bC{ij} \, R 
& = & 
2 \, \Norm \, C_{f_j} \,
{\mc I}^{(i)}_{jr}
\,
\Bn ( \{ k \}_{\slashed i} )
\, ,
\eeq
where $C_{f_j} \equiv C_F \, f_j^{q,\bar q} + C_A \, f_j^g$ is the quadratic Casimir operator for parton $j$, we have set $f_j^{q,\bar q} \equiv f_j^q + f_j^{\bar q}$, and ${\mc I}^{(i)}_{jr}$ is the eikonal kernel defined in \eq{eq:SiR}.

Building on the collinear and soft-collinear expressions just presented, we can finally define the hard-collinear limit $\bHC{ij} = \bC{ij} \, \big( 1 - \bS{i} - \bS{j} \big)$ of the real matrix element $R$ 
as
\beq
\label{eq:HCijR1} 
\bHC{ij} \, R
& \equiv & 
\frac{\Norm}{s_{ij}} \, 
\bigg\{
\inF{i} \, \inF{j} \,
P_{ij, \sF}^{\mu\nu, \hc}(z_i) \,
B_{\mu\nu}
\big( \{k\}_{\slashed i \slashed j}, k_{[ij]} \big)
\\
&&
\hspace{10mm}
+ \,
\Big[
\inF{i} \, \inI{j} \,
\frac{P_{[ij]i, \sI}^{\mu\nu, \hc}(x_{[ij]})}{x_{[ij]}}\,
B_{\mu\nu}
\big( \{k\}_{\slashed i \slashed j}, x_{[ij]} \, k_{j} \big)
+
(i\leftrightarrow j)
\Big]
\bigg\}
\, .
\nnb
\eeq
The hard-collinear kernels $P_{ab,\sX}^{\mu\nu, \hc}$ are deduced from the corresponding Altarelli-Parisi kernels $P_{ab,\sX}^{\mu\nu}$ upon subtraction of all soft enhancements, and their explicit expressions are detailed in \appn{app: coll ker}.

Based on this structure, we are in position to define the hard-collinear counterterm as
\beq
\bbHC{ij} \, R 
& \equiv &
\bbHC{ij}^{(0)} R 
+
\bbHC{ij}^{(\Fm)} R 
\, ,
\eeq
where $\bbHC{ij}^{(0)}R$ is the counterterm introduced in~\cite{Bertolotti:2022ohq} (see Eq.~(2.47)) dealing with the case of a massless recoiler, while $\bbHC{ij}^{(\Fm)} R$ is the new hard-collinear counterterm in presence of a massive recoiler:
\beq
\label{eq:HCijRdef}
\bbHC{ij}^{(\Fm)} R
& \equiv &
\frac{\Norm}{s_{ij}} \,
\inFm{r} \, 
\bigg\{
\inF{i} \, \inF{j} \, 
P_{ij,\sF}^{\mu\nu,\hc}(z_i) \,
B_{\mu\nu}(\{\bar k\}^{(ijr)})
\\
&&
\qquad\qquad
+ \,
\Big[
\inF{i} \, \inI{j} \, 
\frac{P_{[ij]i, \sI}^{\mu \nu,\hc}(x)\!}{x} \,
B_{\mu\nu}(\{\bar k\}^{(irj)})
+
(i\leftrightarrow j)
\Big]
\bigg\}
\ ,
\nnb
\eeq
where $x = (s_{ij} + s_{jr} - s_{ir})/(s_{ij} + s_{jr})$, and the mappings are defined in~\appn{app: mappings}.


\subsection{Integration of the soft counterterm}

The integration of the soft counterterm over the radiative phase space generates poles up to $1/\eps^2$. Double poles stem from configurations where soft gluons are collinear to massless partons. As such, the coefficient of double poles is explicitly mass-independent. However, this coefficient is proportional to the number of available massless colour sources, as only those give rise to soft-collinear enhancements. One then expects double poles from an integrated massive-massless eikonal kernel to be half of the double poles stemming from a fully massless one. Single poles of soft origin are conversely expected to be explicitly sensitive to the presence of masses.

Evaluating the soft integrated counterterm $\bbS{i}^{(\Fm)} R$ entails to compute
\beq
\label{eq:intSimass}
\int d\Phi_\npo \,
\bbS{i}^{(\Fm)} R
& \equiv &
- \, 2 \, 
\frac{\varsi_\npo}{\varsi_n} 
\sum_{k \neq i}  \sum_{\substack{l \neq i \\ l < k }} \,
I_{\rm s,(M)}^{ikl}
\, .
\eeq
The structure of the soft integrated counterterm
$I_{\rm s,(M)}^{ikl}$ closely follows the one outlined in~\cite{Bertolotti:2022ohq} (see Eq.~(3.5) there), where now
\beq
\label{eq:IsMassive}
I_{\rm s,(M)}^{ikl}
& \equiv &
\int d\Phi_n^{(ikl)}  \, 
\bigg[
\inFm{k} \, \inFm{l} \,
I_{\rm s, \sFm\sFm}^{ikl} \,
+
\inFm{k} \, \inF{l} \,
I_{\rm s, \sFm\sF}^{ikl} \,
+
\inF{k} \, \inFm{l} \,
I_{\rm s, \sF\sFm}^{ikl} \,
\bigg] \,
\bar{B}^{(ikl)}_{kl} 
\nnb\\
&&
+
\sum_{ab \, = \, kl,\, lk} \,
\inFm{a} \,\inI{b} \,
\bigg[
\,
\int d\Phi_{n}^{(iab)}(k_b) \,
I_{\rm s, \sFm\sI}^{iab} \,
\bar{B}^{(iab)}_{ab}(k_b)
\nnb\\
&&
\hspace{30mm}
+ \,
\int_0^1 \frac{dx}{x} 
\int d\Phi_{n}^{(iab)}(x k_b) \,
J_{\rm s, \sFm\sI}^{iab}(x) \,
\bar{B}^{(iab)}_{ab}(xk_b)
\bigg]
\, .
\eeq
The computation of the fully massive integrated soft counterterm 
$I_{\rm s,\sFm\sFm}^{ikl}$ is performed in \appn{app: IsFmFm}, 
based on the phase-space parametrisations detailed in Appendix 
\ref{app: mappings}.
The result reads
\beq
\label{eq:ISMM}
I_{\rm s,\sFm\sFm}^{ikl}
& = &
\frac{\as}{2\pi} \, 
\left(
\frac{\bar s_{kl}}{\mu^2}
\right)^{-\eps} \!
f_i^g \,
\frac{1}{\eps}\,
\bigg[
\frac{\bar s_{kl}}{2\sqrt{\lambda}} \,
\ln\frac{\bar s_{kl}-\sqrt{\lambda}}{\bar s_{kl} + \sqrt{\lambda}} + 1
\bigg]
+ \mc O(\eps^0)
\, ,
\eeq
with 
$\bar s_{kl}\equiv\sk{kl}{ikl} \equiv 2\,\kk{k}{ikl}\cdot \kk{l}{ikl}$ 
and $\lambda = \bar s_{kl}^2 - 4 \, m_k^2 \, m_l^2$.
The complete result, including finite terms in $d\to4$, can be found 
in \eq{eq: IsMM}. 
Upon inserting this result into \eq{eq:intSimass} and applying colour algebra, it is straightforward to check that the known structure of one-loop poles is reproduced (up to a global sign) in the case of QCD amplitudes with no massless partons \cite{Catani:2000ef}.

The massive-massless integrals $I_{\rm s,\sFm\sF}^{ikl}$, and $I_{\rm s,\sF\sFm}^{ikl}$ 
depend on a single mass and, thanks to the symmetry of the eikonal kernel, have the same expression, see also Appendix \ref{app: Fm-F and F-Fm mappings}. The detail of their computation is reported in \appn{app: IsFmF, IsFFm} and gives  
\beq
\label{eq:two massive massless}
I_{\rm s,\sFm\sF}^{ikl}
\, = \,
f_i^g \,
I_{\rm s}^{\sFm\sF}\big(\sk{kl}{ikl},m_k\big)
\, ,
\qquad
I_{\rm s,\sF\sFm}^{ikl}
\, = \,
f_i^g \,
I_{\rm s}^{\sFm\sF}\big(\sk{kl}{ikl},m_l\big)
\, ,
\eeq
with
\beq
\label{eq:ISMF}
I_{\rm s}^{\sFm\sF}(s,m)
& = &
\frac{\as}{2\pi}
\left(
\frac{s}{\mu^2}
\right)^{-\eps}
\,
\bigg[\,
\frac{1}{2\eps^2}
+
\frac{1}{2\eps}
\left( 3 + \ln\frac{m^2}{s} \right)
\bigg]
+
{\cal O}(\eps^0)
\, .
\eeq
The complete result, including finite terms in $d\to4$, can be found 
in \eq{eq: IsMF}. 

Finally, the final-initial integrated counterterms $I_{\rm s,\sFm\sI}^{iab}$, 
$J_{\rm s,\sFm\sI}^{iab}$, with $ab=kl, lk$, as well depend on a single mass and have an identical expression. 
Their explicit computation can be found in \appn{app: IsFmF, IsFFm} and the 
result reads 
\beq
&&
I_{\rm s,\sFm\sI}^{iab} 
\, = \,
f_i^g \,
I_{\rm s}^{\sFm\sI}\big(\sk{kl}{iab},m_a\big)
\, ,
\qquad
J_{\rm s,\sFm\sI}^{iab}(x) 
\, = \, 
f_i^g \,
J_{\rm s}^{\sFm\sI}\big(\sk{kl}{iab},x,m_a\big)
\, ,
\eeq
where
\beq
\label{eq:ISMI}
I_{\rm s}^{\sFm\sI}(s,m)
& = &
\frac{\as}{2\pi}
\left(
\frac{s}{\mu^2}
\right)^{-\eps}
\,
\bigg[\,
\frac{1}{2\eps^2}
+
\frac{1}{2\eps}
\left( 3 + \ln\frac{m^2}{s} \right)
\bigg]
+
{\cal O}(\eps^0)
\, ,
\\
J_{\rm s}^{\sFm\sI}(s,x,m)
& = &
\frac{\as}{2\pi}
\left(
\frac{s}{\mu^2}
\right)^{-\eps}
\,
\bigg(\!\!
-\frac{1}{\eps}
\bigg)
\bigg[ \frac{x}{1-x} \bigg]_+
+ {\cal O}(\eps^0)
\, ,
\eeq
and we have defined the usual `plus' distribution as
\beq
\bigg[ \frac{g(x)}{1-x} \bigg]_+
h(x)
\, \equiv \,
\bigg[ \frac{g(x)}{1-x} \bigg]
\Big(
h(x)-h(1)
\Big)
\, ,
\eeq
being $h(x)$ an arbitrary test function which is regular in $x=1$.
The complete expressions for $I_{\rm s}^{\sFm\sI}$ 
and $J_{\rm s}^{\sFm\sI}$, including the finite terms in $d\to4$, are 
given in \eq{eq: IsMI}.

We notice that the pole structures of $I_{\rm s}^{\sFm\sF}$ and of $I_{\rm s}^{\sFm\sI}$ are identical, and their double poles are half of those found in the fully massless case, as anticipated, see for instance Eqs.~(D.5) and (D.6) of \cite{Bertolotti:2022ohq}.
The single pole in $J_{\rm s}^{\sFm\sI}(s,x,m)$ is equal in the massive as 
well as in the massless case, see for instance Eq.~(D.8) of \cite{Bertolotti:2022ohq}. This is again expected, as such a pole is directly related to the factorisation of initial-state collinear singularities and PDF renormalisation, which is by definition mass-independent.


\subsection{Integration of the hard-collinear counterterm}

As hard-collinear NLO counterterms do not feature soft singularities, by construction, their phase-space integral results in single $1/\eps$ poles at most. Moreover, since collinear limits for massive particles are non-singular, such single poles of collinear origin are not affected by the presence of masses. It follows then that the integration of hard-collinear terms in presence of massive recoilers must produce the same poles as in the fully massless case.

The integral of the hard-collinear counterterm with a massive recoiler 
$r$ is
\beq
\label{eq:intHC}
\int \! d\Phi_\npo \,
\bbHC{ij}^{(\Fm)} R
& \equiv &
\frac{\varsi_\npo}{\varsi_n} \,
\inFm{r} \, 
\bigg\{
\inF{i} \, \inF{j}
\int d\Phi_n^{(ijr)} \,
I_{\rm hc, \sFm}^{ijr}(m_r) \,
\bar{B}^{(ijr)}
\\
&&
+ \!
\sum_{ab \, = \, ij, \, ji}
\Big[ 
\inF{a} \, \inI{b}
\int_0^1 \frac{dx}{x} \int d\Phi_n^{(arb)}(xk_b) \,
J_{\rm hc, \sFm}^{arb}(x,m_r) \,
\bar{B}^{(arb)}(xk_b)
\Big]
\bigg\}
\, .
\nnb
\eeq
In writing \eq{eq:intHC} we have exploited that the azimuthal tensor $T^{\mu\nu}_{\sX}$ defined in \eq{eq:APkernels_munu} averages to 0 when integrated over the radiative phase space, see \cite{Catani:1996vz} or Appendix D of \cite{Bertolotti:2022aih}. This allows to reduce the integration of hard-collinear counterterms to that of spin-averaged Altarelli-Parisi kernels $P_{ab,\sX}^{\rm hc}$, which simply multiply the Born matrix element $B = -g^{\mu\nu}B_{\mu\nu}$.

The constituent integrals $I_{\rm hc, \sFm}^{ijr}(m_r)$ and $J_{\rm hc, \sFm}^{arb}(x,m_r)$ can be decomposed according to their flavour content as
\beq
\label{eq:IHCandJHCdecomp}
I_{\rm hc, \sFm}^{ijr}(m_r)
& \equiv &
I_{\rm hc, \sFm}^{\zg}\big(\sk{jr}{ijr},m_r\big) \,
f_{ij}^{q \bar q}
+
I_{\rm hc, \sFm}^{\og}\big(\sk{jr}{ijr},m_r\big) \,
\big[
f_{i}^{q,\bar q} \,f_{j}^{g}
+
f_{j}^{g} \, f_{i}^{q,\bar q}
\,\big]
\nnb\\
&&
+ \,
I_{\rm hc, \sFm}^{\tg}\big(\sk{jr}{ijr},m_r\big) \,
f_{ij}^{gg}
\, ,
\nnb\\[3pt]
J_{\rm hc, \sFm}^{arb}(x,m_r) 
& = &
J_{\rm hc, \sFm}^{\zg}\big(\sk{br}{arb},x,m_r\big) \,
f_{ab}^{q \bar q}
+
J_{\rm hc, \sFm}^{\qg}\big(\sk{br}{arb},x,m_r\big) \,
f_{a}^{q,\bar q}\,f_{b}^{g}
\nnb\\
&&
+ \,
J_{\rm hc, \sFm}^{\gq}\!\big(\sk{br}{arb}\!\!,x,m_r\big) \,
f_{a}^{g} \,f_{b}^{q,\bar q}
+
J_{\rm hc, \sFm}^{\tg}\!\big(\sk{br}{arb}\!\!,x,m_r\big) \,
f_{ab}^{gg}
\, ,
\eeq
where we have set 
$f_{ij}^{q \bar q} \equiv f_i^q f_j^{\bar q} + f_i^{\bar q} f_j^q$, 
and $f_{ij}^{gg} \equiv f_i^g f_j^g$. 
Their computation is described in detail in \appn{app: IhcFmFandI}. 
For $I_{\rm hc, \sFm}^{ijr}$ we get 
\beq
\label{eq:Ihcpoles}
I_{\rm hc,\sFm}^{\zg}(s,m)
& = &
\frac{\as}{2\pi}
\left( \frac{s}{\mu^2}\right)^{-\eps} 
\frac23 \,
T_R \, 
\bigg( \!\!
- 
\frac{1}{\eps} 
\bigg)
+
{\cal O}(\eps^0)
,
\nnb\\
I_{\rm hc,\sFm}^{\og}(s,m)
& = &
\frac{\as}{2\pi}
\left( \frac{s}{\mu^2}\right)^{-\eps} 
\frac12 \,
C_F \, 
\bigg( \!\!
- 
\frac{1}{\eps} 
\bigg)
+
{\cal O}(\eps^0)
\, ,
\nnb\\
I_{\rm hc,\sFm}^{\tg}(s,m)
& = &
\frac{\as}{2\pi}
\left( \frac{s}{\mu^2}\right)^{-\eps} 
\frac13 \,
C_A \, 
\bigg( \!\!
- 
\frac{1}{\eps} 
\bigg)
+
{\cal O}(\eps^0)
\, ,
\eeq
whose finite terms in $d \to 4$ dimensions can be found in \eq{eq: IhcM}. 
The poles of $I_{\rm hc,\sFm}^{(n\rm g)}(s,m)$ are explicitly mass-independent, and coincide with the fully massless case, as anticipated, see for instance Eqs.~(D.12)--(D.14) of \cite{Bertolotti:2022ohq}.
As for the final-initial counterterms $J_{\rm hc,\Fm}^{arb}$ we have 
\beq
J_{\rm hc,\sFm}^{\zg}(s,x,m)
& = &
\frac{\as}{2\pi}
\left( \frac{s}{\mu^2}\right)^{-\eps}
T_R \,
\big[x^2 + (1 - x)^2 \big] \,
\bigg( \!\!
-
\frac{1}{\eps}
\bigg)
+
{\cal O}(\eps^0)
\, ,
\nnb\\
J_{\rm hc,\sFm}^{\qg}(s,x,m)
& = &
\frac{\as}{2\pi}
\left( \frac{s}{\mu^2}\right)^{-\eps} 
C_F \, (1-x) \,
\bigg( \!\!
-
\frac{1}{\eps}
\bigg)
+
{\cal O}(\eps^0)
\, ,
\nnb\\
J_{\rm hc,\sFm}^{\gq}(s,x,m)
& = &
\frac{\as}{2\pi}
\left( \frac{s}{\mu^2}\right)^{-\eps} 
C_F \,
\bigg[
\frac{1+(1-x)^2}{x}
\bigg] \,
\bigg( \!\!
-
\frac{1}{\eps}
\bigg)
+
{\cal O}(\eps^0)
\, ,
\nnb\\
J_{\rm hc,\sFm}^{\tg}(s,x,m)
& = &
\frac{\as}{2\pi}
\left( \frac{s}{\mu^2}\right)^{-\eps}
2 \, C_A \,
\bigg[
\frac{1-x}{x} + x(1-x)
\bigg]
\,
\bigg( \!\!
-
\frac{1}{\eps}
\bigg)
+
{\cal O}(\eps^0)
\, .
\eeq
The complete expressions, up to finite parts in $d \to 4$ dimensions, are 
collected in \eq{eq: JhcM}.
As expected, we notice mass-independence in the pole structure of 
$J_{\rm hc,\sFm}^{(n\rm g)}(s,x,m)$ as well. 
The pole coefficients just reported coincide with those relevant for the 
fully massless case, see for instance Eqs.~(D.24)--(D.28) of 
\cite{Bertolotti:2022ohq}.


\subsection{Cancellation of poles}

In the previous sections we have presented the poles stemming from the integration of soft and hard-collinear counterterms, respectively. We are now in position to check that such poles organise to match the ones of massive one-loop virtual contributions, achieving the desired subtraction.

We start with the $x$-independent part of the integrated soft counterterm, reading its  poles from Eqs.~(\ref{eq:ISMM}), (\ref{eq:ISMF}), and (\ref{eq:ISMI}), together with Eqs.~(3.5) and (D.5)--(D.7) of \cite{Bertolotti:2022ohq}:
\beq
\label{eq:intSoftmass_ples}
\hspace{-8mm}
\sum_{i} \, \inF{i}
\int d\Phi_\npo \,
\bbS{i} \, R \, 
\Big|_{x-\rm indep.}
& \equiv &
\int d\Phi_n \,
I_{\rm s}
\, ,
\eeq
where
\beq
I_{\rm s}
& = &
-
\frac{\as}{2\pi}
\sum_{k,l\neq k}
\left(\frac{\bar s_{kl}}{\mu^2}\right)^{-\eps}
\bigg\{
\inFm{k} \, \inFm{l} \,
\frac{1}{\eps}\,
\bigg[
\frac{\bar s_{kl}}{2\sqrt{\lambda}}\ln\eta + 1
\bigg]
+
\inFI{k} \, \inFI{l} \,
\bigg[\,
\frac{1}{\eps^2}
+
\frac{2}{\eps}
\bigg]
\\
&&
\hspace{30mm}
+ \,
\inFm{k} \, \inFI{l} \,
\bigg[\,
\frac{1}{\eps^2}
+
\frac{1}{\eps}
\Big(
3 + \ln\frac{m_k^2}{\bar s_{kl}}
\Big)
\bigg]
\bigg\}
\, B_{kl}
+ \mc O(\eps^0)
\nnb\\
& = &
-
\frac{\as}{2\pi}
\sum_{k,l\neq k}
\left(\frac{\bar s_{kl}}{\mu^2}\right)^{-\eps}
\bigg\{
\inFm{k} \, \inFm{l} \,
\frac{1}{\eps}\,
\frac{\bar s_{kl}}{2\sqrt{\lambda}}\ln\eta
+
\inFI{k} \, \inFI{l} \,
\frac{1}{\eps^2}
\nnb\\
&&
\hspace{30mm}
+ \,
\inFm{k} \, \inFI{l} \,
\bigg[\,
\frac{1}{\eps^2}
+
\frac{1}{\eps}
\ln\frac{m_k^2}{\bar s_{kl}}
\bigg]
+
\inFm{k} \, \frac1\eps
+
\inFI{k} \, \frac2\eps
\bigg\}
\, B_{kl}
+ \mc O(\eps^0)
\, .
\nnb
\eeq
The terms where $k$ is massive precisely reconstruct, up to a sign, the analogous terms in the virtual one loop contribution, see Section 3.1 of \cite{Catani:2000ef}. In particular, one can use colour algebra to cast the last one as a colour-uncorrelated Born:
\beq
\sum_{k,l\neq k}
\inFm{k} \, \frac1\eps \,
B_{kl}
\, = \,
-
\sum_{k}
\inFm{k} \, \frac1\eps \,
C_F \, B
\, .
\eeq
In order to proceed with the check of pole cancellation, one should combine the remaining fully massless soft terms with the hard-collinear poles, and add $x$-dependent poles. However, since 
$x$-dependent soft contributions are identical to their fully massless counterparts, and hard-collinear integrated poles are insensitive to the mass of the chosen recoiler, the cancellation for these terms works in the same way as in the fully massless case. This was shown in detail in \cite{Bertolotti:2022ohq}, and we refrain from repeating the same argument here.
As a consequence, the singularity structure of Section 3.1 of \cite{Catani:2000ef} is reproduced for an arbitrary pattern of QCD particle masses. The same happens for the PDF counterterm, which is proportional to regularised Altarelli-Parisi splitting functions.


\section{Automated Local Analytic Sector Subtraction in \mad}
\label{sec:implementation}

A milestone in the completion of the Local Analytic Sector Subtraction programme is the numerical implementation and validation of the algorithm. The universality of the subtraction scheme makes it naturally suited to be incorporated in an automated Monte Carlo event generator.

A first effort in this direction was undertaken in~\cite{Bertolotti:2022ohq},
where the NLO subtraction formula for massless QCD radiation was implemented within \mad \cite{Lionetti:2018gko,Hirschi:2019fkz,Becchetti:2020wof}. The latter is a Python-based framework  designed to automate the generation and handling of local subtraction terms at higher orders in perturbation theory. It builds on the \aMC package\footnote{Specifically, \mad relies on \aMC for the generation of tree-level and one-loop matrix elements, the latter being handled by the {\sc MadLoop} module~\cite{Hirschi:2011pa}.}~\cite{Alwall:2014hca,Frederix:2018nkq}, with which it shares the same philosophy of full automation.

The work in~\cite{Bertolotti:2022ohq} allowed on one hand to thoroughly validate the NLO subtraction formula at the numerical level. On the other hand, numerical performances were severely affected by intrinsic limitations of the computational environment, namely the absence of a low-level code implementation, and of optimised phase-space integration routines\footnote{Originally, the \mad integration was steered by the Python version of {\sc Vegas3}~\cite{PETERLEPAGE1978192,Lepage:1980dq,Lepage:2020tgj}.}.
As an example, the NLO validation collected in Table 1 of \cite{Bertolotti:2022ohq}, just concerning inclusive cross-section results, required cluster resources.

To overcome such limitations, we have developed a new, improved version of the \mad framework. Key novelties include the construction of Python-to-Fortran meta-coding structures, and the introduction of an optimised phase-space integrator. Specifically, we highlight the following features of the new code.
\begin{itemize}
\item The original, fully general, \mad Python framework is employed to fill Fortran templates for matrix elements and counterterms at process-generation level. This allows to rely uniquely on fast Fortran routines at run time, in the same spirit of what is done in \aMC.
\item Different contributions (e.g.~Born, real radiation, etc.) to the cross sections, as well as different phase-space sectors within the same contribution, are treated in a fully independent manner. This allows for a fully parallelised execution of all of these elements.
\item A single-diagram-enhanced multi-channelling strategy \cite{Maltoni:2002qb} is adopted, based on Born diagrams. This is combined with a sector-aware parametrisation of the radiative phase space, ensuring efficient integration of subtraction terms.
\item Within a given cross-section contribution/sector, the fraction of phase-space points attributed to a single channel is adapted to the relative contribution of the latter to the total error budget.
\item Full support for differential distributions is now built-in.
\end{itemize}

As a first step, we have implemented in the new \mad environment the NLO subtraction formula for final-state QCD radiation, including the massive case detailed in Section \ref{sec:massive NLO}.
The mentioned optimisations have allowed to gain insight about the computational performances of the new implementation, as compared to other available event generators. New results for fiducial and differential cross sections, as well as their comparison with \aMC benchmark results, are presented in Section \ref{sec:numNLO}. 

As the numerical framework is constructed to support higher-order extensions, we have also started implementing in \mad the first elements of Local Analytic Sector Subtraction at NNLO, as provided in \cite{Bertolotti:2022aih}. We have considered leptonic collisions with only massless QCD partons in the final state. As a first case study, we provide in Section \ref{sec:massless NNLO} a proof-of-concept numerical analysis of the subtraction algorithm, focusing on a specific double-real channel of di-jet production in electron-positron annihilation at NNLO.


\subsection{Numerical validation at NLO}
\label{sec:numNLO}

We validate the NLO Local Analytic Sector Subtraction and its \mad implementation through a comparison with \aMC \cite{Alwall:2014hca,Frederix:2018nkq}, which is based \cite{Frederix:2009yq} on the FKS subtraction scheme \cite{Frixione:1995ms,Frixione:1997np}. We focus on leptonic scatterings, considering the production of two or three hadronic jets, as well as of a pair of heavy quarks at NLO.
Collisions are simulated for a centre-of-mass energy $\sqrt{s}=1$ TeV, setting the renormalisation scale as $\mu_R=M_Z=91.188$ GeV, with $\alpha_s(M_Z)=0.118$.
Jets are defined via the anti-$k_t$ clustering algorithm \cite{Cacciari:2008gp} as implemented in {\sc Fastjet} \cite{Cacciari:2011ma}, with a radius $R=0.4$ and requiring the jets' transverse momenta and pseudo-rapidities to satisfy $p_{tj}>20$ GeV, $|\eta_j|<5$.

In Table \ref{tab:valid} 
\begin{table}
\begin{center}
\begin{tabular}{|c|c|c|c|c|}
\hline
Process &  LO   & LO       & NLO corr. & NLO corr. \\
        &  \mad & MG5\_aMC & \mad      & MG5\_aMC  \\
\hline
$e^+e^- \to jj$          & 0.53149(3)   & 0.5314(2)  & 0.02000(3)    & 0.0201(4)     \\     
\hline
$e^+e^- \to jjj$         & 0.4743(2)    & 0.4736(5)  & -- \!0.146(1) & -- \!0.144(2) \\     
\hline
$e^+e^- \to t \bar{t}$   & 0.16644(1)   & 0.1663(2)  & 0.010203(3)   & 0.0101(2)     \\ 
\hline
$e^+e^- \to b \bar{b}$   & 0.0923258(6) & 0.09233(3) & 0.003471(2)   & 0.00348(9)    \\ 
\hline
$e^+\nu_e \to t \bar{b}$ & 0.341187(2)  & 0.34118(9) & 0.018226(5)   & 0.0180(2)     \\
\hline
\end{tabular}
\end{center}
\caption{\label{tab:valid}
Inclusive results obtained for LO cross sections and NLO corrections with \mad and \aMC.
Numbers are in pb. Integration errors, affecting the last digit, are shown in parentheses.}
\end{table}
we report inclusive cross sections for the considered processes. We display separately Born results and relative ${\cal O}(\as)$ NLO corrections, as obtained with both \mad and \aMC using the same setup. In the table and in the figures below, the latter code is referred to as MG5\_aMC, for brevity. Full agreement is observed between the results of the two numerical frameworks.
We emphasise that the conditions under which the two codes were executed to obtain the results in Table \ref{tab:valid} were not identical. First, LO \aMC results were obtained from the NLO \aMC interface, in which the integration of the Born cross section is not as optimised as in the pure-LO interface. As for NLO results, while in \mad all real-radiation sectors, as well as virtual corrections, are integrated separately, \aMC treats different terms together. This makes it non-trivial to perform a fair comparison in terms of number of sampled random points per contribution.
Still, the numbers in Table \ref{tab:valid} are obtained with a comparable run time for the two codes, on the same multi-core machine.
\mad is seen to provide results of a similar, if not superior, statistical quality with respect to \aMC, for all the simple processes considered.
Next, we document a validation at the level of differential distributions, focusing on observables sensitive to QCD radiation. In the following figures we display NLO corrections only, namely we turn off all Born contributions, in order to highlight the comparison between different NLO subtraction schemes and implementations.

\begin{figure}
\centering
\includegraphics[width=0.45\textwidth]{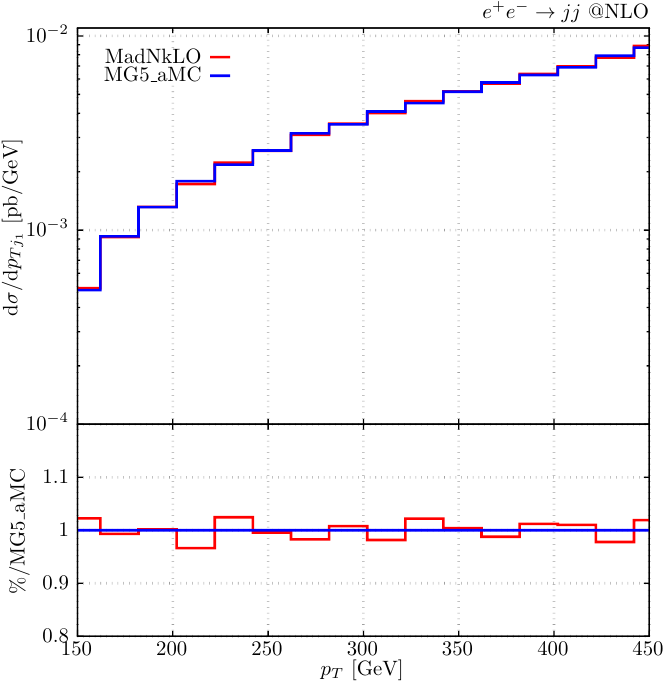}
\includegraphics[width=0.45\textwidth]{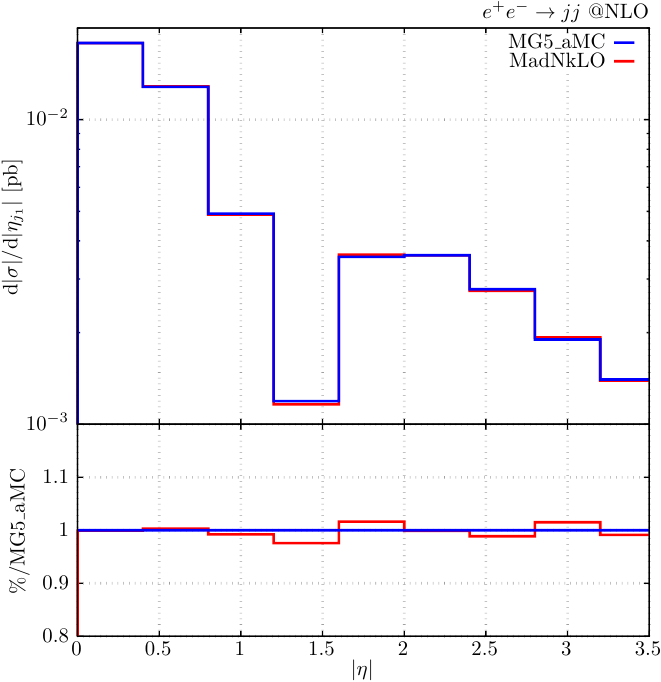}
\includegraphics[width=0.45\textwidth]{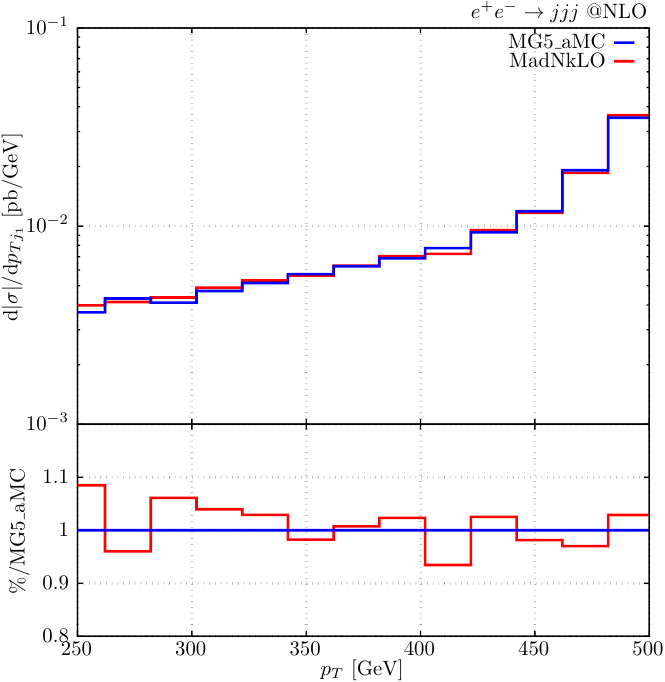}
\includegraphics[width=0.45\textwidth]{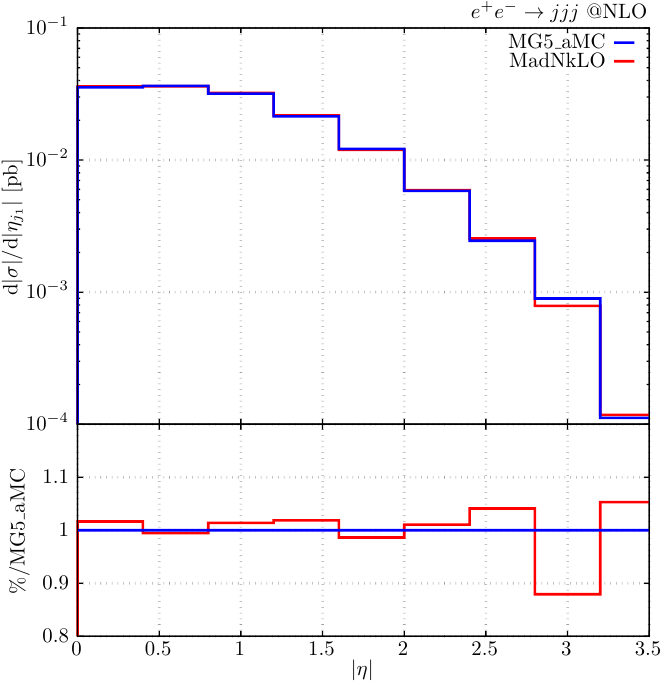}
\caption{NLO correction to the transverse momentum (left column) and pseudo-rapidity (right column) of the leading jet for two-jet (upper row) and three-jet (bottom row) production in $e^+e^-$ annihilation. Blue and red curves are obtained with \aMC and \mad, respectively. The insets display bin-by-bin ratios to the \aMC baseline.}
\label{fig:MKvsMG-2j}  
\end{figure}

We start with the leptonic production of two and three jets at NLO, presenting in Fig.~\ref{fig:MKvsMG-2j} 
the transverse momentum ($p_{Tj_1}$) and pseudo-rapidity ($\eta_{j_1}$) of the leading jet. Since we are considering NLO corrections alone, distributions are not necessarily positive-definite, in which case we plot the absolute value of the differential cross sections.
Full agreement is observed between \mad and \aMC over the whole ranges displayed. Statistical fluctuations are comparable for two-jet production, while slightly more pronounced for \mad in the three-jet case. This may be related to the coexistence of different soft phase-space mappings within each partition, which will be the subject of a future dedicated analysis, beyond the scope of this article.

We then turn to the validation of massive QCD processes at NLO. We consider the leptonic production of $t\bar t$, $b \bar b$, and $t \bar b$ final states, which allows to probe the massive subtraction of Section \ref{sec:massive NLO} in different energy regimes.

\begin{figure}[ht]
  \centering
  \includegraphics[width=0.45\textwidth]{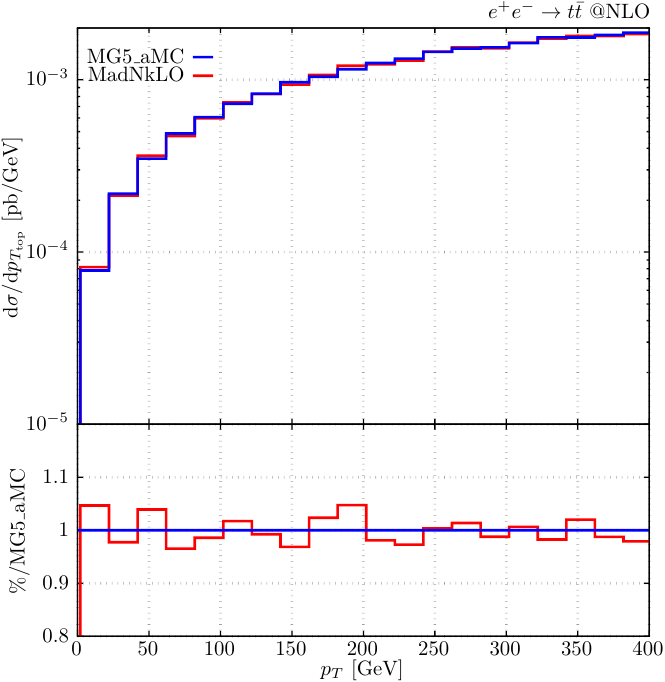}
  \includegraphics[width=0.45\textwidth]{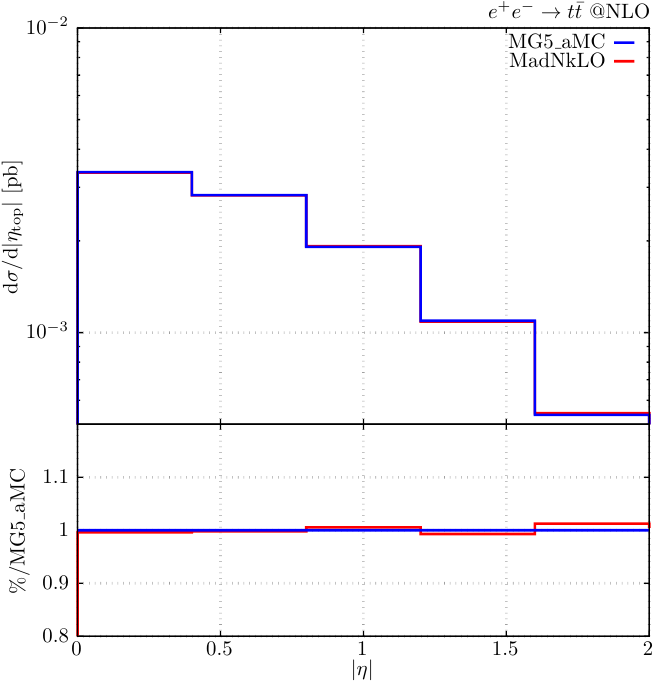}
    \includegraphics[width=0.45\textwidth]{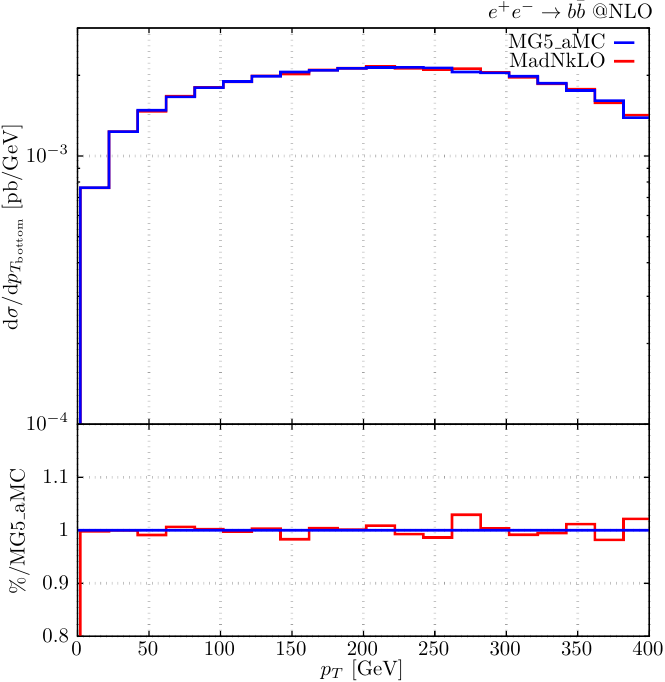}
    \includegraphics[width=0.45\textwidth]{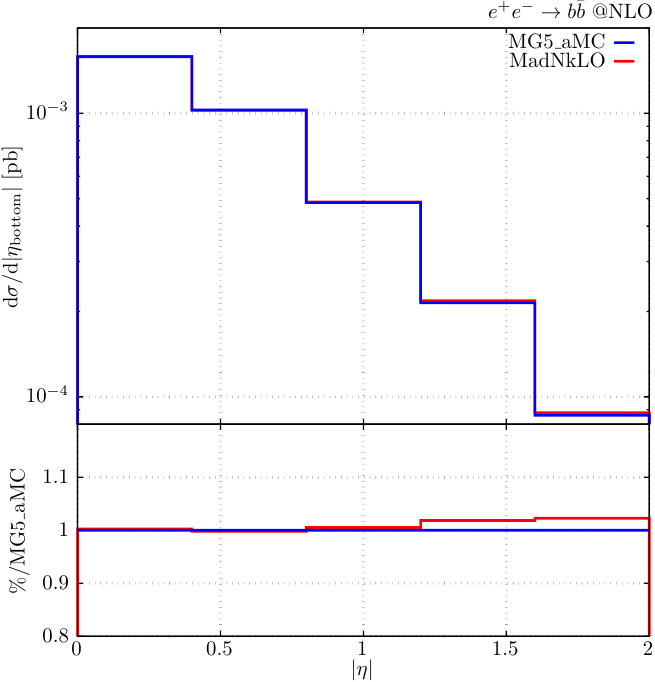}
    \caption{Same as in Fig.~\ref{fig:MKvsMG-2j} for $t\bar{t}$ (upper row) and $b\bar{b}$ (bottom row) leptonic production.}  \label{fig:MKvsMG-ttb}  
\end{figure}

\begin{figure}
  \centering
    \includegraphics[width=0.45\textwidth]{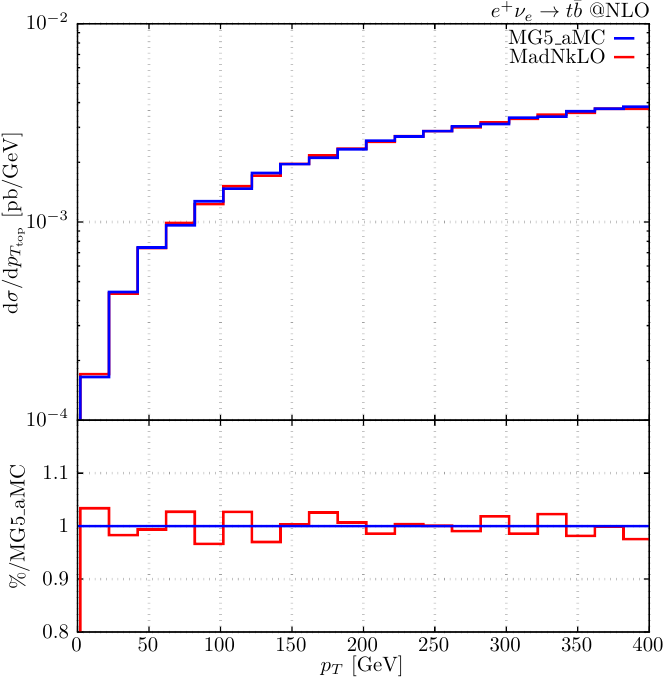}
    \includegraphics[width=0.45\textwidth]{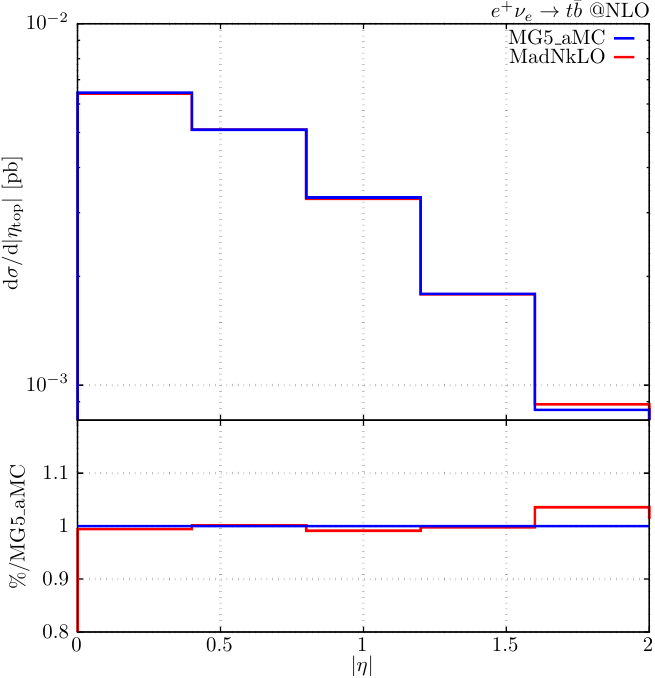}
    \caption{Same as in Fig.~\ref{fig:MKvsMG-2j} for the charged-current process $e^+\nu_e\to t\bar{b}$.}
\label{fig:MKvsMG-CC}  
\end{figure}

In Fig.~\ref{fig:MKvsMG-ttb} we collect same-flavour production processes, $t\bar t$, $b \bar b$, and display the transverse momentum and pseudo-rapidity of the heavy quark (as opposed to anti-quark). 
We find consistent agreement with respect to \aMC. 
An even more stringent test of the massive subtraction is provided by $t\bar b$ production.
The large mass hierarchy between top and bottom quarks validates the subtraction in presence of a multi-scale problem, as well as in asymmetric-mass conditions. In Fig.~\ref{fig:MKvsMG-CC} we display the transverse momentum and pseudo-rapidity of the top quark, again witnessing full agreement between \mad and \aMC.
For the considered massive processes, we observe a remarkable numerical stability of the \mad results: the \mad curves displayed in Figs.~\ref{fig:MKvsMG-ttb} and \ref{fig:MKvsMG-CC}
 were obtained using only a fraction of the phase-space points employed in \aMC, resulting in systematically shorter run times.

Given the evidence collected in this section, we consider our \mad validation fully satisfactory, and move on to NNLO developments.


\subsection{Numerical validation at NNLO}
\label{sec:massless NNLO}

In \cite{Bertolotti:2022aih} an analytic formula was obtained, within Local Analytic Sector Subtraction, for the cancellation of NNLO IR singularities in the case of massless final-state QCD radiation. 
In this section we take the first steps towards a thorough numerical implementation of that formula, and present numerical tests of singularity cancellation in a case study. The relevant definitions are collected in Appendix \ref{app:NNLOnum}.

We consider the process of di-jet production in $e^+ e^-$ annihilation, and focus on the $e^+ e^- \rightarrow q \bar{q} q' \bar{q}'$ radiative channel contributing to the NNLO double-real differential cross section. 
Moreover, we select the 3-particle symmetrised sector $\Z{\{q \bar{q} q'\}}$, defined in Eq.~\eqref{eq:Z}, in which the subtracted double-real correction schematically reads
\beq
\label{eq:RRsub}
RR^{\,\rm sub}_{\{q \bar{q} q'\}}
\, = \, 
\RR \; \Z{\{q \bar{q} q'\}} 
- 
K^{\one}_{\{q \bar{q} q'\}} 
- 
 K^{\two}_{\{q \bar{q} q'\}} + K^{\otwo}_{\{q \bar{q} q'\}} 
\, .
\eeq
Here, $RR$ stands for the squared double-real matrix element, while $K^{\one}_{\{q \bar{q} q'\}}$, $K^{\two}_{\{q \bar{q} q'\}}$, and $K^{\otwo}_{\{q \bar{q} q'\}}$ collect the single-unresolved, the uniform double-unresolved, and the strongly-ordered double-unresolved subtraction counterterms, respectively.

For the selected sector, and given the particle content, phase-space singularities stem from the single-unresolved collinear limit, $\bC{q\bar{q}}$, from the double-unresolved double-soft limit $\bS{q\bar{q}}$, and from the uniform triple-collinear limit $\bC{q\bar{q}q'}$. Correspondingly, the underlying real-emission and Born processes are $e^+ e^- \rightarrow g q' \bar{q}'$, and $e^+ e^- \rightarrow q' \bar{q}'$, respectively. 
The flavour of the particles involved in the sector also determines the physical content of the counterterms, which read
\beq
\label{eq:NNLOCTs}
K^{\one}_{\{q \bar{q} q'\}} 
& = &
\bbC{q\bar{q}} \,
RR \; \Z{\{q \bar{q} q'\}}
\, ,
\nnb
\\
K^{\two}_{\{q \bar{q} q'\}} 
& = &
\left(
\,
\bbS{q\bar{q}} 
+
\bbHC{q\bar{q} q'}
\right)
RR \; \Z{\{q \bar{q} q'\}}
\, ,
\nnb
\\
K^{\otwo}_{\{q \bar{q} q'\}}
& = &
\bbC{q\bar{q}} \,
\left(
\,
\bbS{q\bar{q}} 
+ 
\bbHC{q\bar{q} q'}
\right)
RR \; \Z{\{q \bar{q} q'\}}
\, ,
\eeq
where the hard-triple-collinear counterterm is defined as
\beq
 \bbHC{q\bar{q} q'}
 \, \equiv \,
  \bbC{q\bar{q} q'}
\left(1 - \bbS{q\bar{q}} \right)
\, ,
\eeq
and explicit expressions for all of these subtraction terms are provided in Eq.~\eqref{eq:appNNLOCTs}.

The tests to be performed to confirm a succesful singularity cancellation can be grouped into  two sets:
\beq
\label{eq:conrel}
\big\{ 
\bC{ij}, \; \bS{ij}, \; \bC{ijk}
\big\}\,\RR^{\,\rm sub}_{\{ijk\}}
& \to & 
\mbox{integrable}
\, ,
\\
\label{eq:conspur}
\big\{ 
\bC{kr}, \;
\bC{ijr}, \; \bC{ikr}, \; \bC{jkr}
\big\}\,\RR^{\,\rm sub}_{\{ijk\}}
& \to & 
\mbox{integrable}
\, ,
\eeq
where $(ijk) = (q \bar{q} q')$, and $r=\bar{q}'$ sets the recoiler assignment for this sector.
Eq.~\eqref{eq:conrel} collects the singular configurations of $\RR \; \Z{\{q \bar{q} q'\}}$, that are supposed to be tamed by the presence of counterterms in $\RR^{\,\rm sub}_{\{ijk\}}$, by construction. A detailed account of the expected cancellation pattern is reported in Eqs.~\eqref{eq:c_in_pieces}--\eqref{eq:cc_in_pieces}.
Eq.~\eqref{eq:conspur} contains collinear limits involving the recoiler particle $r$.
The necessity for checking such configurations originates from the presence of the recoiler in the definition of the collinear splitting kernels: this could in principle give rise to a spurious singular behaviour of the counterterms when the recoiler is collinear to one or more of the partons defining the sector.

The cancellation test proceeds as follows.  We obtain an $(n+2)$-body phase-space point by building two consecutive radiations on top of a generated Born configuration. Adopting Catani-Seymour parametrisations, the integration measure corresponding to the double-real radiation is written in terms of six kinematic variables, $w', y', z'$, and $\phi/\pi, y, z$, each with support in $[0,1]$. Unprimed variables parametrise the first radiation, passing from $n$ to $(n+1)$ final-state particles, while primed quantities describe the radiation of the $(n+2)$-th parton.
We report in Eq.~\eqref{eq:para} the specific double-radiative phase space that was used for the numerical test reported in the following.

As the azimuthal variables, $w'$ and $\phi$, are not connected to IR singularities, the cancellation test focuses on the other four Catani-Seymour variables, without loss of generality. Given their expressions in terms of kinematic invariants, reported in Eq.~\eqref{eq:zyzy}, different IR-singular configurations are reached by scaling appropriate combinations of such variables to 0 or 1. As an example, the single-collinear limit $\bC{ij}$ is obtained by scaling $y'$ to 0, leaving the other variables unchanged; conversely, a double-soft configuration $\bS{ij}$ implies $y'$, $z$, and $y$ to approach 0 at the same rate. The scalings required to probe all singular configurations of Eqs.~\eqref{eq:conrel}, \eqref{eq:conspur} are collected in Table \ref{table:limits}.

\begin{table}[h]
\begin{center}
\begin{tabular}{|c|c|c|c|c|c|c|c|}
\hline
Limit & $\bC{ij}$ & $\bS{ij}$ & $\bC{ijk}$ & $\bC{kr}$ & 
$\bC{ijr}$ & $\bC{ikr}$ & $\bC{jkr}$ \\
\hline
$z'$ & -- & -- & -- & -- & -- &  0 &  1 \\
$y'$ &  0 &  0 &  0 & -- & -- & -- & -- \\
$z $ & -- &  0 & -- &  1 &  0 &  1 &  1 \\
$y $ & -- &  0 &  0 & -- & -- & -- & -- \\
\hline
\end{tabular}
\end{center}
\caption{\label{table:limits}
IR scaling behaviour of the kinematic variables $(z',y',z,y)$ as defined by the Catani-Seymour mapping of Eq.~\eqref{eq:zyzy}. Numerical entries (0 or 1) indicate the limiting value of the variables in the given IR configuration, while the `--' symbol means that the corresponding variable is not scaled.}
\end{table}
We assign a common scaling parameter $\lambda$ to all phase-space variables relevant to a given IR limit, so that the limit is reached in all cases when $\lambda\to0$. For instance, considering Table \ref{table:limits}, in the collinear $\bC{ij}$ limit one sets $y'\sim\lambda$, while to probe the spurious $\bC{ikr}$ configuration, one sets $z'\sim\lambda$, and $z\sim 1-\lambda$.

\begin{figure}[ht]
\centering
\includegraphics[width=0.48\textwidth]{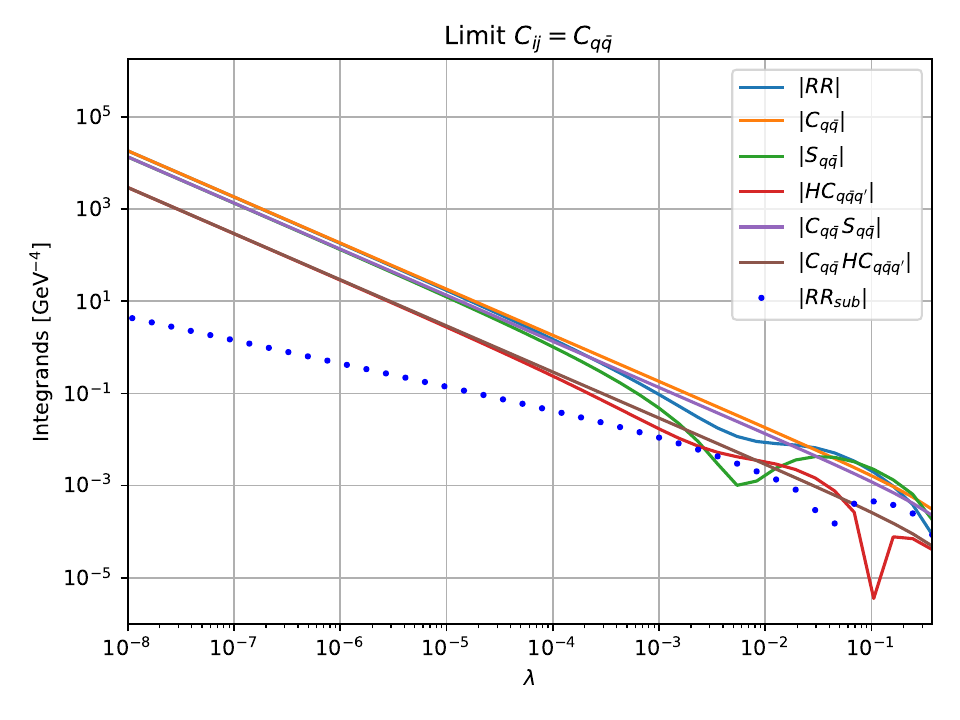}
\includegraphics[width=0.48\textwidth]{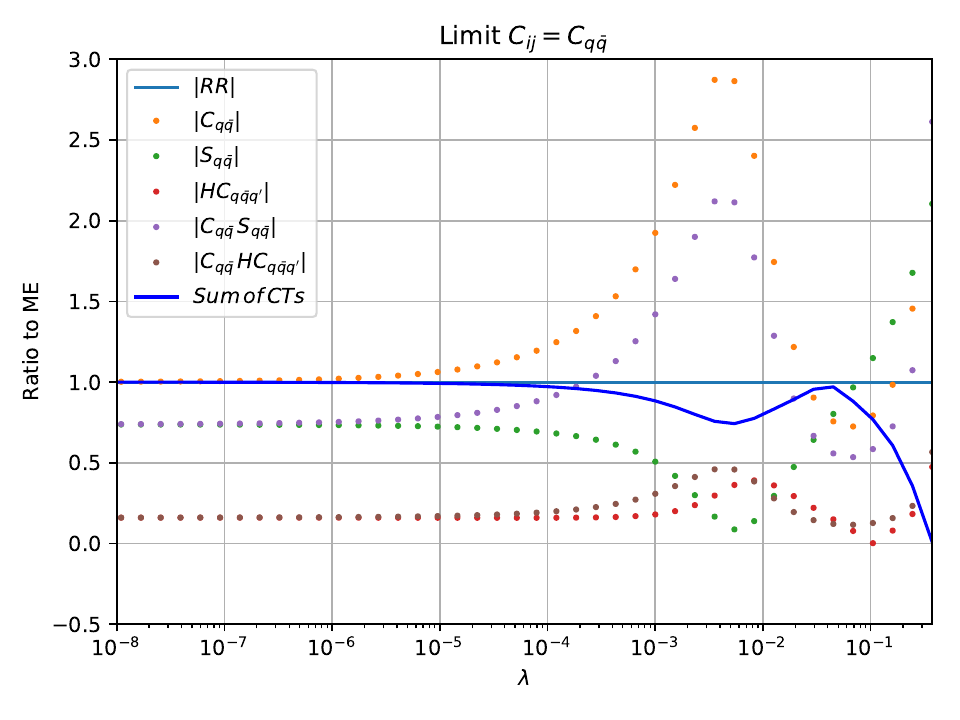}
\includegraphics[width=0.48\textwidth]{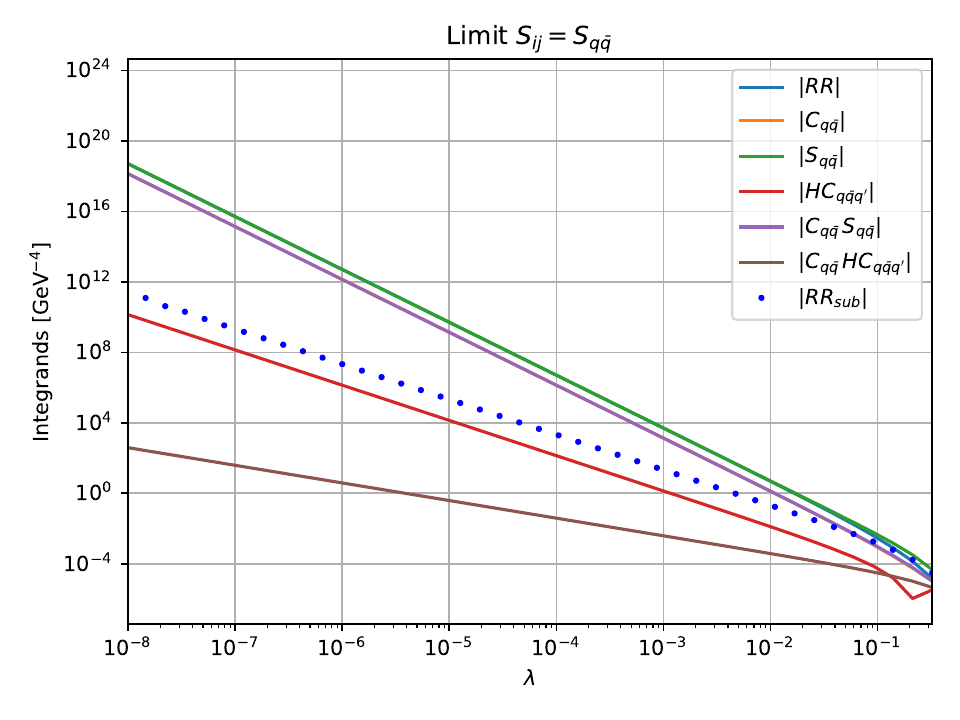}
\includegraphics[width=0.48\textwidth]{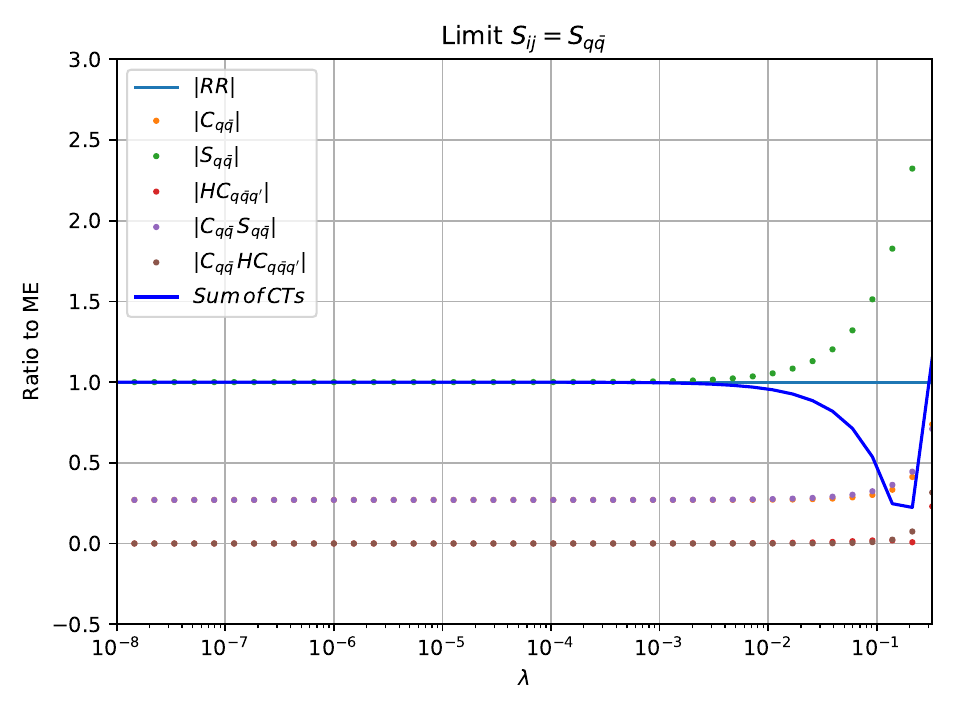}
\includegraphics[width=0.48\textwidth]{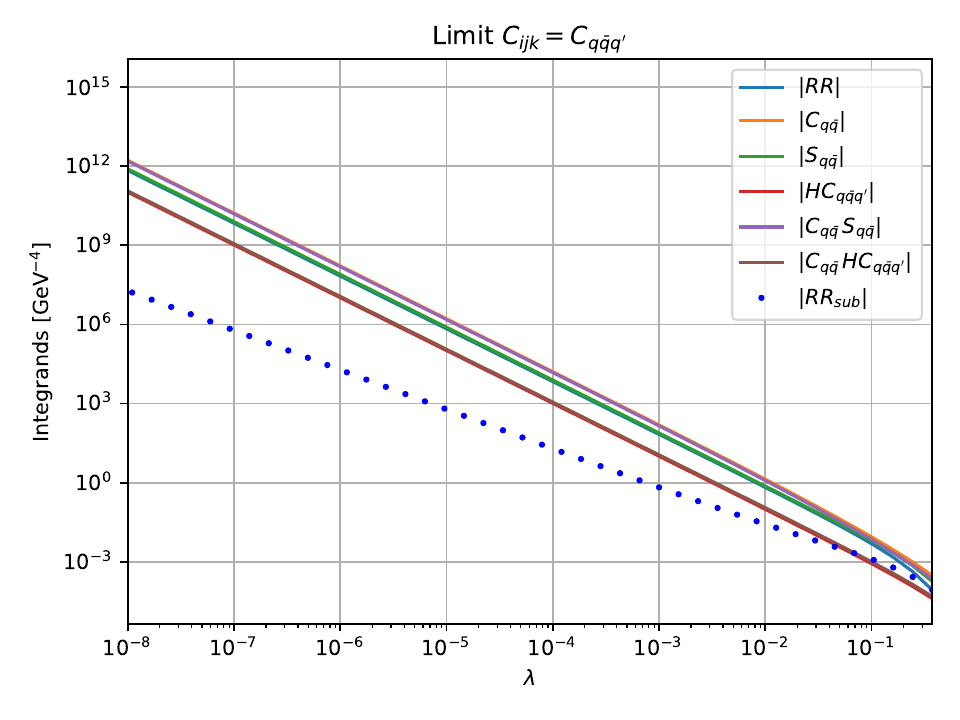}
\includegraphics[width=0.48\textwidth]{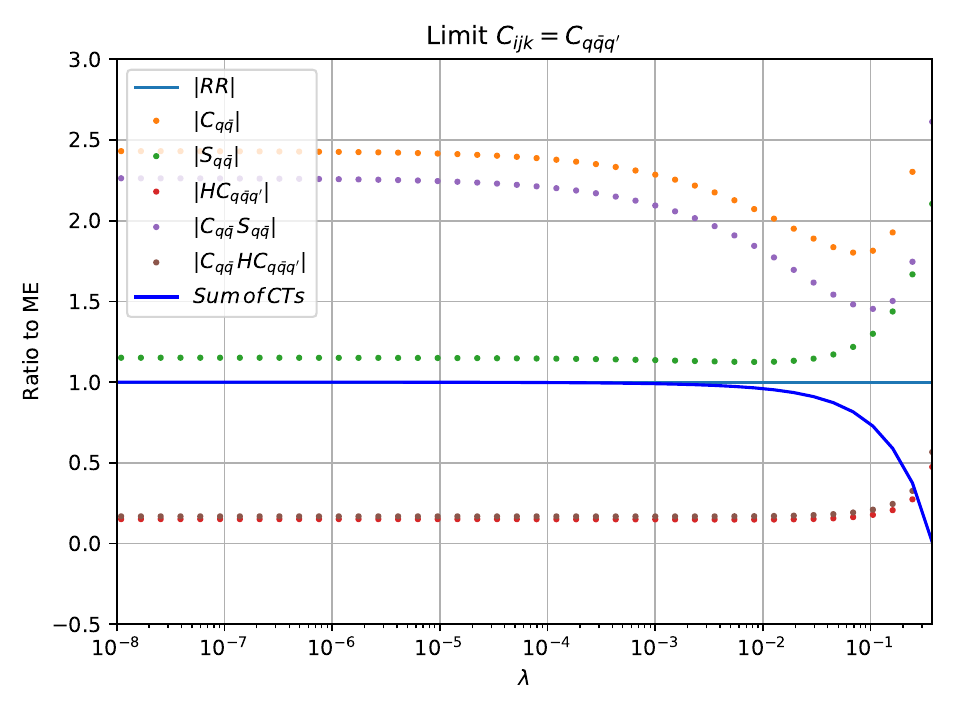}
\caption{\label{fig:primary}
Scaling behaviour of the double-real matrix element and of the subtraction counterterms for the process $e^+ e^- \rightarrow q \bar{q} q' \bar{q}'$, in sector $\Z{\{q \bar{q} q'\}}$. Top row: single-collinear limit $\bC{ij}$. Middle row: double-soft limit $\bS{ij}$. Bottom row: triple-collinear limit $\bC{ijk}$. Further details are in the main text.}
\end{figure}

The results of running the cancellation checks of Eq.~\eqref{eq:conrel} on the subtracted double-real contribution $RR^{\,\rm sub}_{\{q \bar{q} q'\}}$ are displayed in Fig.~\ref{fig:primary}.
The figure collects six panels, organised in two columns and three rows. Each row is dedicated to a different singular configuration in Eq.~\eqref{eq:conrel}, corresponding to the following physical assignments to variable $\lambda$:
\beq
\bC{ij} 
\, : \,  
\lambda \sim \theta_{ij}^2
\rightarrow 0
\, ,
\qquad
\bS{ij} 
\, : \,  
\lambda \sim E_i, \, E_j
\rightarrow 0
\, ,
\qquad
\bC{ijk} 
\, : \,  
\lambda \sim \theta_{ij}^2, \, \theta_{ik}^2
\rightarrow 0 
\, .
\eeq
For each of these IR limits, the left panels in Fig.~\ref{fig:primary} represent the absolute scaling of each contribution to $RR^{\,\rm sub}_{\{q \bar{q} q'\}}$ as the limit is approached ($\lambda\to0$). All terms are multiplied by the common phase-space jacobian (second line of Eq.~\eqref{eq:para}), and plotted as a function of $\lambda$.
The right panels in Fig.~\ref{fig:primary} collect the various counterterms, as well as their sum, again as functions of $\lambda$, normalised to the unsubtracted double-real matrix element $RR$.

From the plots on the left column one immediately evinces that the subtracted double-real matrix element (blue dotted) has a visibly milder scaling with respect to the unsubtracted one (teal solid), as $\lambda$ approaches 0. In particular, in the $\bC{ij}$, $\bS{ij}$, $\bC{ijk}$ limits, respectively, the $RR$ behaviour $(\lambda^{-1},\lambda^{-3}, \lambda^{-2})$ is reduced to $(\lambda^{-1/2},\lambda^{-2}, \lambda^{-3/2})$ after subtraction. This confirms the successful cancellation of leading-power singularities by means of the counterterms defined in Eq.~\eqref{eq:NNLOCTs}. The leftover behaviour of $RR^{\,\rm sub}_{\{q \bar{q} q'\}}$ as a function of $\lambda$ can be assessed considering the scaling of the phase-space measure, which in the three singular limits is $d\lambda \times (\lambda^0, \lambda^2, \lambda^1)$, respectively. Combining these informations, one can conclude that the subtracted double-real cross section is finite in the double-soft limit, while it features square-root integrable singularities in the collinear limits.

The cancellation of the leading IR singularities is similarly appreciated in the right plots of Fig.~\ref{fig:primary}, where all counterterm contributions (dotted), together with their sum (blue solid -- labelled as `Sum of CTs'), are displayed as ratios to the double-real matrix element (teal solid). 
As $\lambda$ decreases, each individual quantity stabilises towards a constant value. Notably, the overall sum (blue solid) correctly converges to unity, indicating that the coefficient of the leading-power singularity in $\RR$ is precisely reproduced by the combined subtraction counterterms.

\newpage

\begin{figure}[ht]
\centering
\includegraphics[width=0.48\textwidth]{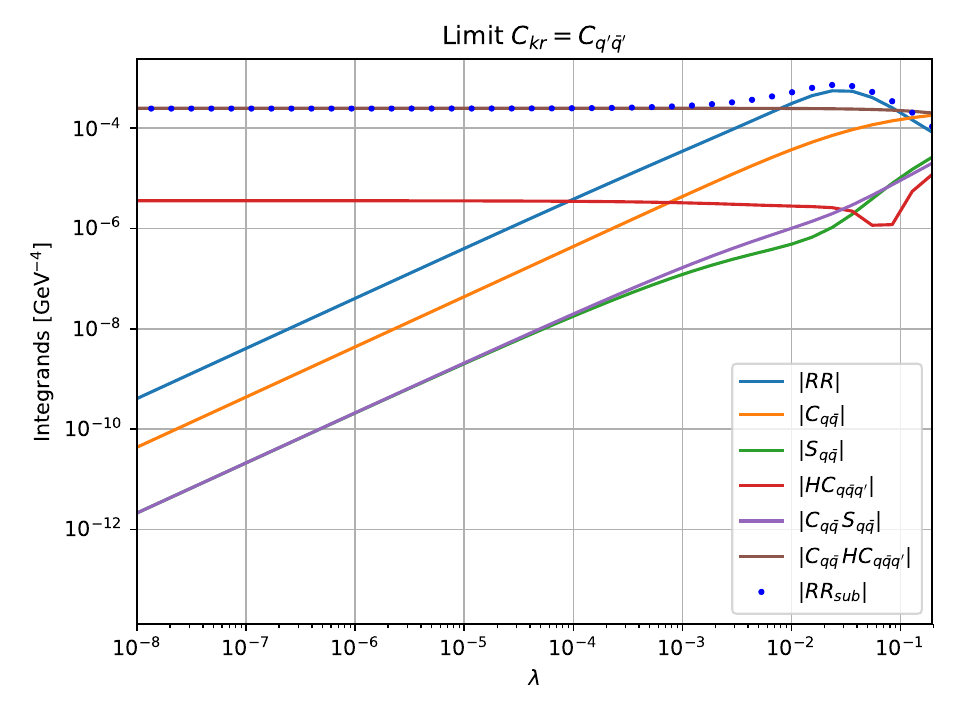}
\includegraphics[width=0.48\textwidth]{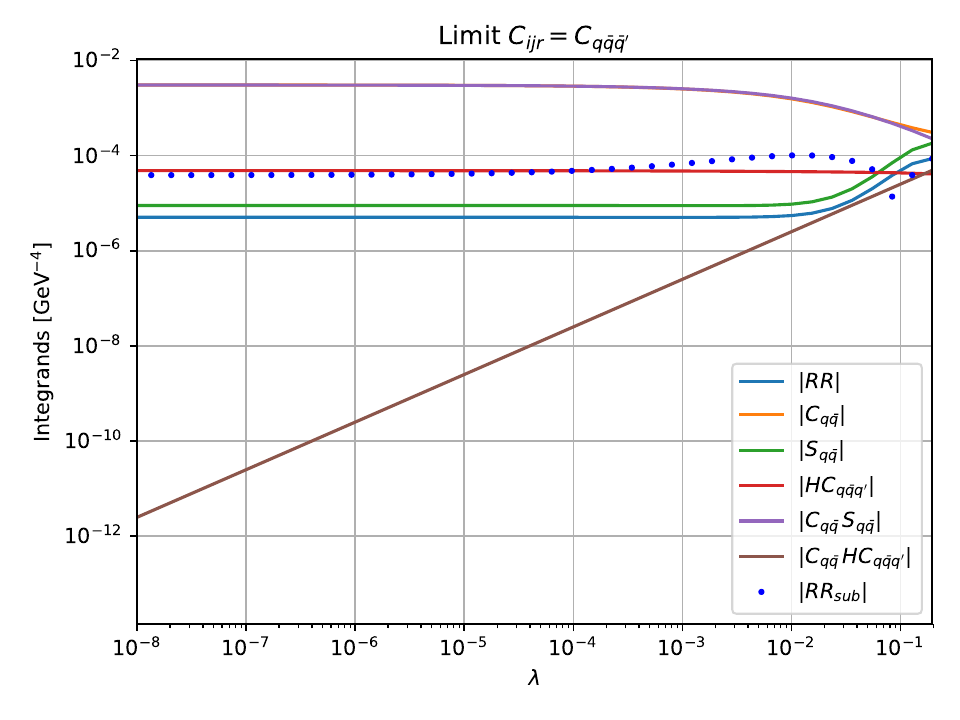}
\includegraphics[width=0.48\textwidth]{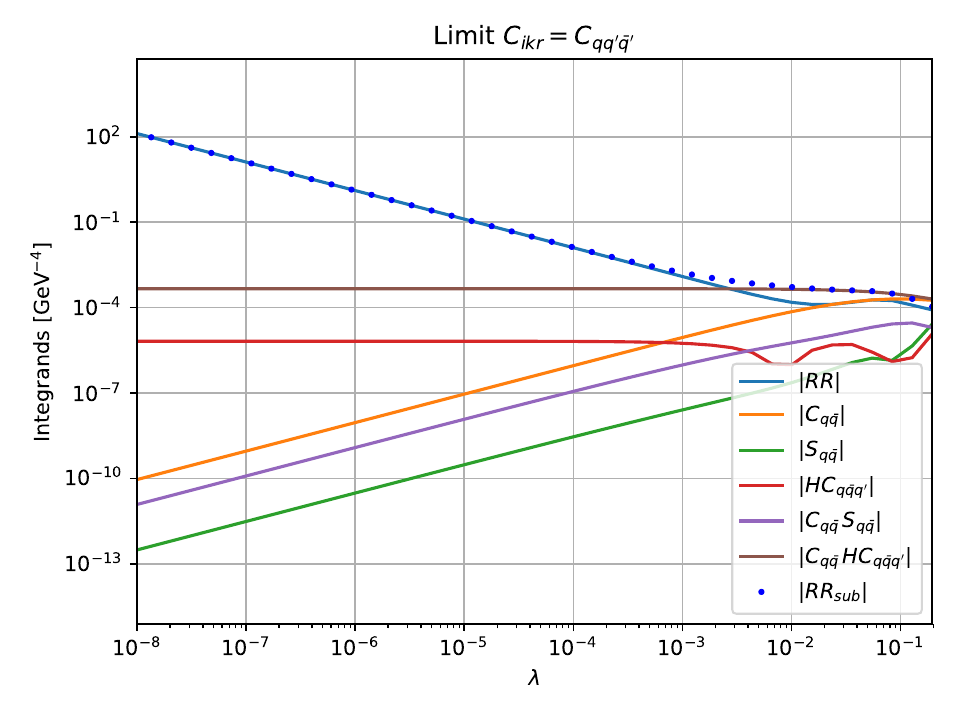}
\includegraphics[width=0.48\textwidth]{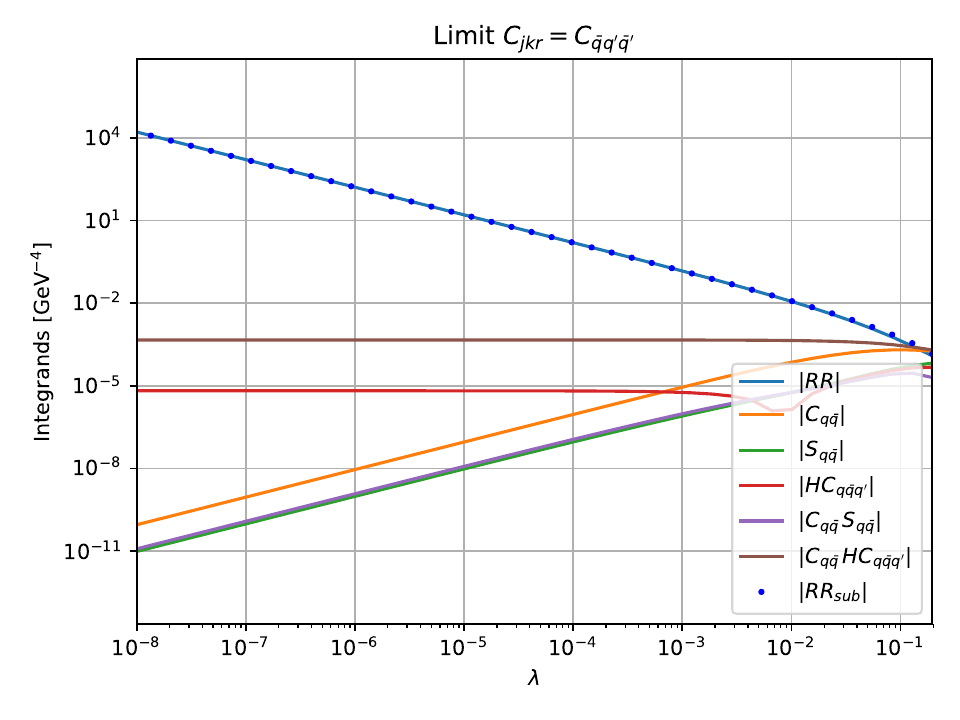}	
\caption{Scaling behaviour of the double-real matrix element and of the subtraction counterterms for the process $e^+ e^- \rightarrow q \bar{q} q' \bar{q}'$, in sector $\Z{\{q \bar{q} q'\}}$. Top row: single-collinear limit $\bC{kr}$, and triple-collinear limit $\bC{ijr}$. Bottom row: triple-collinear limits $\bC{ikr}$ and $\bC{jkr}$. Further details are in the main text.}
\label{fig:spurious}  
\end{figure}

The same reasoning just described can be applied to the study of the spurious collinear limits listed in Eq.~\eqref{eq:conspur}.
The plots presented in the top row of Fig.~\ref{fig:spurious} reveal a regular behaviour for each individual counterterm, as well as for the double-real matrix element, in the $\bC{kr}$ and $\bC{ijr}$ limits. As for the $\bC{ikr}$ and $\bC{jkr}$ limits, displayed in the bottom row of Fig.~\ref{fig:spurious}, both the unsubtracted and the subtracted double-real matrix elements have a singular $\lambda^{-1/3}$ behaviour. This is integrable in nature, whence we conclude that all spurious limits are as well under control.

\begin{figure}[ht]
\centering
\includegraphics[width=0.85\textwidth]{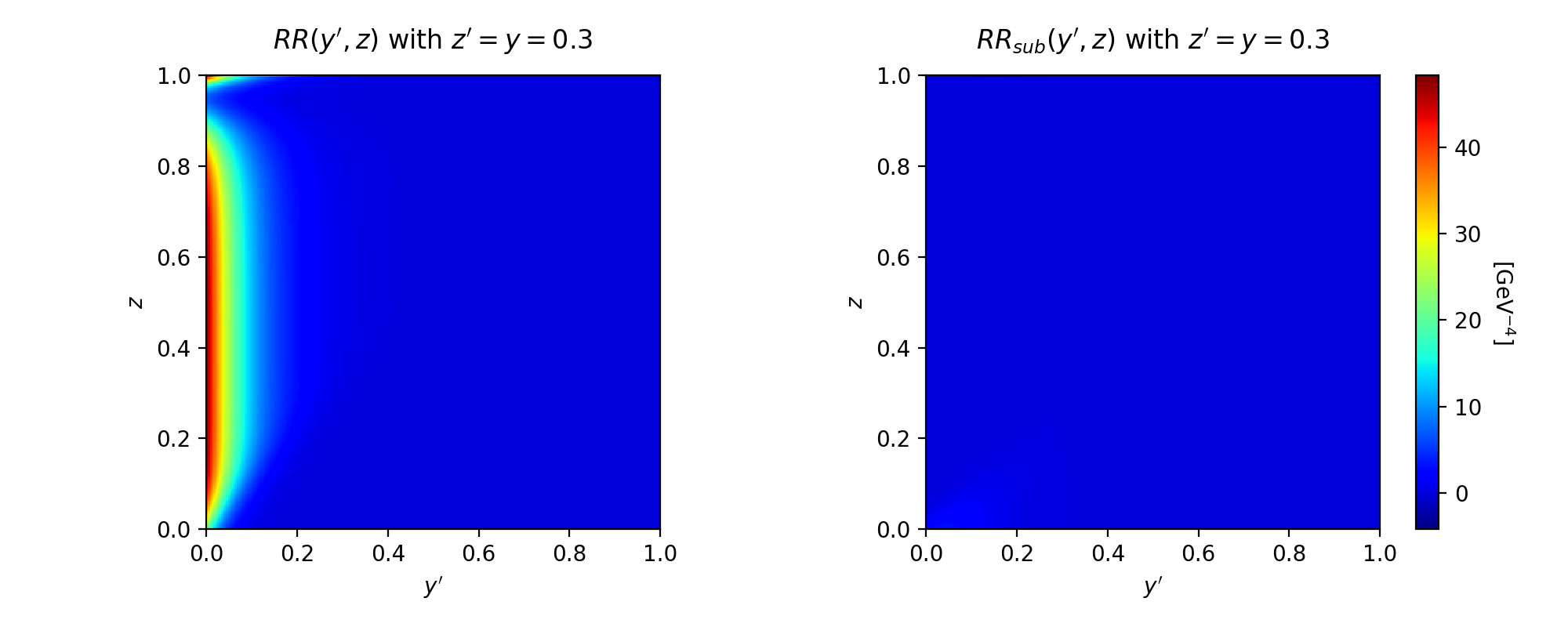}
\includegraphics[width=0.85\textwidth]{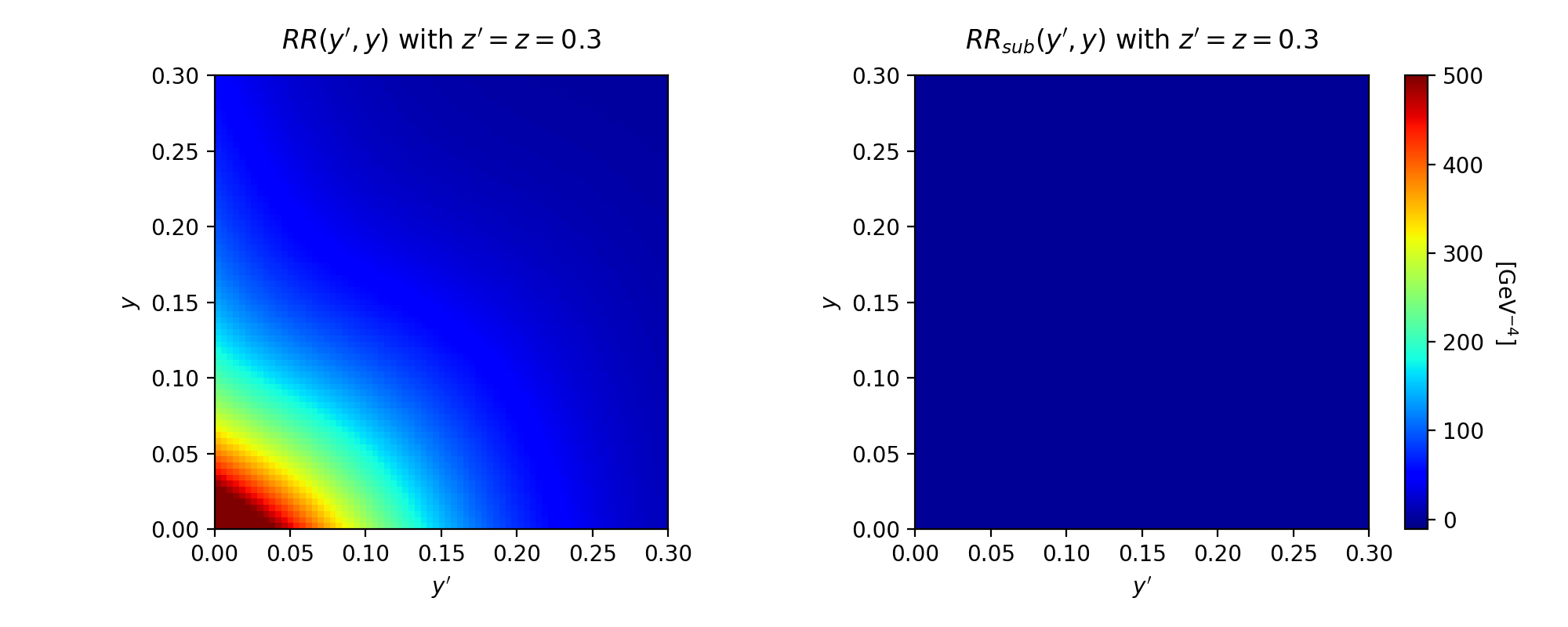}	
\caption{Density plots of the unsubtracted (left column) and subtracted (right column) double-real matrix element for two representative phase-space regions, in the process $e^+ e^- \rightarrow q \bar{q} q' \bar{q}'$, within sector $\Z{\{q \bar{q} q'\}}$. Further details are in the main text.}
\label{fig:2D}  
\end{figure}

As a further supporting evidence for singularity cancellation, we probed the double-real correction in representative portions of the phase space where IR divergences are expected to arise. The density plots in Fig.~\ref{fig:2D} present the result of such an analysis, where we sample the unsubtracted (left columns) and the subtracted (right columns) double-real matrix elements, respectively. In the top row, we fix $z' = y = 0.3$, and vary the remaining $y',z$ variables over their entire $[0,1]\times[0,1]$ support. 
The upper left plot clearly shows an enhancement in $RR$ extending over the entire $z$ range, at $y' \to 0$. As indicated in Table \ref{table:limits}, this corresponds to the presence of the single-collinear singularity $\bC{ij}$ in the unsubtracted double-real correction. In contrast, the corresponding density plot for the subtracted contribution $RR^{\,\rm sub}$ (top right) is regular in the whole region, confirming a successful cancellation.

The phase-space configuration shown in the bottom row of Fig.~\ref{fig:2D} is obtained by setting $z' = z = 0.3$,  while varying $y'$ and $y$. In the lower left plot, a strong divergent peak appears in the $y' \sim y \rightarrow 0$ limit, which corresponds to the triple-collinear singularity $\bC{ijk}$, as detailed in Table \ref{table:limits}. The latter is effectively mitigated by the subtraction counterterms, as demonstrated by the stable behaviour of $RR^{\,\rm sub}$ in the lower right plot. 

These considerations conclude the analysis of IR singularities, demonstrating the finiteness of the subtracted double-real matrix element for the considered channel in sector $\Z{\{q \bar{q} q'\}}$.


\section{Conclusion}
\label{sec:end}

In this article we have presented a number of developments in the framework of Local Analytic Sector Subtraction.
First, we have considered the NLO IR subtraction in presence of massive QCD particles in the final state. We have defined and analytically integrated all massive IR counterterms, reproducing the known singularity structure of one-loop massive matrix elements. As a designing feature, Local Analytic Sector Subtraction uses different phase-space mappings and parametrisations for different contributions to a given IR counterterm. This ensures the phase space to be optimally aligned with the natural Lorentz invariants appearing in each contribution, and in turn entails a comparatively simpler analytic counterterm integration. In the massive NLO case considered here, this structural simplicity is particularly evident in the finite parts of the integrated counterterms, which feature at worst di-logarithms for some of the soft contributions, and up to simple logarithms for hard-collinear terms. 
Such a simplicity has been attained via a thorough  exploration of the phase-space symmetries, and their translation into new parametrisations of the integration measures. This analysis not only allowed us to simplify the integrations, as in the case of the soft NLO counterterm with two massive colour sources, but also gives the freedom of choosing phase-space mappings independently of the analytic integration.
The latter feature constitutes a relevant starting point in view of the integration of NNLO double-soft massive kernels, which we leave for future work.

After the analytic discussion of NLO massive subtraction, we have turned our attention to the numerical implementation of the subtraction scheme, starting at NLO and documenting the validation of a variety of processes featuring QCD in the final state. The implementation is carried out in the \mad framework, which offers the level of generality required for an automated generation of NLO differential cross sections. We have described the steps we have taken to overcome the original technical limitations of \mad, in terms of phase-space generation, integration of the matrix elements, execution speed, and stability of the results. These features and the newly introduced massive NLO subtraction have been validated through a successful comparison with benchmark \aMC results, for processes with either massless or massive QCD particles in the final state.

Since \mad is designed to host subtraction schemes at NNLO (and beyond), we have reported the first elements of the automated implementation of Local Analytic Sector Subtraction at NNLO within the \mad framework. The implementation has so far been limited to the counterterms relevant to $e^+e^-\to q\bar q q' \bar q'$, namely a contribution to the quark channel in the production of two hadronic jets via leptonic collisions at NNLO in QCD. We have shown the complete cancellation of the single- and double-unresolved singularities of the double-real matrix elements, providing first non-trivial evidence of the stability performances of our NNLO implementation. The natural future development in this direction is the 
complete \mad implementation of NNLO Local Analytic Sector Subtraction for final-state QCD radiation, of which this article represents a first important step.


\section*{Acknowledgements}

We are grateful to V.~Hirschi for his help with the \mad framework, to E.~Maina and M.~Zaro for collaboration in the early stages of the implementation, to L.~Magnea for a careful reading of the manuscript, and to G.~Passarino for useful discussions. We thank G.~Bevilacqua, B.~Chargeishvili, A.~Kardos, S.-O.~Moch, Z.~Trocsanyi, for useful exchanges regarding the consistency of the Local Analytic Sector Subtraction method at NNLO. 
GB is supported by the Science and Technology Research Council (STFC) under the Consolidated Grant ST/X000796/1. 
The work of GL has been partly supported by the Alexander von Humboldt foundation.
PT has been partially supported by the Italian Ministry of University and Research (MUR) through grant
PRIN 2022BCXSW9, and by Compagnia di San Paolo through grant TORP\_S1921\_EX-POST\_21\_01. 


\appendix


\section{Mappings and radiative phase space with massive particles}
\label{app: mappings}

In this appendix we detail the phase-space mappings relevant for the case of massive particles in the final state. Moreover, we explore different parametrisations of the radiative phase space, exploiting its symmetries to simplify analytic counterterm integration.


\subsection{Final-final mapping with two massive final-state particles}
\label{app: Fm-Fm mapping}

Given three different final-state momenta $k_a$, $k_b$, $k_c$ with squared masses
\beq
k_a^2 = 0 \, ,
\qquad
k_b^2 = m_b^2
\, ,
\qquad
k_c^2 = m_c^2
\, ,
\eeq
we construct the mapped momenta 
\beq
\{\bar k\}^{(abc)}
\, \equiv \,
\Big\{ 
\{k\}_{\slashed{a}\slashed{b}\slashed{c}}, 
\kk{b}{abc}, \kk{c}{abc} 
\Big\}
\, .
\eeq
Momenta $\bar{k}_b\equiv\kk{b}{abc}$ and 
$\bar{k}_c\equiv\kk{c}{abc}$ are given by
\beq
\inFm{b} \, \inFm{c} :
\quad
\bar{k}_c^{\mu}
& = &
\sqrt{
\frac{\bs^2-4m_b^2m_c^2}{s_{[ab]c}^2 - 4m_c^2(s_{ab}+m_b^2)}
}
\left(k_c^{\mu} - \frac{s_{[ab]c}+2m_c^2}{2Q^2}Q^{\mu} \right)
+
\frac{\bs+2m_c^2}{2Q^2} \, Q^{\mu}
\, ,
\nnb\\[2mm]
\bar{k}_b^{\mu}
& = &
Q^{\mu}-\bar{k}_c^{\mu}
\, ,
\eeq
where $s_{ij} = 2 \, k_i\cdot k_j$ and
\beq
\bs
\, = \,
s_{abc}
\, = \,
s_{ab} + s_{bc} + s_{ac}
\, = \,
Q^2 - m_b^2 - m_c^2
\, ,
\qquad\quad
s_{[ab]c} \, = \, s_{ac} + s_{bc}
\, .
\eeq
All other momenta are left unchanged by the mapping, namely $\bar{k}_i^{\mu} = k_i^{\mu}$ with $i \neq a,b,c$.
Such definitions guarantee that the Born-level momenta $\bar{k}_b$ and $\bar{k}_c$ satisfy mass-shell conditions and total momentum conservation, as
\beq
\bar{k}_b^2
\, = \,
k_b^2
\, = \,
m_b^2
\, ,
\qquad
\bar{k}_c^2
\, = \,
k_c^2
\, = \,
m_c^2
\, ,
\qquad
Q^{\mu}
\, = \,
k_a^{\mu} 
+
k_b^{\mu}
+
k_c^{\mu}
\, = \,
\bar{k}_b^{\mu}
+
\bar{k}_c^{\mu}
\, .
\eeq
The mapping $\{\bar k\}^{(abc)}$ induces an exact phase-space factorisation
\beq
\int d \Phi_\npo  
\, = \, 
\frac{\varsi_\npo}{\varsi_n}
\int d\Phi_n^{(abc)}
\int d\Phi_{\rm rad,\sFm\sFm}^{(abc)}
\, ,
\label{npo1phsp}
\eeq
where $d\Phi_n^{(abc)} \equiv d\Phi_n ( \{\bar k\}^{(abc)} )$ and 
we explicitly extracted the ratio of the relevant 
symmetry factors $\varsi_\npo$ and $\varsi_n$.
The radiative phase-space measure associated with the unresolved particle $k_a$ can be parametrised in terms of the usual Catani-Seymour kinematic variables \cite{Catani:1996vz,Catani:2002hc}
\beq
\label{eq:CSvariables}
z
\, = \,
\frac{s_{ac}}{s_{ac}+s_{bc}}
\, ,
\qquad \qquad
y
\, = \,
\frac{s_{ab}}{s_{ab}+s_{bc}+s_{ac}}
\, ,
\eeq
yielding
\beq
\label{eq:CSinvariants}
s_{ab} 
\, = \, 
y \, \bs
\, , 
\qquad
s_{ac} 
\, = \, 
z (1 - y) \, \bs
\, , 
\qquad
s_{bc} 
\, = \, 
(1 - z)(1 - y) \, \bs
\, .
\eeq
According to \cite{Catani:2002hc}, the parametrised radiative integration measure is
\beq
\label{eq:dPhirad1MM}
\int d \Phi_{\rm rad,\sFm\sFm}^{(abc)} 
& = &
\frac{(2 \pi)^{-3+2\eps}}{4} \, \bs^{2-2\eps} \,
\left( \sqrt{\lambda}\right)^{-1+2\eps}
\int d^{d-3} \Omega
\nnb\\
&&
\hspace{-4mm}
\times \, 
\int_0^{y_+} dy \, (1-y)^{1-2\eps} (\bs y+m_b^2)^{-\eps}
\int_{z_{-}(y)}^{z_+(y)} dz \, (z_+ - z)^{-\eps} (z - z_{-})^{-\eps}
\, ,
\qquad
\eeq
where the integration boundaries are given by 
\beq
y_+
\, = \,
1
+
\frac{2m_c^2}{\bs}
-
\frac{2m_c Q}{\bs}
\, ,
\qquad
z_{\pm}(y)
\, = \,
\frac{\bs y}{2(\bs y+m_b^2)} \,
\left(1 \pm \frac{\sigma}{\bs(1-y)} \right)
\, ,
\eeq
and 
\beq
\lambda
& = &
\bs^2 - 4m_b^2m_c^2
\, ,
\nnb\\
\sigma
& \equiv &
\sigma(y)
\, = \,
\sqrt{\lambda + \bs^2y^2 - 2y\bs(\bs+2m_c^2)}
\, = \,
\sqrt{\bs^2(1-y)^2 - 4m_c^2(\bs y+m_b^2)}
\, .
\eeq
The integration over the solid angle can be carried out independently, yielding the factor
\beq
\label{eq:solidangle}
\int d^{d-3} \Omega
\, = \,
\frac{2 \pi^{1-\eps}}{\Gamma(1-\eps)}
\, ,
\eeq
so that
\beq
\label{eq: dPhirad FmFm yz}
\int d\Phi_{\rm rad,\sFm\sFm}^{(abc)} 
& = &
\frac{(4 \pi)^{-2+\eps}\,\bs}{\Gamma(1-\eps)} 
\left(\frac{\bs}{\sqrt{\lambda}}\right)^{\!\!1-2\eps}
\\
&&
\qquad
\times \,
\int_0^{y_+} dy \, (1-y)^{1-2\eps} (\bs y + m_b^2)^{-\eps}
\int_{z_{-}}^{z_+} dz \, (z_+ - z)^{-\eps} (z - z_{-})^{-\eps}
\, .
\nnb
\eeq
It is easy to verify that 
\beq
\sigma(0) = \sqrt{\lambda}
\, ,
\qquad
\sigma(y_+) = 0
\, ,
\qquad
z_{\pm}(0) = 0
\, ,
\qquad
z_{\pm}(y_+) = \frac{\bs y_+}{2(\bs y_+ + m_b^2)}
\, ,
\eeq
so that the integration domain
is the region between the two curves $z=z_\pm(y)$, which 
meet at the endpoints 
\beq
P_1 
& \equiv & 
\big(y,z_{\pm}(y)\big)|_{y=0} = 
\big(0,z_{\pm}(0)\big) = 
(0,0)
\, ,
\nnb\\
P_2 
& \equiv &
\big(y,z_{\pm}(y)\big)|_{y=y_+} = 
\big(y_+,z_{\pm}(y_+)\big) = 
\bigg(y_+,\frac{\bs y_+}{2(\bs y_+ + m_b^2)}\bigg)
\, .
\eeq
The two curves satisfy the following equation: 
\beq
\label{eq:zpmfirst}
&&
z
\, = \,
z_\pm(y)
\, = \,
\frac{\bs y}{2(\bs y+m_b^2)} \, 
\bigg( 1 \pm \frac{\sigma(y)}{\bs(1-y)} \bigg)
\nnb\\[2mm]
& \quad\Longrightarrow\quad &
\bs\, y(1-y)^2 z(1-z)
- 
m_b^2\,(1-y)^2 z^2
-
m_c^2\,y^2
\, = \,
0
\, .
\eeq
We now check how these curves transform under the 
exchange of $k_b$ with $k_c$. 
We introduce variables $Y,Z$ to re-parametrise the invariants in  Eq.~(\ref{eq:CSinvariants}) under the  
exchange $k_b \leftrightarrow k_c$, as
\beq
s_{ab} 
& = &
(1-Y) Z \, \bs
\, ,
\qquad 
s_{ac} 
\; = \;
Y \, \bs
\, ,
\qquad
s_{bc} 
\; = \;
(1-Y) (1-Z) \, \bs
\, .
\eeq
The change of variable $y,z \to Y,Z$ is thus defined by
the relations $Z (1-Y) = y$, and $Y = z (1-y)$, which give
\begin{align}
Y & = \; z(1-y)
\, ,
&
Z & = \; \frac{y}{1-z(1-y)}
\, ,
\nnb\\
y & = \; Z(1-Y)
\, ,
&
z & = \; \frac{Y}{1-Z(1-Y)}
\, .
\end{align}
In therms of the new variables, Eq.~(\ref{eq:zpmfirst}), defining the boundaries of the integration domain, becomes
\beq
\bs\, Y(1-Y)^2 Z(1-Z)
-
m_b^2\,Y^2
- 
m_c^2\,(1-Y)^2 Z^2
\, = \,
0
\, .
\eeq
This means that the re-parametrisation $y,z \to Y,Z$ is 
equivalent to the exchange $m_b \leftrightarrow m_c$. 
It is easy to verify that this holds in general for the radiative phase-space 
$d\Phi_{\rm rad,\sFm\sFm}^{(abc)}$. 
Indeed the Jacobian of the transformation reads 
\beq
J(Y,Z) 
& = &
\left| 
\frac{\de y}{\de Y}\,\frac{\de z}{\de Z} 
- 
\frac{\de y}{\de Z}\,\frac{\de z}{\de Y} 
\right|
=
\frac{1-Y}{1-Z(1-Y)} 
\, ,
\qquad
\eeq
so that  
\beq
dy \, dz \, (1-y) 
& = &
dY \, dZ \, J(Y,Z) [1-Z(1-Y)] 
\; = \;
dY \, dZ \, (1-Y)
\, .
\eeq
Furthermore, we can easily verify that 
\beq
(1-y)^2(\bs y+m_b^2)
(z_+-z)(z-z_-)
& = &
\bs y(1-y)^2 z(1-z)
-
m_b^2(1-y)^2 z^2
-
m_c^2 y^2
\nnb\\
& = &
\bs Y(1-Y)^2 Z(1-Z)
-
m_b^2 Y^2
- 
m_c^2(1-Y)^2 Z^2
\nnb\\
& = &
(1-Y)^2(\bs Y+m_c^2)
(Z_+-Z)(Z-Z_-)
\, ,
\nnb
\eeq
where we have set
\beq
&&
Y_+
\, = \,
1
+
\frac{2m_b^2}{\bs}
-
\frac{2m_b Q}{\bs}
\, ,
\qquad\quad
Z_{\pm}
\, = \,
\frac{\bs Y}{2(\bs Y+m_c^2)} \, 
\bigg( 1 \pm \frac{\Sigma(Y)}{\bs(1-Y)} \bigg)
\, ,
\nnb\\[2mm]
&&
\Sigma(Y)
\, = \,
\sqrt{\lambda + \bs^2Y^2 - 2Y\bs(\bs+2m_b^2)}
\, = \,
\sqrt{\bs^2(1-Y)^2 - 4m_b^2(\bs Y+m_c^2)}
\, .
\eeq
The radiative phase-space in Eq.~(\ref{eq: dPhirad FmFm yz}) can then be equivalently rewritten as 
\beq
\label{eq: dPhirad FmFm YZ}
\int d\Phi_{\rm rad,\sFm\sFm}^{(abc)} 
& = &
\frac{(4 \pi)^{-2+\eps}\,\bs}{\Gamma(1-\eps)} 
\left(\frac{\bs}{\sqrt{\lambda}}\right)^{\!\!1-2\eps}
\\
&&
\qquad
\times \,
\int_0^{Y_+}  dY \, (1-Y)^{1-2\eps} (\bs Y + m_c^2)^{-\eps}
\int_{Z_{-}}^{Z_+}  dz \, (Z_+ - Z)^{-\eps} (Z - Z_{-})^{-\eps}
\, .
\nnb
\eeq

A parametrisation that makes the $b \leftrightarrow c$ symmetry more evident is the one in terms of variables $u,v$ defined by $(1-v) (1-u)
 = y$ and $u = (1-z)(1-y)$, which gives 
\begin{align}
u & = \; (1-z)(1-y)
\, ,
&
v & = \; \frac{z(1-y)}{y+z-yz}
\, ,
\nnb\\
y & = \; (1-u)(1-v)
\, ,
&
z & = \; \frac{v(1-u)}{u+v-uv}
\, .
\end{align}
The Jacobian of the transformation reads 
\beq
J(u,v) 
& = &
\left| 
\frac{\de y}{\de u}\,\frac{\de z}{\de v} 
- 
\frac{\de y}{\de v}\,\frac{\de z}{\de u} 
\right|
\, = \,
\frac{1-u}{u+v-uv} 
\, ,
\eeq
yielding
\beq
dy \, dz \, (1-y) 
& = &
du \, dv \, J(u,v) \, (u+v-uv)
\; = \;
du \, dv \, (1-u)
\, .
\eeq
In addition, we have 
\beq
(1-y)^2 (\bs y+m_b^2)(z_+ - z)(z - z_{-})
& = &
\bs\, y(1-y)^2 z(1-z)
- 
m_b^2\,(1-y)^2 z^2
-
m_c^2\,y^2
\nnb\\
& = &
(1-u)^2
\Big[
\bs\,u\,v(1-v)
-
m_b^2\,v^2
- 
m_c^2\,(1-v)^2
\Big]
\nnb\\
& = &
(1-u)^2
(\bs u+m_b^2+m_c^2)
(v_+ - v)(v - v_{-})
\, ,
\qquad\qquad
\eeq
where $v_\pm$ are the solutions of the equation 
$\bs uv(1-v) - m_b^2 v^2 - m_c^2(1-v)^2 = 0$, and read 
\beq 
v_\pm
=
v_\pm(u)
& = &
\frac
{\bs u+2m_c^2 \pm \sqrt{\bs^2u^2 - 4 m_b^2 m_c^2}}
{2(\bs u+m_b^2+m_c^2)}
\, .
\eeq
In terms of $u$ and $v$, Eq.~(\ref{eq:zpmfirst}) becomes
\beq
(1-u)^2
\Big[
\bs\,u\,v(1-v)
-
m_b^2\,v^2
- 
m_c^2\,(1-v)^2
\Big]
\, = \,
0
\, ,
\eeq
or equivalently 
\beq
(1-u)^2
(\bs u+m_b^2+m_c^2)
(v_+(u) - v)(v - v_-(u))
=
0
\, .
\eeq
That is to say that the two curves defining the integration domain of $v$ 
are $v=v_\pm(u)$. One has that $v_+(u)\geq v_-(u)$ for all values of $u \in [u_-,1]$, the two curves being equal only for
\beq
u = u_- = \frac{2 \, m_b \, m_c}{\bs}
\, .
\eeq
We thus conclude that the integration domain in the $u,v$ parametrisation reads 
\beq
\int_0^{y_+} dy \, \int_{z_-(y)}^{z_+(y)} dz \, (1-y)
& = &
\int_{u_-}^1 du \, \int_{v_-(u)}^{v_+(u)} dv \, (1-u)
\, .
\eeq
The radiative phase space can thus be rewritten as 
\beq
\label{eq: dPhirad FmFm uv}
\int \! d\Phi_{\rm rad,\sFm\sFm}^{(abc)} 
& = &
\frac{(4 \pi)^{-2+\eps}\,\bs}{\Gamma(1-\eps)} 
\left(\frac{\bs}{\sqrt{\lambda}}\right)^{\!\!1-2\eps}
\\
&&
\qquad
\times \,
\int_{u_-}^1 du \, (1-u)^{1-2\eps} (\bs u + m_b^2 + m_c^2)^{-\eps}
\int_{v_-}^{v_+} dv (v_+ \!-\! v)^{-\eps} (v - v_-)^{-\eps}
\, ,
\nnb
\eeq
with
\beq
\label{eq: invariants uv}
s_{ab} 
=
(1-u)(1-v) \, \bs
\, , 
\qquad
s_{ac} 
= 
(1 - u) v \, \bs
\, , 
\qquad
s_{bc} 
= 
u \, \bs
\, .
\eeq


\subsection{Final-final mapping with one massive particle}
\label{app: Fm-F and F-Fm mappings}

The cases involving three final-state momenta, only one of which being massive, are special cases of the fully massive mapping considered in \appn{app: Fm-Fm mapping}. We report them explicitly in the following, for completeness.
\\ \\
{\bf Case $m_b=0$}
\\[5pt]
In the $m_b=0$ case, we have 
\beq
\inF{b} \, \inFm{c} :
\quad
\bar{k}_c^{\mu}
& = &
\frac{\bs}{\sqrt{s_{[ab]c}^2 - 4 m_c^2 s_{ab}}} 
\left(k_c^{\mu} - \frac{s_{[ab]c}+2m_c^2}{2(\bs+m_c^2)}Q^{\mu}\right)
+
\frac{\bs+2m_c^2}{2(\bs+m_c^2)} Q^{\mu}
\, ,
\nnb\\[2mm]
\bar{k}_b^{\mu}
& = &
Q^{\mu}-\bar{k}_c^{\mu}
\, .
\eeq
The relevant quantities defining the radiative integration measure in the three provided parametrisations are
\beq
&&
y_+
\, = \,
1
+
\frac{2m_c^2}{\bs}
-
\frac{2m_c \sqrt{\bs+m_c^2}}{\bs}
\, ,
\qquad
Y_+
\, = \,
1
\, ,
\nnb\\[5pt]
&&
\lambda = \bs^2
\, ,
\qquad
\sigma(y)
\, = \,
\sqrt{\bs^2(1-y)^2 - 4m_c^2\,\bs\,y}
\, ,
\qquad
\Sigma(Y)
\; = \;
\bs\,(1-Y)
\, ,
\nnb\\[5pt]
&&
z_{\pm}
\, = \,
\frac{1}{2} \, \bigg( 1 \pm \frac{\sigma}{\bs(1-y)} \bigg)
\, ,
\qquad
Z_{-}
\, = \,
0
\, ,
\qquad
Z_{+}
\, = \,
\frac{\bs Y}{\bs Y+m_c^2}
\, ,
\nnb\\[5pt]
&&
u_- = 0
\, ,
\qquad
v_- 
=
\frac{m_c^2}{\bs u+m_c^2}
\, ,
\qquad
v_+
=
1
\, .
\eeq
The phase-space measure then reads
\beq
\label{eq:dPhirad FFm}
\int d\Phi_{\rm rad,\sF\sFm}^{(abc)} 
& = &
\frac{(4 \pi)^{-2+\eps}}{\Gamma(1-\eps)} \,
\bs^{1-\eps} \,
\int_0^{y_+} dy \, y^{-\eps} \, (1-y)^{1-2\eps}
\int_{z_-}^{z_+} dz \, (z_+ - z)^{-\eps} (z - z_{-})^{-\eps}
\nnb\\
& = &
\frac{(4 \pi)^{-2+\eps}\,\bs}{\Gamma(1-\eps)} \,
\int_0^1  dY \, (1-Y)^{1-2\eps} (\bs Y + m_c^2)^{-\eps}
\int_0^{Z_+} dZ \, Z^{-\eps} (Z_+ - Z)^{-\eps}
\nnb\\
& = &
\frac{(4 \pi)^{-2+\eps}\,\bs}{\Gamma(1-\eps)} 
\int_0^1 du \, (1-u)^{1-2\eps} (\bs u + m_c^2)^{-\eps}
\int_{v_-}^1 dv \, (1 - v)^{-\eps} (v - v_-)^{-\eps}
\, .
\nnb\\
\eeq
We notice that in this case $ z_\pm = 1 - z_\mp$: the phase-space measure is symmetric under the exchange $a\leftrightarrow b$, which corresponds to the variable transformation  
$z\leftrightarrow 1-z$. 
\\ \\
{\bf Case $m_c=0$}
\\[5pt]
Similarly, in the $m_c=0$ case, we have 
\beq
\inFm{b} \, \inF{c} :
\quad
\bar{k}_c^{\mu}
& = &
\frac{\bs}{s_{[ab]c}}\, k_c^{\mu}
\; = \;
\frac{s_{abc}}{s_{[ab]c}}\, k_c^{\mu}
\, ,
\nnb\\[2mm]
\bar{k}_b^{\mu}
& = &
k_a^{\mu}
+
k_b^{\mu}
-
\frac{s_{ab}}{s_{[ab]c}}\, k_c^{\mu}
\, ,
\eeq
which are identical to the fully massless case.
The relevant quantities defining the integration measure in the three provided parametrisations are
\beq
&&
y_+
\, = \,
1
\, ,
\qquad
Y_+
\, = \,
1
+
\frac{2m_b^2}{\bs}
-
\frac{2m_b \sqrt{\bs+m_b^2}}{\bs}
\, ,
\qquad
\nnb\\[5pt]
&&
\lambda = \bs^2
\, ,
\qquad
\sigma(y)
\, = \,
\bs\,(1-y)
\, ,
\qquad
\Sigma(Y)
\; = \;
\sqrt{\bs^2(1-Y)^2 - 4m_b^2\,\bs\,Y}
\, ,
\qquad
\nnb\\[5pt]
&&
z_{-}
\, = \,
0
\, ,
\qquad
z_{+}
\, = \,
\frac{\bs y}{\bs y+m_b^2}
\, ,
\qquad
Z_{\pm}
\, = \,
\frac{1}{2} \, \bigg( 1 \pm \frac{\Sigma(Y)}{\bs(1-Y)} \bigg)
\, ,
\qquad
\nnb\\[5pt]
&&
u_- = 0
\, ,
\qquad
v_-
=
0
\, ,
\qquad
v_+
=
\frac{\bs u}{\bs u+m_b^2}
\, .
\eeq
The phase-space measure becomes
\beq
\label{eq:dPhirad FmF}
\int d \Phi_{\rm rad,\sFm\sF}^{(abc)} 
& = & 
\frac{(4 \pi)^{-2+\eps}\,\bs}{\Gamma(1-\eps)} \,
\int_0^1 dy \, (1-y)^{1-2\eps} (\bs y + m_b^2)^{-\eps}
\int_0^{z_+} dz \, z^{-\eps} (z_+ - z)^{-\eps}
\nnb\\
& = & 
\frac{(4 \pi)^{-2+\eps}}{\Gamma(1-\eps)} \,
\bs^{1-\eps} \,
\int_0^{Y_+} dY \, Y^{-\eps} \, (1-Y)^{1-2\eps}
\int_{Z_-}^{Z_+} dZ \, (Z_+ - Z)^{-\eps} (Z - Z_{-})^{-\eps}
\nnb\\
& = &
\frac{(4 \pi)^{-2+\eps}\,\bs}{\Gamma(1-\eps)} 
\int_0^1 du \, (1-u)^{1-2\eps} (\bs u + m_b^2)^{-\eps}
\int_0^{v_+} dv \, v^{-\eps} (v_+ - v)^{-\eps}
\, .
\eeq
We stress that the relation $d\Phi_{\rm rad, \sF\sFm}(Y,Z,m_c) = d\Phi_{\rm rad, \sFm\sF}(y,z,m_b)$, valid in the massive-massless case, directly stems from the equivalence between Eq.~(\ref{eq: dPhirad FmFm yz}) and Eq.~(\ref{eq: dPhirad FmFm YZ}). This in turn is what allows to get the same expression for the two integrated massive-massless soft kernels in Eq.~(\ref{eq:two massive massless}).


\subsection{Initial-final mapping with one massive particle}
\label{app: Fm-I mapping}

Given two final-state momenta $k_a$, $k_b$ and one initial-state momentum 
$k_c$, with masses
\beq
k_a^2 = 0 \, ,
\qquad
k_b^2 = m_b^2
\, ,
\qquad
k_c^2 = 0
\, ,
\eeq
we construct the mapped momenta 
\beq
\{\bar k\}^{(abc)}
\, = \,
\Big\{ 
\{k\}_{\slashed{a}\slashed{b}\slashed{c}}, 
\kk{b}{abc}, \kk{c}{abc} 
\Big\}
\, ,
\eeq
where $\bar{k}_b\equiv\kk{b}{abc}$ and 
$\bar{k}_c\equiv\kk{c}{abc}$ are given by
\beq 
\inFm{b} \inI{c} :
\quad
\bar{k}^{(abc)}_{b} 
& = &
k_a + k_b - \frac{s_{ab}}{s_{ac} + s_{bc}} \, k_c 
\, ,
\nnb\\
\bar{k}^{(abc)}_{c} 
& = &
\frac{s_{ac} + s_{bc} - s_{ab}}{s_{ac} + s_{bc}} \, k_c 
\, , 
\eeq
while all other momenta are left unchanged, $\bar{k}_i^{\mu} = k_i^{\mu}$ with 
$i \neq a,b,c$. 
Such transformations guarantee that the Born-level momenta $\bar{k}_b$ and 
$\bar{k}_c$ satisfy mass-shell conditions and total momentum conservation, as
\beq
\bar{k}_b^2
\, = \,
k_b^2
\, = \,
m_b^2
\, ,
\qquad
\bar{k}_c^2
\, = \,
k_c^2
\, = \,
0
\, ,
\qquad
Q^{\mu}
\, = \,
k_a^{\mu} 
+
k_b^{\mu}
-
k_c^{\mu}
\, = \,
\bar{k}_b^{\mu}
-
\bar{k}_c^{\mu}
\, .
\eeq
The mapping $\{\bar k\}^{(abc)}$ induces the phase-space factorisation
\beq
\int d \Phi_\npo ( k_c)  
\, = \, 
\frac{\varsi_\npo}{\varsi_n}
\int \int d \Phi_n^{(abc)} (x k_c ) \,
d \Phi_{\rm rad}^{(abc)}
\, ,
\eeq
where $d\Phi_n^{(abc)}(x k_c ) \equiv d\Phi_n ( \{\bar k\}^{(abc)} )$, and 
we explicitly extracted the ratio of the relevant 
symmetry factors $\varsi_\npo$ and $\varsi_n$.
The radiative phase-space measure associated with the unresolved particle $k_a$ can be parametrised in terms of the following kinematic variables:
\beq
x \, = \, \frac{s_{ac} + s_{bc} - s_{ab}}{s_{ac} + s_{bc}} \, , 
\qquad \qquad
z \, = \, \frac{s_{ac}}{s_{ac} + s_{bc}} \, .
\label{eq:CSparam2}
\eeq
The invariants constructed with momenta $k_a$, $k_b$, and $k_c$ can be written in terms of these integration variables, 
as
\beq
s_{ab} \, = \, (1-x)  \, \bs
\, ,
\qquad \quad
s_{ac} \, = \, z  \, \bs
\, ,
\qquad \quad
s_{bc} \, = \, (1-z) \, \bs
\, ,
\eeq
where 
\beq
\bs
\, = \,
\bs_{bc}^{(abc)} 
\, = \,
2 \bar{k}_{b}^{(abc)} \cdot k_c 
\, = \,
\frac{2 \bar{k}_{b}^{(abc)} \cdot  \bar{k}_c^{(abc)}}{x}
\, = \,
\frac{m_b^2 - Q^2}{x}
\, = \,
\frac{s_{ac} + s_{bc} - s_{ab}}{x}
\, .
\eeq
The parametrised radiative integration measure \cite{Catani:2002hc} reads
\beq
\int d \Phi_{\rm rad,\sFm\sI}^{(abc)} 
& = &
\frac{(2 \pi)^{-3+2\eps}\,\bs}{4} \,
\int d^{d-3} \Omega \,
\int_0^1 \! dx \, \big[\bs(1-x)+m_b^2\big]^{-\eps}
\int_0^{z_+} \! dz \, (z_+ - z)^{-\eps} z^{-\eps}
\, ,
\qquad
\eeq
with
\beq
z_{+}
\, = \,
\frac{\bs(1-x)}{\bs(1-x)+m_b^2}
\, .
\eeq
Upon integration over the solid angle, see Eq.~(\ref{eq:solidangle}), one gets
\beq
\label{eq:dPhirad FmI}
\int d \Phi_{\rm rad,\sFm\sI}^{(abc)} 
& = &
\frac{(4 \pi)^{-2+\eps}\,\bs}{\Gamma(1-\eps)} \,
\int_0^1 \! dx \, \big[\bs(1-x)+m_b^2\big]^{-\eps}
\int_0^{z_+} \! dz \, (z_+ - z)^{-\eps} z^{-\eps}
\, .
\qquad
\eeq


\section{Collinear kernels}
\label{app: coll ker}

We report here the Altarelli-Parisi kernels relevant for the final-state splitting of a parent parton $[ij]$ into a pair of collinear siblings $ij$:
\beq
\label{eq:APkernels_munu1 F}
P_{ij,\sF}^{\mu\nu}(z_i)
& = &
-
P_{ij,\sF}(z_i) \,
g^{\mu\nu}
+
Q_{ij,\sF}(z_i) \,
\bigg[ \!
- g^{\mu\nu} 
+ 
(d - 2) \, 
\frac{\widetilde{k}_{\sF}^\mu\,\widetilde{k}_{\sF}^\nu}{\widetilde{k}_{\sF}^2}
\bigg] 
\, ,
\eeq
with
\beq
P_{ij,\sF}(z_i)
& = &
T_R \bigg( 1 - \frac{2 z_i z_j}{1-\eps} \bigg) 
f_{ij}^{q \bar q}
+
C_F \bigg[ \frac{2z_i}{z_j} + (1-\eps)z_j \bigg] \,
f_{i}^{q,\bar q} \, f_{j}^{g}
\nnb\\
&&
+ \,
C_F \bigg[ \frac{2z_j}{z_i} + (1-\eps)z_i \bigg] \,
f_{i}^{g} \, f_{j}^{q,\bar q}
+
C_A \bigg( \frac{z_i}{z_j} + \frac{z_j}{z_i} + z_i z_j \bigg)
f_{ij}^{gg}
\, ,
\nnb\\
Q_{ij,\sF}(z_i)
& = &
T_R \, \frac{2 z_i z_j}{1-\eps} \,
f_{ij}^{q \bar q}
-
2 \, C_A \, z_i z_j \,
f_{ij}^{gg}
\, .
\eeq
The longitudinal momentum fractions of partons $i$, $j$, and their transverse momenta with respect to the collinear direction are
\beq
z_i
\, = \,
\frac{s_{ir}}{s_{ir}+s_{jr}}
\, = \,
1-z_j
\, ,
\qquad
\widetilde{k}_{\sF}^{\mu}
\, = \,
z_j \, k_i ^{\mu} 
- 
z_i \, k_j^{\mu}
-  
(z_j-z_i)
\frac{s_{ij}}{s_{ir}+s_{jr}}  k_r^{\mu}
\, .
\eeq
Upon subtracting all soft enhancements from the kernels, one obtains the final-state hard-collinear splitting functions:
\beq
\label{eq:APkernels_munu_hc F}
P_{ij,\sF}^{\mu\nu, \hc}(z_i)
& = &
-
P_{ij,\sF}^{\hc}(z_i) \,
g^{\mu\nu}
+
Q_{ij,\sF}(z_i) \,
\bigg[ \!
- g^{\mu\nu} 
+ 
(d - 2) \, 
\frac{\widetilde{k}_{\sF}^\mu\,\widetilde{k}_{\sF}^\nu}{\widetilde{k}_{\sF}^2}
\bigg] 
\, ,
\eeq
with 
\beq
\label{eq: PhcF}
P_{ij,\sF}^\hc(z_i)
& = &
P_{ij,\sF}^{{\rm hc},\zg}(z_i) \,
f_{ij}^{q \bar q}
+
P_{ij,\sF}^{{\rm hc},\og}(z_i) \,
f_{i}^{q,\bar q} \,f_{j}^{g}
\nnb\\
&&
\hspace{24mm}
+ \,
P_{ij,\sF}^{{\rm hc},\og}(z_j) \,
f_{i}^{g} \, f_{j}^{q,\bar q}
+
P_{ij,\sF}^{{\rm hc},\tg}(z_i) \,
f_{ij}^{gg}
\, ,
\nnb\\
P_{ij,\sF}^{{\rm hc},\zg}(z) 
& = &
T_R \bigg[ 1 - \frac{2 z(1-z)}{1-\eps} \bigg]
\, ,
\nnb\\
P_{ij,\sF}^{{\rm hc},\og}(z) 
& = &
C_F (1-\eps)(1-z)
\, ,
\nnb\\
P_{ij,\sF}^{{\rm hc},\tg}(z) 
& = &
2 C_A \, z(1-z)
\, .
\eeq

The kernels for the initial-state splitting of a parent parton $c=j$ (or $c=i$) into a pair of collinear siblings $ab = [ij]i$ (or $ab = [ji]j$) are
\beq
\label{eq:APkernels_munu2 I}
P_{ab,\sI}^{\mu\nu}(x_a)
& = &
-
P_{ab,\sI}(x_a) \,
g^{\mu\nu}
+
Q_{ab,\sI}(x_a) \,
\bigg[  
- g^{\mu\nu} 
+ 
(d \!-\! 2) \, 
\frac{\widetilde{k}_{\sI}^\mu\,\widetilde{k}_{\sI}^\nu}{\widetilde{k}_{\sI}^2}
\bigg] 
\, ,
\eeq
with
\beq
P_{ab,\sI}(x_a)
& = &
P_{ab,\sF}(x_a)
\, ,
\nnb\\
Q_{ab,\sI}(x_a)
& = &
-
2 \, \frac{x_b}{x_a} \,
C_F \, f_{a}^{g} \, f_{b}^{q,\bar q}
-
2 \, \frac{x_b}{x_a} \,
C_A \, f_{ab}^{gg}
\, .
\eeq
In this case, longitudinal fractions and transverse momenta are defined as 
\beq
x_b
\, = \,
\frac{s_{br}}{s_{cr}}
\, = \,
1-x_{a}
\, ,
\qquad
\widetilde{k}_{\sI}^\mu 
\, = \, 
k_b^{\mu} 
-
x_b \, k_c ^{\mu}
-
\frac{s_{bc}}{s_{cr}} \, k_r^{\mu}
\, .
\qquad
\eeq
The hard-collinear version of the kernels reads
\beq
\label{eq:APkernels_munu_hc I}
P_{ab,\sI}^{\mu\nu, \hc}(x_a)
& = &
-
P_{ab,\sI}^\hc(x_a) \,
g^{\mu\nu} 
+
Q_{ab,\sI}(x_a) \,
\bigg[  
- \! g^{\mu\nu} 
+ 
(d \!-\! 2) \, 
\frac{\widetilde{k}_{\sI}^\mu\,\widetilde{k}_{\sI}^\nu}{\widetilde{k}_{\sI}^2}
\bigg] 
\, ,
\nnb
\eeq
where we have defined
\beq
\label{eq: PhcI}
P_{ab,\sI}^\hc(x_a)
& = &
P_{ab,\sI}^{{\rm hc},\zg}(x_a) \,
f_{ab}^{q \bar q}
+
P_{ab,\sI}^{{\rm hc},\qg}(x_a) \,
f_{a}^{q,\bar q} \,f_{b}^{g}
\nnb\\
&&
\hspace{24mm}
+ \,
P_{ab,\sI}^{{\rm hc},\gq}(x_a) \,
f_{a}^{g} \, f_{b}^{q,\bar q}
+
P_{ab,\sI}^{{\rm hc},\tg}(x_a) \,
f_{ab}^{gg}
\, ,
\nnb\\
P_{ab,\sI}^{ \hc,\zg}(x)
& = &
T_R 
\bigg[ 1 - \frac{2 \ x(1-x)}{1-\eps} \bigg]
\, ,
\nnb\\
P_{ab,\sI}^{ \hc,\qg}(x)
& = &
C_F  (1- \eps) \, (1-x)
\, ,
\nnb\\
P_{ab,\sI}^{ \hc,\gq}(x)
& = &
C_F 
\bigg[ \frac{1+(1-x)^2}{x} - \epsilon \, x \bigg]
\, ,
\nnb\\
P_{ab,\sI}^{ \hc,\tg}(x)
& = &
2 C_A 
\bigg[ \frac{1-x}{x} + x(1-x) \bigg]
\, .
\eeq


\section{Integration of massive counterterms}
\label{app: IntMassiveApp} 

In this appendix we outline the integration of all NLO counterterms in 
presence of massive particles in the final state.


\subsection{Integration of the fully massive soft counterterm}
\label{app: IsFmFm} 

We start by analytically integrating the soft counterterm 
\beq
I_{\rm s,\sFm\sFm}^{ikl}
& = &
\Norm
\int d\Phi_{\rm rad,\sFm\sFm}^{(ikl)} \, \mc E_{kl}^{(i)}
\, ,
\eeq
over the radiative phase space $d\Phi_{\rm rad,\sFm\sFm}^{(ikl)}$. This is obtained with the massive mapping of \eq{eq:mapping}, with the identifications $a\to i$, $b\to k$, $c\to l$, recalling that parton $i$ is a gluon, thus massless. 
In \appn{app: Fm-Fm mapping} we have introduced different equivalent parametrisations for $\int d\Phi_{\rm rad,\sFm\sFm}^{(abc)}$. 
To integrate the eikonal kernel $\mc E_{kl}^{(i)}$, which is symmetric under $k \leftrightarrow l$ exchange, it is best to use the symmetric parametrisation of \eq{eq: dPhirad FmFm uv}. 
With the invariants of \eq{eq: invariants uv}, the eikonal kernel reads 
\beq
\mc E_{kl}^{(i)}
& = &
\mc I_{kl}^{(i)} 
-
\frac{1}{2}\, \mc I_{kk}^{(i)} 
-
\frac{1}{2}\, \mc I_{ll}^{(i)} 
=
f_i^g
\bigg(
\frac{s_{kl}}{s_{ik} s_{il}}
-
\frac{m_k^2}{s_{ik}^2}
-
\frac{m_l^2}{s_{il}^2}
\bigg)
\nnb\\
& = &
f_i^g \,
\frac1{\bs(1-u)^2}
\bigg[
\frac{u}{v(1-v)}
-
\frac{m_k^2}{\bs (1-v)^2}
-
\frac{m_l^2}{\bs v^2}
\bigg]
\, ,
\eeq
with $\bs=\sk{kl}{ikl}$.
It is interesting to notice that the numerator of the eikonal kernel 
reproduces the $\eps$-dependent part of the phase-space measure: 
\beq
\mc E_{kl}^{(i)}
& = &
f_i^g \,
\frac
{\bs\,u\,v(1-v) - m_k^2\,v^2 - m_l^2\,(1-v)^2}
{\bs^2 (1-u)^2 v^2 (1-v)^2}
\nnb\\
& = &
f_i^g \,
\frac
{(\bs u+m_k^2+m_l^2)(v_+-v)(v-v_-)}
{\bs^2 (1-u)^2 v^2 (1-v)^2}
\, .
\eeq
The fully massive soft integral $I_{\rm s,\sFm\sFm}^{ikl}$
is then given by 
\beq
I_{\rm s,\sFm\sFm}^{ikl}
& = &
f_i^g \,
\Norm \,
\frac{(4 \pi)^{-2+\eps}}{\Gamma(1-\eps)} \!
\left(\!\frac{\bs}{\sqrt{\lambda}}\!\right)^{\!\!1-2\eps} \!\!
\int_{u_-}^1 \!\!\!\! du \, (1-u)^{-1-2\eps} 
(\bs u \!+\! m_k^2 \!+\! m_l^2)^{-\eps} \!
\nnb\\
&&
\qquad
\times \, 
\int_{v_-}^{v_+} \!\!\!\! dv \, 
(v_+ \!-\! v)^{-\eps} (v \!-\! v_-)^{-\eps}
\bigg[
\frac{u}{v}
+
\frac{u}{1-v}
-
\frac{m_k^2}{\bs(1-v)^2}
-
\frac{m_l^2}{\bs v^2}
\bigg]
\, ,
\qquad
\eeq
with 
\beq
\lambda = \bs^2 - 4 m_k^2 m_l^2
\, ,
\qquad
u_- = \frac{2 m_k m_l}{\bs}
\, ,
\qquad
v_\pm
=
\frac
{\bs u+2m_l^2 \pm \sqrt{\bs^2u^2 - 4 m_k^2 m_l^2}}
{2(\bs u+m_k^2+m_l^2)}
\, .
\qquad
\eeq
The only singular behaviour giving raise to a pole in $\eps$ is for 
$u\to1$. 
We thus subtract and add back the integrand at $u=1$:
\beq
I_{\rm s,\sFm\sFm}^{ikl}
& = &
I_{\rm s,\sFm\sFm}^{ikl(0)}
+
I_{\rm s,\sFm\sFm}^{ikl(1)}
+
\mathcal{O}(\eps)
\, ,
\\[2mm]
I_{\rm s,\sFm\sFm}^{ikl(0)}
& = &
f_i^g \,
\frac{\as}{2\pi} 
\left( \frac{\bs}{\mu^2}\right)^{\!\!-\eps} \!\!
\left(\!\frac{\bs}{\sqrt{\lambda}}\!\right)^{\!\!1-2\eps} \!\!
\bigg(\frac{Q^2}{\bs} \bigg)^{\!\!-\eps} \!
\int_{u_-}^1 \!\!\!\! du \, (1-u)^{-1-2\eps} 
\nnb\\
&&
\qquad \qquad
\times \, 
\int_{V_-}^{V_+} \!\!\!\! dv \, 
(V_+ \!-\! v)^{-\eps} (v \!-\! V_-)^{-\eps}
\bigg[
\frac{1}{1-v}
+
\frac{1}{v}
-
\frac{m_k^2}{\bs(1-v)^2}
-
\frac{m_l^2}{\bs v^2}
\bigg]
\, ,
\nnb\\[2mm]
I_{\rm s,\sFm\sFm}^{ikl(1)}
& = &
f_i^g \,
\frac{\as}{2\pi} \left( \frac{\bs}{\mu^2}\right)^{\!\!-\eps} 
\frac{\bs}{\sqrt{\lambda}}
\int_{u_-}^1 \!\! \frac{du}{1-u} \,
\bigg\{ \,
\int_{v_-(u)}^{v_+(u)} \!\!\!\! dv \, 
\bigg[
\frac{u}{1-v}
+
\frac{u}{v}
-
\frac{m_k^2}{\bs(1-v)^2}
-
\frac{m_l^2}{\bs v^2}
\bigg]
\nnb\\
&&
\hspace{47mm}
-
\int_{v_-(1)}^{v_+(1)} \!\!\!\! dv \, 
\bigg[
\frac{1}{1-v}
+
\frac{1}{v}
-
\frac{m_k^2}{\bs(1-v)^2}
-
\frac{m_l^2}{\bs v^2}
\bigg]
\bigg\}
\, ,
\qquad
\eeq
where we have set 
\beq
V_\pm 
\, \equiv \, 
v_\pm(1) 
\, = \,
\frac{\bs+2m_l^2 \pm \sqrt{\lambda}}{2(\bs+m_k^2+m_l^2)}
\, .
\eeq
To compute $I_{\rm s,\sFm\sFm}^{ikl(0)}$, it is useful to introduce 
the change of variable 
\beq
v'
\, = \,
\frac{v-V_-}{V_+-V_-}
\, ,
\qquad\quad
v = V_- + (V_+ - V_-) v'
\, ,
\eeq
thus obtaining
\beq
I_{\rm s,\sFm\sFm}^{ikl(0)}
& = &
f_i^g \,
\frac{\as}{2\pi} \left( \frac{\bs}{\mu^2}\right)^{-\eps} 
\frac{\bs}{\sqrt{\lambda}}\,
\left(\frac{\bs\,Q^2}{\lambda}\right)^{-\eps} 
\int_{u_-}^1 du \, (1-u)^{-1-2\eps} 
(V_+-V_-)^{1-2\eps}
\nnb\\
&&
\hspace{-4mm}
\times \, 
\int_0^1 dv' \, 
[v'(1-v')]^{-\eps}
\bigg\{
\frac{1}{1 - V_- - (V_+-V_-) v'}
+
\frac{1}{V_- + (V_+-V_-) v'}
\nnb\\
&&
\hspace{29mm}
- \,
\frac{m_k^2}{\bs [1 - V_- - (V_+-V_-) v']^2}
-
\frac{m_l^2}{\bs [V_- + (V_+-V_-) v']^2}
\bigg\}
\, .
\qquad
\eeq
The $u$-integration generates a single $\eps$ pole, as anticipated
\beq
\int_{u_-}^1 du \, (1-u)^{-1-2\eps} 
& = &
\frac{(1-u_-)^{-2\eps}}{-2\eps}
\, = \, 
- \frac{1}{2\eps} + \ln\frac{\bs - 2 m_k m_l}{\bs} + \cal O(\eps)
\, ,
\eeq
while the $v$-integration can be performed using the standard results
\beq
&&
\int_0^1 dv' \, 
[v'(1-v')]^{-\eps}
\frac{1}{a v'+b}
\, = \,
\frac{1}{a} \,
\ln\frac{a+b}{b}
+
\frac{\eps}{a} \,
\bigg[
-
\Li_2\bigg( - \frac{a}{b} \bigg)
+
\Li_2\bigg( \frac{a}{a+b} \bigg)
\bigg]
+ 
{\cal O}(\eps^2)
\, ,
\nnb\\[2mm]
&&
\int_0^1  dv' \, 
[v'(1-v')]^{-\eps}
\frac{1}{(a v'+b)^2}
\, = \,
\frac{1}{b(a+b)} \,
\bigg[
1
+
\eps \,
\frac{a+2b}{a} \,
\ln\frac{a+b}{b}
\bigg]
+ 
{\cal O}(\eps^2)
\, .
\eeq
Considering that 
\beq
&&
V_+ - V_-
\, = \,  
\frac{\sqrt{\lambda}}{\bs+m_k^2+m_l^2}
\, ,
\qquad
V_+ + V_-
\, = \,  
\frac{\bs+2m_l^2}{\bs+m_k^2+m_l^2}
\, ,
\nnb\\
&&
2 - V_+ -  V_-
\, = \,  
\frac{\bs+2m_k^2}{\bs+m_k^2+m_l^2}
\, ,
\qquad
V_+ V_-
\, = \, 
\frac{m_l^2}{\bs+m_k^2+m_l^2}
\, ,
\nnb\\
&&
(1-V_+)(1-V_-)
\, = \, 
\frac{m_k^2}{\bs+m_k^2+m_l^2}
\, ,
\qquad
\frac{V_-(1-V_+)}{V_+(1-V_-)}
\, = \,
\frac{\bs-\sqrt{\lambda}}{\bs+\sqrt{\lambda}}
\, ,
\qquad
\eeq
and using the di-logarithmic identity
\beq
\Li_2\bigg( - \frac{2\sqrt{\lambda}}{\bs+2m^2-\sqrt{\lambda}} \bigg)
& = &
-
\Li_2\bigg( \frac{2\sqrt{\lambda}}{\bs+2m^2+\sqrt{\lambda}} \bigg)
-
\frac12 \,
\ln^2\frac{\bs+2m^2+\sqrt{\lambda}}{\bs+2m^2-\sqrt{\lambda}}
\, ,
\qquad
\eeq
we get 
\beq
I_{\rm s,\sFm\sFm}^{ikl(0)}
& = &
- \,
f_i^g \,
\frac{\as}{2\pi}
\left( \frac{\bs}{\mu^2}\right)^{-\eps}
\frac{\bs}{2\sqrt{\lambda}} \,
\Bigg\{
-
\bigg[ \frac{1}{\eps} + \ln\frac{Q^2\,\bs}{(\bs - 2 m_k m_l)^2} \bigg]
\bigg[
\ln\frac{\bs-\sqrt{\lambda}}{\bs+\sqrt{\lambda}}
+
\frac{2\sqrt{\lambda}}{\bs}
\bigg]
\nnb\\
&&
\hspace{15mm}
+ \,
\frac12 \,
\ln^2
\frac
{\bs + 2m_k^2 + \sqrt{\lambda}}
{\bs + 2m_k^2 - \sqrt{\lambda}}
-
\frac{2m_k^2}{\bs} \,
\ln
\frac
{\bs + 2m_k^2 + \sqrt{\lambda}}
{\bs + 2m_k^2 - \sqrt{\lambda}}
\nnb\\
&&
\hspace{15mm}
+
\frac12 \,
\ln^2
\frac
{\bs + 2m_l^2 + \sqrt{\lambda}}
{\bs + 2m_l^2 - \sqrt{\lambda}}
-
\frac{2m_l^2}{\bs} \,
\ln
\frac
{\bs + 2m_l^2 + \sqrt{\lambda}}
{\bs + 2m_l^2 - \sqrt{\lambda}}
+
\ln\frac{\bs-\sqrt{\lambda}}{\bs+\sqrt{\lambda}}
\nnb\\
&&
\hspace{15mm}
+ \,
2 \,
\Li_2\bigg( \frac{2\sqrt{\lambda}}{\bs+2m_k^2+\sqrt{\lambda}} \bigg)
+
2 \,
\Li_2\bigg( \frac{2\sqrt{\lambda}}{\bs+2m_l^2+\sqrt{\lambda}} \bigg)
+ 
{\cal O}(\eps)
\Bigg\}
\, .
\qquad
\eeq
Next, we introduce the convenient notation 
\beq
\eta_{k,l}
=
\frac
{\bs+2m_{k,l}^2-\sqrt{\lambda}}
{\bs+2m_{k,l}^2+\sqrt{\lambda}}
\, ,
\qquad\qquad
\eta 
=
\eta_k \, \eta_l
=
\frac{\bs-\sqrt{\lambda}}{\bs+\sqrt{\lambda}}
\, ,
\qquad
\eeq
and arrive at our final expression for $I_{\rm s,\sFm\sFm}^{ikl(0)}$:
\beq
I_{\rm s,\sFm\sFm}^{ikl(0)}
& = &
f_i^g \,
\frac{\as}{2\pi}
\left(\frac{\bs}{\mu^2}\right)^{-\eps}
\Bigg\{
\bigg[ 
\frac{1}{\eps} + \ln\frac{Q^2\,\bs}{(\bs - 2 m_k m_l)^2} 
\bigg]
\bigg[
\frac{\bs}{2\sqrt{\lambda}} \ln\eta
+
1
\bigg]
\nnb\\
&&
\hspace{30mm}
- \,
\frac{\bs}{\sqrt{\lambda}} \,
\bigg[
\Li_2(1-\eta_k)
+
\Li_2(1-\eta_l)
+
\frac18
\ln^2\frac{\eta_k}{\eta_l}
+
\frac{\ln^2\eta}{8}
\bigg]
\nnb\\
&&
\hspace{30mm}
- \,
\frac{m_k^2-m_l^2}{2\sqrt{\lambda}} 
\ln\frac{\eta_k}{\eta_l} 
-
\frac{Q^2}{2\sqrt{\lambda}} \ln\eta 
+ 
{\cal O}(\eps)
\Bigg\}
\, .
\eeq

We now turn to the computation of $I_{\rm s,\sFm\sFm}^{ikl(1)}$. We first perform the $v$-integration, making use of the following relations:
\beq
&&
v_+ v_-
=
\frac{m_l^2}{\bs u+m_k^2+m_l^2}
\, ,
\qquad
(1-v_+)(1-v_-)
=
\frac{m_k^2}{\bs u+m_k^2+m_l^2}
\, ,
\nnb\\
&&
v_+ - v_-
= 
\frac{\sqrt{\bs^2u^2-4m_k^2m_l^2}}{\bs u+m_k^2+m_l^2}
\, ,
\qquad
\frac{v_-(1-v_+)}{v_+(1-v_-)}
=
\frac
{\bs u-\sqrt{\bs^2u^2-4m_k^2m_l^2}}
{\bs u+\sqrt{\bs^2u^2-4m_k^2m_l^2}}
\, ,
\eeq
and get
\beq
I_{\rm s,\sFm\sFm}^{ikl(1)}
& = &
f_i^g \,
\frac{\as}{2\pi} \left( \frac{\bs}{\mu^2}\right)^{-\eps} 
\frac{\bs}{\sqrt{\lambda}}
\int_{u_-}^1 \frac{du}{1-u} \,
\bigg[
- \,
u\,
\ln
\frac
{\bs u-\sqrt{\bs^2u^2-4m_k^2m_l^2}}
{\bs u+\sqrt{\bs^2u^2-4m_k^2m_l^2}}
+
\ln\eta
\nnb\\
&&
\hspace{47mm}
- \,
2 \,
\frac{\sqrt{\bs^2u^2-4m_k^2m_l^2}}{\bs} 
+ 
2 \,
\frac{\sqrt{\lambda}}{\bs} 
\bigg]
\, .
\eeq
At this point, we perform a change of variables to get rid 
of the square root. 
Denoting as $u_1$ and $u_2$ the two roots of the quadratic 
equation $\bs^2u^2 - 4m_k^2m_l^2 = 0$, namely 
\beq
u_1 
\, = \,
\frac{2 m_k m_l}{\bs}
\, = \,
u_-
\, ,
\qquad\qquad
u_2
\, = \,
- \, \frac{2 m_k m_l}{\bs}
\, ,
\eeq
we introduce the variables
\beq
t \, = \,
\sqrt{\frac{u-u_1}{u-u_2}}
\, ,
\qquad\qquad
\beta 
& = &
\sqrt{\frac{1-u_1}{1-u_2}} 
\, = \,
\sqrt{\frac{\bs-2m_km_l}{\bs+2m_km_l}}
\, .
\eeq
This modifies the integration measure as follows:
\beq
&&
u 
\, = \, 
\frac{2 m_k m_l}{\bs}\,\frac{1+t^2}{1-t^2}
\, ,
\qquad
1-u 
\, = \,
\frac{\bs+2m_km_l}{\bs}\,\frac{\beta^2-t^2}{1-t^2}
\, ,
\nnb\\
&&
\sqrt{\bs^2u^2-4m_k^2m_l^2}
\, = \,
4 m_k m_l\,\frac{t}{1-t^2}
\, ,
\qquad
\int_{u_-}^1 du
\, = \,
\frac{4 m_k m_l}{\bs} \int_0^{\beta} \frac{2t\,dt}{(1-t^2)^2}
\, .
\eeq
The constants appearing in the $t$-integration can be recast in terms of the sole variable $\beta$, with the help of the following relations: 
\beq
&&
1-\beta^2
\, = \, 
\frac{4m_km_l}{\bs+2m_km_l}
\, ,
\qquad\qquad
\frac{\beta}{1-\beta^2} 
\, = \,
\frac{\sqrt{\lambda}}{4m_km_l}
\, ,
\nnb\\
&&
\frac{1+\beta^2}{1-\beta^2} 
\, = \,
\frac{\bs}{2m_km_l}
\, ,
\qquad\qquad
\bigg(\frac{1-\beta}{1+\beta}\bigg)^{2}
\, = \,
\frac{\bs-\sqrt{\lambda}}{\bs+\sqrt{\lambda}}
\, = \,
\eta
\, .
\eeq
The $I_{\rm s,\sFm\sFm}^{ikl(1)}$ integral then becomes
\beq
I_{\rm s,\sFm\sFm}^{ikl(1)}
& = &
- \,
f_i^g \,
\frac{\as}{2\pi} \left( \frac{\bs}{\mu^2}\right)^{-\eps} 
\frac{\bs}{\sqrt{\lambda}} \,
\frac{4m_km_l}{\bs}\,
\int_0^{\beta} dt \, 
\nnb\\
&&
\qquad
\times \,
\bigg[
\frac{1+\beta^2}{1-\beta^2}
\bigg(
\frac{1}{\beta-t} 
- 
\frac{1}{\beta+t}
- 
\frac{1}{1-t} 
+ 
\frac{1}{1+t}
\bigg)
\ln\frac{1-t}{1-\beta}\frac{1+\beta}{1+t}
\nnb\\
&&
\qquad\qquad
- \,
\frac{1}{\beta}
\frac{2}{1-\beta^2}
\ln\frac{1-\beta}{1+\beta}
+
\frac{4\beta}{1-\beta^2}
\frac{1}{\beta+t} 
- 
\frac{2}{1+\beta} \,
\frac{1}{1-t} 
\nnb\\
&&
\qquad\qquad
- \,
\frac{2}{1-\beta} \,
\frac{1}{1+t}
-
\frac{2}{(1-t)^2} 
- 
\frac{2}{(1+t)^2} 
\bigg]
\, .
\eeq
Its explicit evaluation gives
\beq
I_{\rm s,\sFm\sFm}^{ikl(1)}
& = &
- \,
f_i^g \,
\frac{\as}{2\pi} \left( \frac{\bs}{\mu^2}\right)^{-\eps} 
\frac{\bs}{\sqrt{\lambda}} \,
\frac{4m_km_l}{\bs}\,
\nnb\\
&&
\qquad
\times \,
\bigg\{
\frac{1+\beta^2}{1-\beta^2}
\bigg[
-
\Li_2\bigg[ \bigg(\frac{1-\beta}{1+\beta}\bigg)^{2\,} \bigg]
-
2 
\ln\frac{2\beta}{1+\beta}
\ln\frac{1-\beta}{1+\beta}
+
\zeta_2
\bigg]
\nnb\\
&&
\qquad\qquad
- \,
\frac{4\beta}{1-\beta^2}
+ 
\frac{4\beta}{1-\beta^2}
\ln 2 
- 
\frac{2\beta}{1-\beta^2}
\ln(1-\beta) 
- 
\frac{2\beta}{1-\beta^2}
\ln(1+\beta)
\bigg\}
\nnb\\[5pt]
& = &
f_i^g \,
\frac{\as}{2\pi} \left( \frac{\bs}{\mu^2}\right)^{-\eps} 
\bigg\{
- 
\ln\frac{m_km_l}{\bs-2m_km_l}
\ln\eta
+
2 \,
\Li_2\big(\eta \big)
+ 
\frac12 \,
\ln^2\eta
-
2 \,
\zeta_2
\nnb\\
&&
\hspace{60mm}
+ \,
2 \,
\frac{\sqrt{\lambda}}{\bs}
\bigg[
2
+ 
\ln\frac{m_km_l}{\bs+2m_km_l} 
\bigg] \,
\bigg\}
\, .
\eeq
Collecting all contributions, we finally obtain 
\beq
\label{eq: IsMM}
I_{\rm s,\sFm\sFm}^{ikl}
& = &
I_{\rm s,\sFm\sFm}^{ikl(0)}
+
I_{\rm s,\sFm\sFm}^{ikl(1)}
+ 
{\cal O}(\eps)
\\
& = &
f_i^g \,
\frac{\as}{2\pi}
\left( \frac{\bs}{\mu^2}\right)^{-\eps}
\bigg\{
\bigg[ \frac{1}{\eps} + \ln\frac{Q^2\,\bs}{(\bs  - 2 m_k m_l)^2} \bigg]
\bigg[
\frac{\bs}{2\sqrt{\lambda}} \ln\eta
+
1
\bigg]
\nnb\\
&&
\hspace{20mm}
- \,
\frac{\bs}{\sqrt{\lambda}} \,
\bigg[
\Li_2(1-\eta_k)
+
\Li_2(1-\eta_l)
+
\frac18
\ln^2\frac{\eta_k}{\eta_l}
+
\frac18
\ln^2\eta
\bigg]
\nnb\\
&&
\hspace{20mm}
- \,
\frac{m_k^2-m_l^2}{2\sqrt{\lambda}}
\ln\frac{\eta_k}{\eta_l} 
-
\frac{Q^2}{2\sqrt{\lambda}} \ln\eta 
- 
\ln\frac{m_km_l}{\bs-2m_km_l}
\ln\eta
\nnb\\
&&
\hspace{20mm}
+ \,
2 \,
\Li_2(\eta)
+
\frac12 
\ln^2\eta
-
2 \,
\zeta_2
+
2 \,
\frac{\sqrt{\lambda}}{\bs}
\bigg[
2
+ 
\ln\frac{m_km_l}{\bs+2m_km_l} 
\bigg]
+ 
{\cal O}(\eps)
\bigg\}
\, ,
\nnb
\eeq
\beq
\bs = \sk{kl}{ikl}
\, ,
\qquad\qquad
\eta_{k,l}
=
\frac
{\bs+2m_{k,l}^2-\sqrt{\lambda}}
{\bs+2m_{k,l}^2+\sqrt{\lambda}}
\, ,
\qquad\qquad
\eta 
=
\eta_k \, \eta_l
=
\frac{\bs-\sqrt{\lambda}}{\bs+\sqrt{\lambda}}
\, .
\qquad
\eeq


\subsection{Integration of massive-massless soft counterterms}
\label{app: IsFmF, IsFFm}

In case the massive colour source $k$ and the massless source $l$ are both in the final state, the integral of the eikonal kernel is given by 
\beq
I_{\rm s,\sFm\sF}^{ikl}
& = &
\Norm
\int d\Phi_{\rm rad,\sFm\sF}^{(ikl)} \, \mc E_{kl}^{(i)}
\, ,
\eeq
where $d\Phi_{\rm rad,\sFm\sF}^{(ikl)}$ is the radiative phase-space measure of \eq{eq:dPhirad FmF} with the replacement $a \to i$, $b\to k$, $c\to l$. 
Similarly, if $k$ is massless and $l$ is massive, we have
\beq
I_{\rm s,\sF\sFm}^{ikl}
& = &
\Norm
\int d \Phi_{\rm rad,\sF\sFm}^{(ikl)} \, \mc E_{kl}^{(i)}
\, ,
\eeq
where $d\Phi_{\rm rad,\sF\sFm}^{(ikl)}$ is the radiative phase-space measure of \eq{eq:dPhirad FFm} with $a \to i$, $b\to k$, $c\to l$.
The eikonal kernel is
\beq
\mc E_{kl}^{(i)}
\, = \,
\mc I_{kl}^{(i)} 
-
\frac{1}{2}\, \mc I_{kk}^{(i)} 
-
\frac{1}{2}\, \mc I_{ll}^{(i)} 
\, , 
\eeq
and the individual soft contributions, with generic masses $m_k$ and 
$m_l$, read ($\bs=\sk{kl}{ikl}$)
\beq
\mc I_{kl}^{(i)}
& = &
f_i^g \,
\frac{s_{kl}}{s_{ik} s_{il}}
\, = \,
f_i^g \,
\frac{1-z}{\bs yz}
\, = \,
f_i^g \,
\frac{1-Z}{\bs\,YZ}
\, ,
\nnb\\
\mc I_{kk}^{(i)}
& = &
f_i^g \,
\frac{s_{kk}}{s_{ik}^2}
\, = \,
f_i^g \,
\frac{2 \, m_k^2}{\bs^2 y^2}
\, = \,
f_i^g \,
\frac{2 \, m_k^2}{\bs^2 Z^2 (1-Y)^2}
\, ,
\nnb\\
\mc I_{ll}^{(i)}
& = &
f_i^g \,
\frac{s_{ll}}{s_{il}^2}
\, = \,
f_i^g \,
\frac{2 \, m_l^2}{\bs^2 z^2 (1-y)^2}
\, = \,
f_i^g \,
\frac{2 \, m_l^2}{\bs^2 Y^2}
\, .
\label{eq:eik2}
\eeq
We note that the symmetry of $\mc I_{kl}^{(i)}$ under exchange 
$k \leftrightarrow l$ implies that its expression is identical both 
in terms of $(Y,Z)$, and of $(y,z)$. 

As for $I_{\rm s,\sFm\sF}^{ikl}$, in which $m_l=0$, the integration measure is simpler when written in terms of the $(y,z)$ variables:
\beq
\label{eq:IiklsMF}
I_{\rm s,\sFm\sF}^{ikl}
& = &
f_i^g \,
\Norm
\frac{(4 \pi)^{-2+\eps}}{\Gamma(1-\eps)} \,
\int_0^1 dy \, (1-y)^{1-2\eps} \,
(\bs y + m_k^2)^{-\eps}
\nnb\\
&&
\qquad\qquad\qquad
\times \,
\int_0^{z_+} dz \, z^{-\eps} (z_+ - z)^{-\eps} 
\bigg[
\frac{1 - z}{yz} - \frac{m_k^2}{\bs y^2}
\bigg]
\nnb\\
& \equiv &
f_i^g \,
I_{\rm s}^{_{\sFm\sF}}(\bs, m_k)
\, ,
\eeq
with $ z_{+} = z_+(m_k) = \bs y/(\bs y+m_k^2)$.

For $I_{\rm s,\sF\sFm}^{ikl}$ the parametrisation in terms of $(Y,Z)$ is more convenient. We obtain an expression which is identical to \eq{eq:IiklsMF}, with the formal replacements $(y,z,m_k)\to(Y,Z,m_l)$.
This in turn means that 
$I_{\rm s,\sF\sFm}^{ikl}=f_i^g\,I_{\rm s}^{_{\sFm\sF}}\big(\sk{kl}{ikl},m_l\big)$, 
hence we can compute a single constituent integral.

We proceed to the computation of $I_{\rm s}^{_{\sFm\sF}}(s,m)$ by performing 
the change of variable $z\to z_+(m) z$, obtaining straightforwardly
\beq
\label{eq: IsMF}
I_{\rm s}^{_{\sFm\sF}}(s,m)
& = &
\Norm
\frac{(4 \pi)^{-2+\eps}}{\Gamma(1-\eps)} \,
s^{\,-\eps}
\int_0^1 dy \, y^{-1-2\eps} \,(1-y)^{1-2\eps} 
(y + \rho)^{\eps} 
\int_0^1 \! dz \, z^{-1-\eps} (1 - z)^{1-\eps}
\nnb\\
& = &
\frac{\as}{2\pi}
\left( \frac{s}{\mu^2}\right)^{-\eps}
\bigg[\,
\frac{1}{2\eps^2}
+
\frac{1}{2\eps}
(3 + \ln\rho)
-
\Li_2(-\rho)
-
\frac{\ln^2\rho}{4}
+
(1+\rho)\ln(1+\rho)
\nnb\\
&&
\hspace{21mm}
+ \,
\bigg(\frac12-\rho\bigg)\ln\rho
-
\frac{15}{4}\,\zeta_2
+
5
+
{\cal O}(\eps)
\bigg]
\, ,
\eeq
with $\rho = m^2/s$.


If we now consider the case of a massive source $k$ in the final state, and a massless source $l$ in the initial state, the integral of the final-initial eikonal kernel of \eq{eq:IsMassive} is given by 
\beq
\label{eq:intSoftMI}
I_{\rm s,\sFm\sI}^{ikl}
+ 
\int_0^1 \frac{dx}{x} 
J_{\rm s,\sFm\sI}^{ikl}(x)
& = &
\Norm
\int d\Phi_{\rm rad,\sFm\sI}^{(ikl)} \,
\mc E_{kl}^{(i)}
\, \equiv \,
{\cal J}_{\rm rad}^{\sFm\sI}\big(\sk{kl}{ikl},m_k\big)
\, ,
\eeq
where $d\Phi_{\rm rad,\sFm\sI}^{(ikl)}$ is the radiative phase-space measure of \eq{eq:dPhirad FmI} with the replacements $a \to i$, $b\to k$, 
$c\to l$. The case where $k$ and $l$ are exchanged, also relevant for \eq{eq:IsMassive}, gives the same result upon replacing $m_k\to m_l$, again owing to the symmetry of the eikonal kernel.
In the chosen parametrisation, the integrand is given by ($\bs=\sk{kl}{ikl}$)
\beq
\mc E_{kl}^{(i)}
\, = \,
\mc I_{kl}^{(i)} 
-
\frac{1}{2}\, \mc I_{kk}^{(i)} 
\, = \,
f_i^g \,
\bigg[
\frac{1-z}{\bs(1-x)z}
-
\frac{m_k^2}{\bs^2 (1-x)^2}
\bigg]
\, ,
\eeq
whence
\beq
{\cal J}_{\rm rad}^{\sFm\sI}(s,m)
& = &
f_i^g \,
\Norm
\frac{(4 \pi)^{-2+\eps}}{\Gamma(1-\eps)}
\int_0^1 \frac{dx}{1-x} \, 
\big[s(1-x)+m^2\big]^{-\eps}
\nnb\\
& \times &
\int_0^{z_+(m)} dz \,
(z_+(m) - z)^{-\eps} z^{-\eps} 
\bigg[
\frac{1-z}{z} - \frac{m^2}{s(1-x)}
\bigg]
,
\qquad
\eeq
with $z_{+}(m) = s(1-x)/(s(1-x)+m^2)$.
We then perform the usual change of variable $z \to z_+(m)\,z$ and get 
\beq
{\cal J}_{\rm rad}^{\sFm\sI}(s,m)
& = &
f_i^g \,
\Norm
\frac{(4 \pi)^{-2+\eps}\,s^{\,-\eps}}{\Gamma(1-\eps)}
\int_0^1 dx \,
(1-x)^{-1-2\eps} \,
\big(1-x+\rho\big)^{\eps}
\nnb\\
&&
\hspace{40mm}
\times \,
\int_0^1 dz \,z^{-1-\eps} (1-z)^{1-\eps} 
\, ,
\eeq
with $\rho=m^2/s$. 
The $\eps$ pole generated by the integration over $z$ can be easily extracted in closed form, as it happens for final-state soft radiation:
\beq
\int_0^1 dz \,z^{-1-\eps} (1-z)^{1-\eps} 
& = &
\frac{\Gamma(-\eps)\Gamma(2-\eps)}{\Gamma(2-2\eps)}
\, .
\qquad
\eeq
Conversely, when it comes to extracting the $\eps$ pole stemming from the $x$-integration, we have to recall that the Born matrix element $\bar{\Bn}^{(ikl)}_{kl}(xk_l)$ is $x$-dependent. 
We then expand the integral as the sum of an endpoint and a subtracted contribution:
\beq
\int_0^1 dx \,
(1-x)^{-1-2\eps} \,
\big(1-x+\rho\big)^{\eps} \,
f(x)
& = &
\frac{\rho^\eps \,\Gamma(-2\eps)}{\Gamma(2-2\eps)}\,
_2F_1\bigg(-\eps,-2\eps,2-2\eps,-\frac{1}{\rho}\bigg) \,
f(1)
\nnb\\
&&
\qquad + \,
\int_0^1 \frac{dx}{x} \,
\bigg[
\frac{x\,\big(1-x+\rho\big)^{\eps}}{(1-x)^{1+2\eps}}
\bigg]_+ \,
f(x)
\, .
\qquad\;
\eeq
Once combined with the $z$ integration, this allows to write 
\beq
{\cal J}_{\rm rad}^{\sFm\sI}(s,m)
& = &
f_i^g \,
I_{\rm s}^{\sFm\sI}(s,m)
+
f_i^g \,
\int_0^1 \frac{dx}{x} \,
J_{\rm s}^{\sFm\sI}(s,x,m)
\, ,
\eeq
where 
\beq
I_{\rm s}^{\sFm\sI}(s,m)
& = &
\Norm
\frac{(4 \pi)^{-2+\eps}s^{-\eps}}{\Gamma(1-\eps)} \,
\frac{\Gamma(-\eps)\Gamma(2-\eps)}{\Gamma(2-2\eps)} \,
\frac{\rho^\eps\,\Gamma(-2\eps)}{\Gamma(2-2\eps)} \,
_2F_1\bigg(-\eps,-2\eps,2-2\eps,-\frac{1}{\rho}\bigg)
\, ,
\nnb\\
J_{\rm s}^{\sFm\sI}(s,x,m)
& = &
\Norm
\frac{(4 \pi)^{-2+\eps}s^{-\eps}}{\Gamma(1-\eps)} \,
\frac{\Gamma(-\eps)\Gamma(2-\eps)}{\Gamma(2-2\eps)} \,
\bigg[
\frac{x\,\big(1-x+\rho\big)^{\eps}}{(1-x)^{1+2\eps}}
\bigg]_+
\, .
\eeq
These integrals are related to the ones in \eq{eq:intSoftMI} by the following identifications:
\beq
&&
I_{\rm s,\sFm\sI}^{ikl} 
\, = \,
f_i^g \,
I_{\rm s}^{\sFm\sI}\big(\sk{kl}{ikl},m_k\big)
\, ,
\qquad
J_{\rm s,\sFm\sI}^{ikl}(x) 
\, = \, 
f_i^g \,
J_{\rm s}^{\sFm\sI}\big(\sk{kl}{ikl},x,m_k\big)
\, .
\eeq
Expanding in $\eps$, we finally obtain
\beq
\label{eq: IsMI}
I_{\rm s}^{_{\sFm\sI}}(s,m)
& = &
\frac{\as}{2\pi}
\left( \frac{s}{\mu^2}\right)^{-\eps}
\bigg[\,
\frac{1}{2\eps^2}
+
\frac{1}{2\eps}
(3 + \ln\rho)
-
\Li_2(-\rho)
-
\frac{\ln^2\rho}{4}
\\
&&
\hspace{21mm}
+ \,
(1+\rho)\ln(1+\rho)
+
\bigg(\frac12-\rho\bigg)\ln\rho
-
\frac{7}{4}\,\zeta_2
+
3
+
{\cal O}(\eps)
\bigg]
\, ,
\qquad
\nnb\\
J_{\rm s}^{_{\sFm\sI}}(s,x,m)
& = &
\frac{\as}{2\pi}
\left( \frac{s}{\mu^2}\right)^{-\eps}
\bigg\{
\bigg( - \frac{1}{\eps} - 1 \bigg)
\bigg[ \frac{x}{1-x} \bigg]_+
-
\bigg[
\frac{x}{1-x}
\ln\frac{1-x+\rho}{(1-x)^2}
\bigg]_+
+ {\cal O}(\eps)
\bigg\}
\, ,
\nnb
\eeq
with $\rho=m^2/s$.


\subsection{Integration of hard-collinear counterterms with massive recoiler}
\label{app: IhcFmFandI}

The integrals $I_{\rm hc,\sFm}^{A}$ with $A=\zg, \, \og, \, \tg$ have been introduced in \eq{eq:IHCandJHCdecomp} within the flavour decomposition of the integrated final-state hard-collinear counterterms. They are defined as
\beq
I_{\rm hc,\sFm}^{A}\big(\bs,m_r\big)
& = &
\Norm \, 
\int d\Phi_{\rm rad}^{(ijr)} \,
\frac{P_{ij, \sF}^{{\rm hc},A}(z)}{s_{ij}}
\\
& = &
\Norm \, 
\frac{(4 \pi)^{-2+\eps}}{\Gamma(1-\eps)} \,
\bs^{\,-\eps} 
\int_0^{y_+} dy \,
y^{-1-\eps} \, (1-y)^{1-2\eps}
\nnb\\
&&
\qquad
\times \,
\int_{z_-(y)}^{z_+(y)} dz \, 
(z_+ - z)^{-\eps} (z - z_{-})^{-\eps} \,
P_{ij, \sF}^{{\rm hc},A}(z)
\, ,
\nnb
\eeq
where $P_{ij, \sF}^{{\rm hc},A}(z)$ are the hard-collinear kernels of \eq{eq: PhcF}, and 
\beq
&&
y_+
\, = \,
1
+
\frac{2m_r^2}{\bs}
-
\frac{2m_r \sqrt{\bs + m_r^2}}{\bs}
\, ,
\qquad
z_{\pm}(y)
\, = \,
\frac{\bs(1-y) \pm \sigma}{2\bs(1-y)}
\, ,
\nnb\\
&&
\sigma
\, = \,
\sqrt{\bs^2(1 - y)^2 - 4m_r^2 \bs y}
\, ,
\qquad
 \bs
 \, = \,
 \sk{jr}{ijr}
\, .
\eeq 
The single $1/\eps$ poles of these integrals are generated by the behaviour of the integrands at $y=0$. We then add and subtract this endpoint contribution.
Considering that $z_-(0)=0$ and $z_+(0)=1$, we get
\beq
I_{\rm hc,\sFm}^{A}(\bs,m_r)
& = &
\frac{\as}{2\pi}
\left( \frac{\bs}{\mu^2}\right)^{-\eps} 
\int_0^{y_+} dy \, y^{-1-\eps} \, 
\int_{0}^{1} dz \, 
z^{-\eps} \,(1-z)^{-\eps} \,
P_{ij, \sF}^{{\rm hc},A}(z)
\\
&&
+ \,
\frac{\as}{2\pi} \,
\int_0^{y_+} \frac{dy}{y} \,
\bigg[
(1-y) \int_{z_-(y)}^{z_+(y)} dz \, P_{ij, \rm F}^{{\rm hc},A}(z)
-
\int_{0}^{1} dz \, P_{ij, \rm F}^{{\rm hc},A}(z)
\bigg]_{\eps=0}
+
{\cal O}(\eps)
\, .
\nnb
\eeq
Inserting the explicit expressions for $P_{ij, \rm F}^{{\rm hc},A}(z)$ from \eq{eq: PhcF}, and using the relations
\beq
z_+ - z_-
\, = \,
\frac{\sigma}{\bs(1-y)}
\, ,
\qquad
z_+ + z_-
\, = \,
1
\, ,
\qquad
z_+ z_- 
\, = \,
\frac{m_r^2\,y}{\bs(1-y)^2}
\, ,
\qquad
\eeq
we first perform the integrations over $z$, obtaining
\beq
I_{\rm hc,\sFm}^{\zg}(\bs,m_r)
& = &
\frac{\as}{2\pi}
\left( \frac{\bs}{\mu^2}\right)^{-\eps} 
\frac23 \,
T_R \, 
\bigg\{
- 
\frac{y_+^{-\eps}}{\eps} 
\bigg(1 +\frac53\eps \bigg)
\nnb\\
&&
\qquad\qquad
+ \,
\int_0^{y_+} dy \,
\bigg[
\frac{1}{y} \,
\bigg( \frac{\sigma}{\bs} - 1 \bigg)
-
\frac{m_r^2 \sigma}{\bs^2(1 - y)^2}
\bigg]
+
{\cal O}(\eps)
\bigg\}
,
\nnb\\
I_{\rm hc,\sFm}^{\og}(\bs,m_r)
& = &
\frac{\as}{2\pi}
\left( \frac{\bs}{\mu^2}\right)^{-\eps} 
\frac12 \,
C_F \, 
\bigg\{
- 
\frac{y_+^{-\eps}}{\eps}
(1+\eps)
+
\int_0^{y_+} \frac{dy}{y} \,
\bigg( \frac{\sigma}{\bs} - 1 \bigg)
+
{\cal O}(\eps)
\bigg\}
\, ,
\nnb\\
I_{\rm hc,\sFm}^{\tg}(\bs,m_r)
& = &
\frac{\as}{2\pi}
\left( \frac{\bs}{\mu^2}\right)^{-\eps} 
\frac13 \,
C_A \, 
\bigg\{
- 
\frac{y_+^{-\eps}}{\eps}
\bigg(1 + \frac53\eps \bigg)
\nnb\\
&&
\qquad\qquad
+ \,
\int_0^{y_+} dy \,
\bigg[
\frac{1}{y} \,
\bigg( \frac{\sigma}{\bs} - 1 \bigg)
+
\frac{2\,m_r^2 \sigma}{\bs^2(1 - y)^2}
\bigg]
+
{\cal O}(\eps)
\bigg\}
\, .
\eeq
In order to perform the $y$ integration we introduce the change of variables
\beq
t
\, = \,
\sqrt{\frac{y_1-y}{y_2-y}}
\, ,
\qquad
y_{1,2}
\, = \,
\frac{(Q \mp m_r)^2}{\bs}
\, ,
\qquad
\sigma
\, = \,
\bs\,\sqrt{(y_1-y)(y_2-y)}
\, ,
\eeq
where $y_{1}=y_+$ and $y_2$ are the roots of the $\sigma$ radicand.
The integration becomes straightforward, and in terms of $\beta=\sqrt{y_1/y_2}=y_1$ we obtain
\beq
\int_0^{y_+} dy \, 
\frac{\sigma}{\bs(1-y)^2}
& = &
\ln\frac{1+\beta}{1-\beta}
+
\frac{1-\beta}{\sqrt{\beta}}\,\arctan(\sqrt{\beta})
-
1
\, ,
\nnb\\
\int_0^{y_+} \frac{dy}{y} \,
\bigg(
\frac{\sigma}{\bs}
-
1
\bigg)
& = &
\frac{(1-\beta)^2}{2\beta}
\ln\frac{1-\beta}{2}
-
\frac{(1+\beta)^2}{2\beta}
\ln\frac{1+\beta}{2}
-
1
\, .
\eeq
The final results, written in terms of $\rho
\, \equiv \, m_r^2/\bs \, = \, (1-\beta)^2/4\beta$, read
\beq
\label{eq: IhcM}
I_{\rm hc,\sFm}^{\zg}(\bs,m_r)
& = &
\frac{\as}{2\pi}
\left( \frac{\bs}{\mu^2}\right)^{\!\!-\eps} 
\frac23 \,
T_R \, 
\bigg\{
- 
\frac{1}{\eps} 
-
\frac83
+
\rho
+
\frac32\,\rho\ln\rho
-
\bigg(1+\frac32\rho\bigg)\ln(1+\rho)
\nnb\\
&&
\hspace{32mm}
- \,
2\,\rho^{3/2} \arctan\!\Big(\sqrt{1+\rho} - \sqrt{\rho}\Big)
\bigg]
+
{\cal O}(\eps)
\bigg\}
,
\nnb\\
I_{\rm hc,\sFm}^{\og}(\bs,m_r)
& = &
\frac{\as}{2\pi}
\left( \frac{\bs}{\mu^2}\right)^{\!\!-\eps} 
\frac12 \,
C_F \, 
\bigg\{
- 
\frac{1}{\eps} 
-
2
+
\rho\ln\rho
-
(1+\rho)\ln(1+\rho)
+
{\cal O}(\eps)
\bigg\}
\, ,
\nnb\\
I_{\rm hc,\sFm}^{\tg}(\bs,m_r)
& = &
\frac{\as}{2\pi}
\left( \frac{\bs}{\mu^2}\right)^{\!\!-\eps} 
\frac13 \,
C_A \, 
\bigg\{
- 
\frac{1}{\eps} 
-
\frac83
-
2\,\rho
-
\ln(1+\rho)
\nnb\\
&&
\hspace{32mm}
+ \,
4\,\rho^{3/2} \arctan\!\Big(\sqrt{1+\rho} - \sqrt{\rho}\Big)
+
{\cal O}(\eps)
\bigg\}
\, .
\eeq

The integrals $J_{\rm hc,\sFm}^{B}$, with $B=\zg,\, \qg,\, \gq,\, \tg$ were as well introduced in \eq{eq:IHCandJHCdecomp}, and are defined as follows ($ac=ij,ji$):
\beq
\int_0^1 \frac{dx}{x} \,
J_{\rm hc,\sFm}^{B}\big(\bs,x,m_r\big)
& = &
\Norm \, 
\int \, d \Phi_{\rm rad}^{(arc)} \,
\frac{P_{[ac]a, \sI}^B(x)}{s_{ac}\,x}
\nnb\\
& = &
\Norm 
\frac{(4 \pi)^{-2+\eps}}{\Gamma(1-\eps)}
\int_0^1 \frac{dx}{x} \, \big[\bs(1 - x) + m_r^2\big]^{-\eps}
\nnb\\
&&
\qquad
\times \, 
\int_0^{z_+} dz \, (z_+ - z)^{-\eps} z^{-1-\eps} \,
P_{[ac]a, \sI}^B(x)
\, ,
\eeq
where the relevant hard-collinear kernels have been defined in \eq{eq: PhcI}, and
\beq
\bs
\, = \,
\sk{cr}{arc}
\, ,
\qquad \qquad
z_{+}
\, = \,
\frac{\bs(1-x)}{\bs(1-x)+m_r^2}
\, .
\eeq
As the kernels just depends on $x$, their integration over $z$ is trivial, and yields
\beq
J_{\rm hc, \sFm}^{B}\big(\bs,x,m_r\big)
& = &
\frac{\as}{2\pi}
\left( \frac{\bs}{\mu^2}\right)^{\!\!-\eps} 
\bigg[
- \,
\frac{1}{\eps}\,
(1-x)^{-2\eps} \,
(1\!-\!x\!+\!\rho)^{\eps}
P_{[ac]a, \sI}^B(x)
+
{\cal O}(\eps)
\bigg]
\, ,
\qquad
\eeq
where $\rho = m_r^2/\bs$.
Upon expanding in $\eps$ and inserting the explicit expressions for the kernels, we immediately get 
\beq
\label{eq: JhcM}
J_{\rm hc,\sFm}^{\zg}(\bs,x,m_r)
& = &
\frac{\as}{2\pi}
\left( \frac{\bs}{\mu^2}\right)^{-\eps}
T_R \,
\bigg\{
\big[ 1\!-\!2x(1\!-\!x) \big]
\bigg[
- 
\frac{1}{\eps}
-
\ln\frac{1-x+\rho}{(1-x)^2}
\bigg]
\nnb\\
&&
\qquad\qquad\qquad
+ \,
2x(1-x)
+
{\cal O}(\eps)
\bigg\}
,
\\
J_{\rm hc,\sFm}^{\qg}(\bs,x,m_r)
& = &
\frac{\as}{2\pi}
\left( \frac{\bs}{\mu^2}\right)^{-\eps} 
C_F (1-x)
\bigg[
- \,
\frac{1}{\eps}\,
-
\ln\frac{1-x+\rho}{(1-x)^2}
+
1
+
{\cal O}(\eps)
\bigg]
\, ,
\nnb\\
J_{\rm hc,\sFm}^{\gq}(\bs,x,m_r)
& = &
\frac{\as}{2\pi}
\left( \frac{\bs}{\mu^2}\right)^{-\eps} 
C_F \,
\bigg\{
\frac{1+(1-x)^2}{x}
\bigg[
- \,
\frac{1}{\eps}\,
-
\ln\frac{1-x+\rho}{(1-x)^2}
\bigg]
+
x
+
{\cal O}(\eps)
\bigg\}
\, ,
\nnb\\
J_{\rm hc,\sFm}^{\tg}(\bs,x,m_r)
& = &
\frac{\as}{2\pi}
\left( \frac{\bs}{\mu^2}\right)^{-\eps} 
2 \, C_A \,  
\bigg[ \frac{1-x}{x} + x(1-x) \bigg]
\bigg[
- \,
\frac{1}{\eps}\,
-
\ln\frac{1-x+\rho}{(1-x)^2}
+
{\cal O}(\eps)
\bigg]
\, .
\nnb
\eeq


\section{Details of NNLO Local Analytic Sector Subtraction}
\label{app:NNLOnum}

In this appendix we collect a selection of formulae and definitions relevant to the NNLO subtraction algorithm outlined in \cite{Bertolotti:2022aih}. This constitutes the analytic foundation for the numerical IR cancellation test discussed in Section \ref{sec:massless NNLO}. 

The 3-particle symmetrised sector function $\Z{ijk}$ is defined as
\beq
\label{eq:Z}
\Z{ijk} 
& = &
\frac{\sum_{a,b,c \, \in {\rm perm}(i,j,k)}\big(\sigma_{abbc}+\sigma_{abcb}\big)}\sigma
\, ,
\eeq
where ${\rm perm}(i,j,k)$ are the 6 permutations of indices $i,j,k$, while the $\sigma_{abcd}$ functions are written in terms of the $\sigma_{ab}$ quantities introduced in Eq.~\eqref{eq:Zofsigma}, as
\beq
\sigma_{abcd} 
\; = \;
(\sigma_{{ab}})^{\alpha} \,
\left(
\frac{\sigma_{cd}\, \sigma_{ac}}{\sigma_{ac} + \delta_{bc} \, \sigma_{ca}}
\right)
\, , 
\qquad 
\sigma
\, = \,
\sum_{\substack{a, b \neq a}} 
\sum_{\substack{c \neq a \\ d \neq a, c}} \sigma_{abcd} 
\, ,
\qquad
\alpha > 1 
\, .
\eeq

The counterterms relevant for our case study, introduced in Eq.~\eqref{eq:NNLOCTs}, are based on the following improved limits
\beq
\label{eq:appNNLOCTs}
\bbC{ij} \,
RR \; \Z{\{ijk\}}
& = &
\Norm \frac{P_{ij(r)}^{\mu\nu}}{s_{ij}} \, 
\bR_{\mu\nu}^{(ijr)} \,
\bar{\mc{Z}}^{(ijr)}_{jk}
\, ,
\nnb
\\
\bbS{ij} \,
RR \; \Z{\{ijk\}}
& = &
\frac{\Norm^{\,2}}{2} \!\!
\sum_{\substack{c \neq i,j \\ d \neq i,j,c}} \!\!
\mc E^{(ij)}_{cd}
\bB_{cd}^{(ijcd)} \,
\frac{\sigma_{ijjk}+\sigma_{ikjk}+\sigma_{jiik}+\sigma_{jkik}}
{
\sum_{a \neq i}\sum_{b \neq i,j} \sigma_{iajb}
+ 
\sum_{a\neq j}\sum_{b\neq j,i} \sigma_{jaib}
}
\, ,
\nnb\\
\bbHC{ijk} \,
RR \; \Z{\{ijk\}}
& = &
\bbC{ijk}
\left(1 - \bbS{ij} \right)
RR \; \Z{\{ijk\}}
\nnb\\
& = &
\frac{\Norm^{\,2}}{s_{ijk}^2} \,
P_{ijk(r)}^{\mu\nu} \,\bB_{\mu\nu}^{(ijkr)}
+
\Norm^{\,2} \,
C_{f_k} \,
\mc E_{kr}^{(ij)} \,
\bB^{(ijkr)}
\, ,
\nnb
\\
\bbC{ij} \,
\bbS{ij} \,
RR \; \Z{\{ijk\}}
& = &
- \,
\Norm^{2} \!\!\!\!
\sum_{\substack{c \neq i,j \\ d \neq i,j,c}} \!\!\!
\Bigg\{ \!
\frac{P_{ij(r)\!}}{s_{ij}} \,
\bar{\mc E}^{(j)(ijr)}_{cd} \!
\nnb\\
&&
\hspace{1cm}
+
\frac{Q_{ij(r)\!}^{\mu\nu}}{s_{ij}}
\Bigg[\!
\frac{\kk{c,\mu}{ijr}}{\sk{jc}{ijr}}
-
\frac{\kk{d,\mu}{ijr}}{\sk{jd}{ijr}}
\!\Bigg] \!
\Bigg[\!
\frac{\kk{c,\nu}{ijr}}{\sk{jc}{ijr}}
-
\frac{\kk{d,\nu}{ijr}}{\sk{jd}{ijr}}
\!\Bigg]
\Bigg\}
\bB_{cd}^{(ijr,jcd)} \,
\bZ{{\rm s},\,jk}^{\{ijr\}}
\, ,
\nnb\\
\bbC{ij} \,
\bbHC{ijk} \,
RR \; \Z{\{ijk\}}
& = &
\bbC{ij} \,
\bbC{ijk}
\left(1 - \bbS{ij} \right)
RR \; \Z{\{ijk\}}
\nnb\\
& = &
\Norm^{\,2} 
\Bigg\{ \;
\frac{P_{ij(r)}}{s_{ij}}
\frac{\bar P_{jk(r)}^{(ijr)\mu\nu}}{\sk{jk}{ijr}}
\bB^{(ijr,jkr)}_{\mu\nu}
- 
2 \, 
C_{f_k} \, 
\bar{\mc E}^{(j)(ijr)}_{kr} \,
\frac{P_{ij(r)}}{s_{ij}} \,
\bB^{(ijr,jkr)}
\Bigg\}
\, .
\nnb \\
\eeq
The above formulae are derived from the general definitions provided in Appendices (C.1) and (C.3) of Ref.~{\cite{Bertolotti:2022aih}}, considering the flavour content $\{ijkr\} = \{q \bar{q} q' \bar{q}'\}$. Further details on the variables involved in these expressions can be found in those appendices.\\

Next, we detail the expected cancellation pattern for the singular configurations collected in Eq.~\eqref{eq:conrel}.
\begin{itemize}
\item Limit $\bC{ij}$
\beq
\label{eq:c_in_pieces}
\bC{ij} \, ( 1 - \bbC{ij}) \, \RR \, \Z{ijk}
& \to & 
\mbox{integrable}
\, ,
\nnb\\
\bC{ij} \, (\bbS{ij} - \bbC{ij} \, \bbS{ij}) \, \RR \, \Z{ijk}
& \to & 
\mbox{integrable}
\, ,
\nnb\\
\bC{ij} \, ( \bbHC{ijk} 
- \bbC{ij} \, \bbHC{ijk} ) \, \RR \, \Z{ijk}
& \to & 
\mbox{integrable} \, .
\eeq
\item Limit $\bS{ij}$
\beq
\label{eq:ss_in_pieces}
\bS{ij} \, ( 1 -  \bbS{ij} ) \, \RR \, \Z{ijk}
& \to & 
\mbox{integrable}
\, ,
\nnb\\
\bS{ij} \, ( \bbC{ij} - \bbC{ij} \, \bbS{ij} ) \, \RR \, \Z{ijk}
& \to & 
\mbox{integrable}
\, ,
\nnb\\
\bS{ij} \, \bbHC{ijk} \, \RR \, \Z{ijk}
& \to & 
\mbox{integrable}
\, ,
\nnb\\
\bS{ij} \, \bbC{ij} \, \bbHC{ijk} \, \RR \, \Z{ijk}
& \to & 
\mbox{integrable}
\, .
\eeq
\item Limit $\bC{ijk}$
\beq
\label{eq:cc_in_pieces}
\bC{ijk} \, (1 - \bbHC{ijk} - \bbS{ij}) \, \RR \, \Z{ijk}
& \to & 
\mbox{integrable}
\, ,
\nnb\\
\bC{ijk} \, \Big( \bbC{ij} - \bbC{ij} \, \bbS{ij}
- \bbC{ij} \,
\bbHC{ijk}
\Big) \, \RR \, \Z{ijk}
& \to & 
\mbox{integrable}
\, .
\eeq
\end{itemize}

Finally, we give an explicit representation of the double-radiative phase space employed in the numerical tests. This is obtained by applying a nested Catani-Seymour mapping $(ijr,jkr)$, in which the first mapping $(ijr)$ naturally describes a $[ij]\to i+j$ splitting, while the second mapping $(jkr)$ is suited for a $[ijk]\to [ij]+k$ radiation.
In four dimensions we have
\beq
\label{eq:para}
\int d\Phi_{\rm rad,2}^{(ijr,jkr)} 
& = & 
\frac{\big(\bar s^{(ijr,jkr)} \big)^2}
{16\, \pi^3} 
\int_0^{\pi} d\phi
\int_0^1 dy
\int_0^1 dz 
\int_0^1 dw'  
\int_0^1 dy' 
\int_0^1 dz' 
\nnb\\
&&
\qquad\qquad\qquad
\times \, 
\Big[ w' (1- w')\Big]^{-\frac{1}{2}}   
\, (1-y') \, y \, (1-y)
\, .
\eeq
The kinematic variables parametrising the phase space are defined as
\beq
\label{eq:zyzy}
&&
z' 
\, = \,
\frac{s_{ir}}{s_{ir}+s_{jr}}
\, ,
\qquad\qquad
y' 
\, = \,
\frac{s_{ij}}{s_{ij}+s_{ir}+s_{jr}}
\, ,
\nnb\\
&&
z 
\, = \,
\frac{\bs^{\,(ijr)}_{jr}}{\bs^{\,(ijr)}_{jr}\!+\!\bs^{\,(ijr)}_{kr}}
\, ,
\qquad\qquad
y
\, = \,
\frac{\bs^{\,(ijr)}_{jk}}{\bs^{\,(ijr)}_{jk}\!+\!\bs^{\,(ijr)}_{jr}\!+\!\bs^{\,(ijr)}_{kr}}
\, .
\eeq
Further information can be found  in Appendix A.3.1 of Ref.~{\cite{Magnea:2020trj}}.

\newpage
\bibliographystyle{JHEP}
\bibliography{madnklo}


\end{document}